\documentclass[a4paper,11pt]{article}
\usepackage{jheppubMod}
\usepackage{amsmath,amsfonts,amssymb,array}
\usepackage{bm,bbm,bbm,bbold}
\usepackage{float}
\usepackage[T1]{fontenc}
\usepackage{dsfont}
\usepackage{enumerate}
\usepackage{float}
\usepackage{graphicx}
\usepackage{lineno,lipsum}
\usepackage{multirow,multicol}
\usepackage{subfigure,slashed}
\usepackage{tabu}
\usepackage{datetime}
\usepackage{ulem}
\usepackage{booktabs}
\graphicspath{{Plots/}}

\hypersetup{bookmarks=true,unicode=true,pdftoolbar=true,pdfmenubar=true,
  pdffitwindow=false,pdfstartview={FitH},
  pdftitle={The Effective Vector Boson Approximation in High-Energy Muon Collisions -- Ruiz, et al JHEP 2022, 114 (2022) [arXiv:2111.02442]},pdfauthor={Names},
  pdfsubject={Subject},pdfcreator={Creator},pdfproducer={Producer},
  pdfkeywords={Muon Colliders}{Effective W Approximation}{Electroweak Boson PDFs}{Helicity Polarization},
  pdfnewwindow=true,colorlinks=true
}

\renewcommand{\arraystretch}{1.2}

\renewcommand{\arraystretch}{1.2}
\newcolumntype{x}[1]{
{\centering\hspace{0pt}}p{#1}}

\newcommand{\GeV}{{\rm ~GeV}}
\newcommand{\TeV}{{\rm ~TeV}}

\newcommand{\fb}{{\rm ~fb}}

\newcommand{\vareps}{\varepsilon}

\newcommand{\mgFull}{\texttt{MadGraph5\_aMC@NLO}}
\newcommand{\mgamc}{\texttt{mg5amc}}

\newcommand{\mpmm}{\mu^+\mu^-}
\newcommand{\orcid}[1]{\,\href{https://orcid.org/#1}{\includegraphics[width=9pt]{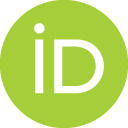}}}
\newcommand{\orcidRR}{0000-0002-3316-2175} 
\newcommand{\orcidAC}{0000-0002-0430-7152} 
\newcommand{\orcidFM}{0000-0003-4890-0676} 

\newcommand{\confirm}[1]{{\color{black}#1}}
\title{\Large  The Effective Vector Boson Approximation\\ in High-Energy Muon Collisions}

\author[a]{Richard Ruiz\orcid{\orcidRR},}
\author[b]{Antonio Costantini\orcid{\orcidAC},}
\author[b,c]{Fabio Maltoni\orcid{\orcidFM},}
\author[b]{and Olivier Mattelaer}

\affiliation[a]{Institute of Nuclear Physics -- Polish Academy of Sciences {\rm (IFJ PAN)},\\ ul. Radzikowskiego, 31-342, Krak{\'o}w, Poland}

\affiliation[b]{Centre for Cosmology, Particle Physics and Phenomenology {\rm (CP3)},\\
Universit\'e Catholique de Louvain, Chemin du Cyclotron, B-1348 Louvain la Neuve, Belgium}

\affiliation[c]{Dipartimento di Fisica e Astronomia, Universit\`a di Bologna e INFN,\\ Sezione di Bologna, via Irnerio 46, I-40126 Bologna, Italy}

\emailAdd{rruiz@ifj.edu.pl}
\emailAdd{antonio.costantini@uclouvain.be}
\emailAdd{fabio.maltoni@uclouvain.be}
\emailAdd{olivier.mattelaer@uclouvain.be}

\abstract{
Due to the inclination for forward gauge radiation, lepton colliders beyond a few TeV are effectively electroweak (EW) boson colliders, suggesting the treatment of EW bosons as constituents of high-energy leptons. In the context of a muon collider, we revisit the validity of $W$ and $Z$ parton distribution functions (PDFs) at leading order in $2\to n$ process. We systematically investigate universal and quasi-universal power-law and logarithmic corrections that arise when deriving (polarized) weak boson PDFs in the collinear limit. We go on to survey a multitude of $2\to n$ processes at $\sqrt{s}=2-30$ TeV via polarized and unpolarized EW boson fusion/scattering. To conduct this study, we report a public implementation of the Effective $W/Z$ and Weizs\"acker-Williams Approximations, which we collectively call the Effective Vector Boson Approximation, into the Monte Carlo event generator \texttt{MadGraph5\_aMC@NLO}. This implementation lays the groundwork for developing matrix-element matching prescriptions involving EW parton showers and renormalized EW PDFs. To further with this agenda, we give recommendations on using $W/Z$ PDFs.
}

\preprint{CP3-21-59, IFJPAN-IV-2021-17, MCNET-21-13, VBSCAN-PUB-07-21}
\arxivnumber{2111.02442}
\begin{document}
\setcounter{page}{1}
\maketitle
\setcounter{page}{2}
\flushbottom


\section{Introduction}\label{sec:intro}

\begin{figure}[!t]
\centering
\includegraphics[width=.6\textwidth]{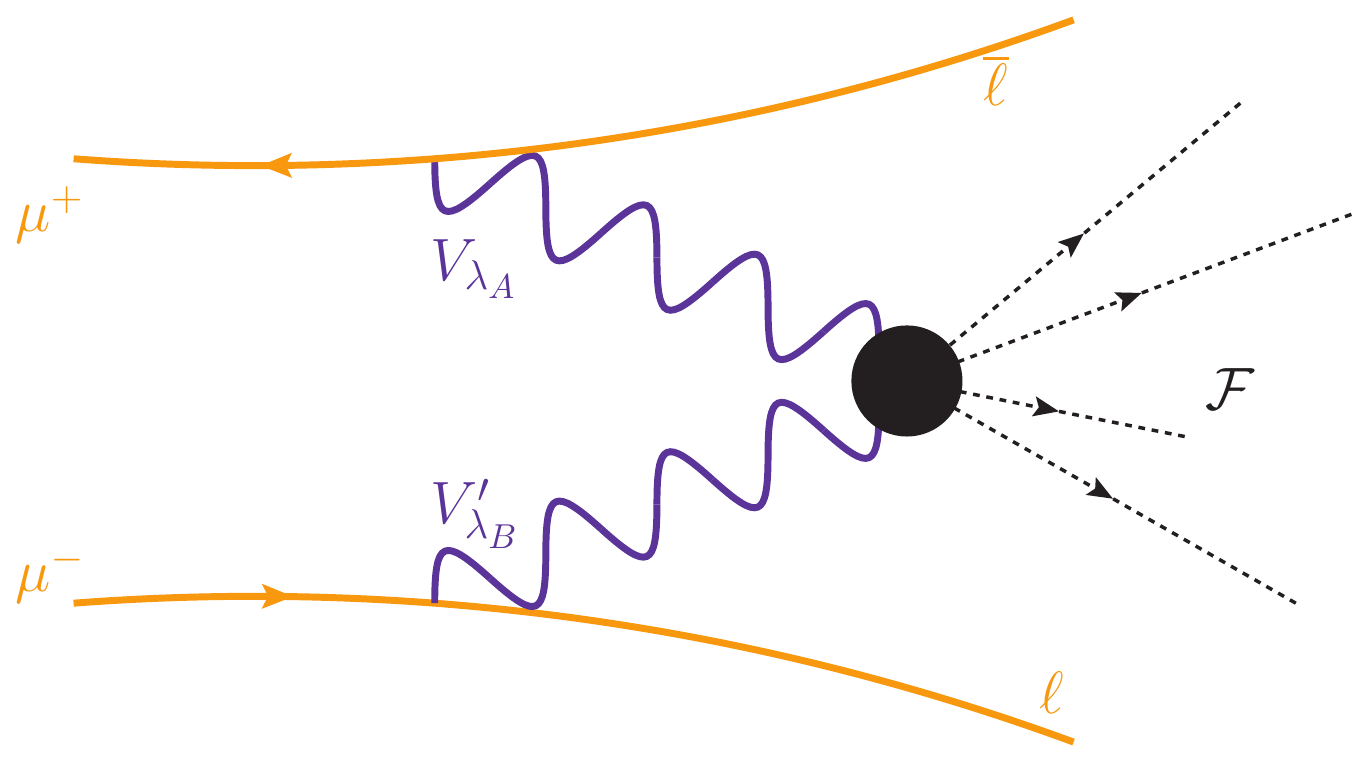}
\caption{Diagrammatic representation of $\mu\to V_\lambda \ell$ splitting and $V V'$ scattering in $\mpmm$ collisions.} 
\label{fig:diagram_MuCo_VBSX}
\end{figure}

Within the Standard Model (SM) of particle physics,  the existence of gauge bosons with nonzero masses is one of the defining characteristics that distinguishes the electroweak (EW) sector from perturbative quantum chromodynamics (pQCD). However, at momentum transfers scales $(Q)$ far above the EW breaking scale, $v=\sqrt{2}\langle \Phi\rangle\approx246\GeV$, weak bosons are effectively massless, thereby softening this distinction. More precisely, at $Q^2\gg M_V^2$, where $M_V=M_W,~M_Z$ are the $W$ and $Z$ boson masses,  process-dependent, power-law terms that scale as {$\delta\sigma\sim(M_V^{2k}/Q^{2k+2})$, with $k>1$,}  become negligible  in $2\to n$ scattering processes and analogously $1\to n'$ decay processes. Consequentially, at sufficiently energetic collider experiments, collinear, $t$-channel emissions of weak bosons from initial-state partons, as shown schematically in Fig.~\ref{fig:diagram_MuCo_VBSX}, can be factorized into a type of weak boson parton distribution
function\footnote{Similarly, $s$-channel splittings of massive weak bosons from final-state partons can be factorized into a type of weak boson fragmentation function. For details, see Refs.~\cite{Nagy:2007ty,Krauss:2014yaa,Hook:2014rka,Chen:2016wkt} and references therein.}
(PDF), and be modeled as \textit{almost massless}, on-shell, initial-state constituents of high-energy lepton and hadron beams~\cite{Dawson:1984gx,Kane:1984bb,Kunszt:1987tk}.

Known as the Effective $W/Z$ Approximation (EWA)~\cite{Dawson:1984gx,Kane:1984bb}, the partitioning of collinear, initial-state $W/Z$ boson emissions out of matrix elements (MEs) and into PDFs has several benefits. Like heavy quark factorization~\cite{Witten:1975bh,Barnett:1987jw,Olness:1987ep,Collins:1998rz,Aivazis:1993kh,Aivazis:1993pi,Dawson:2014pea,Han:2014nja} and the factorization of inelastic photons, i.e., the Weizs\"acker-Williams Approximation~\cite{vonWeizsacker:1934nji,Williams:1934ad}, the EWA significantly simplifies ME computations and phase space integration, particularly in the infrared limits of phase space.
Even in the absence of singularities, such factorization may be necessary to avoid numerical instabilities in real calculations when scale hierarchies are present, e.g., to avoid a large collinear logarithm when $Q^2\gg M_V^2$. Since its inception, the approximation has been used to model numerous scenarios, including weak vector boson fusion/scattering (VBF)~\cite{Duncan:1985vj,Dicus:1987ez,Altarelli:1987ue,Hagiwara:2009wt,
Brehmer:2014pka}, 
heavy quark production from $Wg$-scattering~\cite{Willenbrock:1986cr,Dawson:1986tc}, 
and heavy lepton production~\cite{Dawson:1986tc}.

The EWA, however, also comes at a cost. Like other instances of collinear factorization, invoking the EWA means losing knowledge about:
(i) the recoil kinematics/transverse momentum $(p_T)$ of partons associated with the emission of initial-state weak bosons, and which scale as $\mathcal{O}(M_V^2/Q^2)$ and as $\mathcal{O}(p_T^2/Q^2)$; (ii) the interference between initial-state weak boson polarizations, which scale as $\mathcal{O}(M_V^2/Q^2)$~\cite{Dawson:1984gx};
and (iii) the interference between different EW mass eigenstates, i.e., $\gamma_T/Z_T$ mixing, 
which can have large, $\mathcal{O}(1)$ effects~\cite{Ciafaloni:2005fm,Nagy:2007ty,Chen:2016wkt,Han:2020uid}.
However, in principle, extending highly successful matching and merging techniques pioneered for QCD and QED~\cite{Catani:2001cc,Krauss:2002up,Mangano:2006rw,Lavesson:2008ah} offer a starting path to resolve some of these drawbacks.

Notably, momentum transfers needed to neglect power corrections of the form $(p_T^2/Q^2)^k$ and $(M_V^2/Q^2)^k$, where $k>0$,  are already obtainable with VBF at the LHC~\cite{Aaboud:2017pdi,Sirunyan:2019der,Aaboud:2019nmv,Sirunyan:2019hqb}. Such scales will also be commonplace at energy-frontier colliders proposed to succeed the LHC~\cite{Arkani-Hamed:2015vfh,CEPC-SPPCStudyGroup:2015csa,Abada:2019lih,Benedikt:2018csr,Abada:2019ono,EuropeanStrategyforParticlePhysicsPreparatoryGroup:2019qin,EuropeanStrategyGroup:2020pow}. This rings particularly true for multi-TeV $\mpmm$ colliders. There, the VBF rate is so dominant the collider acts effectively as a ``high-luminosity EW boson collider''~\cite{Costantini:2020stv} and shows promising sensitivity to SM and beyond the SM (BSM) physics~\cite{Buttazzo:2018qqp,Ruhdorfer:2019utl,Han:2020uid,Chiesa:2020awd,Costantini:2020stv,Capdevilla:2020qel,Long:2020wfp,Han:2020pif,Han:2020uak,Buttazzo:2020uzc,Yin:2020afe,Capdevilla:2021rwo,Liu:2021jyc,Bottaro:2021srh,Han:2021udl,Han:2021kes,AlAli:2021let,Franceschini:2021aqd,Buarque:2021dji,Yang:2021zak,Han:2021lnp}.

By virtue of the EW sector obeying a local SU$(2)_L\otimes$U$(1)_Y$ symmetry in the unbroken phase,
some aspect of collinear factorization must survive at these colliders in the limit that 
$(p_T^2/Q^2)^k\rightarrow0$
and
$(M_V^2/Q^2)^k\rightarrow0$.
This is evident from the formulations of factorization theorems and Sudakov exponentiation in 
QED and pQCD, which at times rely more on the presence of multiple, well-separated (hierarchical) mass scales than on being unbroken gauge theories~\cite{Sudakov:1954sw,Yennie:1961ad,Collins:1984kg,Collins:1985ue,Sterman:1986aj,Catani:1989ne,Collins:1989gx,Contopanagos:1996nh,Collins:2011zzd}.
Clearly, being an Abelian/non-Abelian or weakly/strongly coupled theory is less crucial for sufficiently inclusive processes. At the same time, differences between collinear factorization in pQCD and the EW theory must exist since lepton and hadron beams are not composed of weak isospin-averaged states.

More specifically, the fact that muons carry EW quantum numbers implies that their collisions do not represent an inclusive summation over all initial-state weak isospin charges. (This would require $\mu-\nu_\mu$ and $\nu_\mu-\nu_\mu$ beams.) As a result, infrared logarithms beyond lowest order in perturbation theory do not fully cancel, leading to violations of the Block-Nordsieck Theorem~\cite{Ciafaloni:1998xg,Ciafaloni:1999ub,Ciafaloni:2000df,Ciafaloni:2000rp,Bell:2010gi,Manohar:2014vxa,Chen:2016wkt}.
The analogy in pQCD is the violation of the Collinear Factorization Theorem at three-loops when applied to exclusive hadronic final states, e.g., dijet production~\cite{Catani:2011st,Collins:2011zzd,Forshaw:2012bi}. 
However, despite this violation, application of the Collinear Factorization Theorem, which is presently only proved for a handful of processes~\cite{Collins:1984kg,Collins:1985ue,Collins:2011zzd}, to arbitrary processes remains a quantitatively successful paradigm. Motivated by this success, we consider whether such a paradigm can also work for high-energy lepton collisions.

As a step to better understanding collinear factorization in the EW sector
and to further explore the ability of the EWA to predict total and differential cross sections, we consider a framework that combines the EWA for helicity-polarized $W$ and $Z$ bosons with the Weizs\"acker-Williams Approximation
for helicity-polarized photons. We collectively label this the Effective Vector Boson Approximation (EVA).\footnote{Throughout this text, we use the term ``EVA'' when speaking generically about unpolarized or polarized EW boson PDFs, but use ``EWA'' when speaking exclusively about (un)polarized  $W/Z$ PDFs.}
In this framework and in the context of a multi-TeV $\mpmm$ collider, 
we investigate the impact of and validity of (helicity-polarized) $\gamma/W/Z$ PDFs in $2\to n$ process.
To focus on the role of partonic kinematics, we restrict ourselves to leading order (LO) matrix elements and bare, i.e., unrenormalized $\gamma/W/Z$ PDFs, which are finite at LO.
Processes that we consider include: associated and many-Higgs production, many-boson production, as well as associated and multi-top production. We extend recent studies~\cite{Borel:2012by,Brehmer:2014pka,Cuomo:2019siu,Costantini:2020stv,Han:2020uid} by investigating universal and quasi-universal corrections to weak boson PDFs that appear naturally in their derivations. Specifically, we study universal power corrections of the form  $(p_T^2/Q^2)$, which spoil the accuracy of collinear factorization, and quasi-universal power corrections of the form $(M_V^2/Q^2)$, which spoil  the accuracy of the Goldstone Equivalence Theorem~\cite{Lee:1977yc,Lee:1977eg}. Importantly, we also consider the role of universal and quasi-universal logarithmic corrections of the form $\delta\sigma/\sigma\sim\mathcal{O}[\log(\mu_f^2/M_V^2)]$, by exploring scale variation and when the evolution variable in weak boson PDFs is defined in terms of transverse momentum $(p_T)$ or virtuality $(q)$. This is in addition to studying the role of helicity in both total and differential cross sections.  
We note that this study is complementary to extensive studies on uncertainties of the EWA~\cite{Accomando:2006mc,Borel:2012by,Cuomo:2019siu}.

We find strong sensitivity to power corrections when hard-scattering scales $Q$ are below $Q\sim1$ TeV; for larger $Q$, we report agreement between full and approximated MEs when scale uncertainty bands, which can be large, are taken into account. More explicitly: we find that computations with the EWA can reproduce total and differential results within (large) scale uncertainties, so long as factorization-breaking power corrections are sufficiently suppressed. Even for asymptotically large energies, we find scale uncertainties remain large, demonstrating a need for renormalization group (RG) evolution in our factorization theorem for high-energy muon collisions. To strengthen the parallels with pQCD, we give a proof-of-principle demonstration of matrix element matching with transverse weak boson PDFs and full MEs. Given these criteria, we go on to survey nearly two dozen $2\to n$ VBF processes with the EVA in $\mu^+\mu^-$ collisions at $\sqrt{s}=2-30$ TeV. Cross sections and their scale uncertainties are presented for both helicity-polarized and unpolarized initial states. To conduct this study, we report an implementation of the EVA into the general-purpose Monte Carlo event generator {\mgFull} ({\mgamc}). Notably, this public implementation lays the groundwork for developing QCD-like matching prescriptions with initial/final state EW boson radiation as well as (polarized and unpolarized) PDFs that are RG evolved via the EW theory and pQCD. To further with this agenda, we also give some recommendations on using weak boson PDFs in high-energy lepton collisions.

The remainder of this work continues in the following order:
In Sec.~\ref{sec:formalism}, we summarize the EVA formalism used throughout this work and present a formula for EW boson scattering in high-energy $\mpmm$ collisions.  
In Sec.~\ref{sec:setup} we document our computational setup and numerical values for SM inputs. Sec.~\ref{sec:validity} is the first of two principle sections and where we revisit the validity of the EWA. Sec.~\ref{sec:leptons} is the second of two principle sections and where we give a survey of $2\to n$ VBF processes in the EVA. We conclude in Sec.~\ref{sec:conclusions}. There, we give an extended discussion of our findings, reflecting particularly on the parallels we find with more subtle aspects of PDFs in QCD, e.g., the phenomenon of ``precocious scaling.'' Finally, we provide some recommendations on using weak PDFs in high-energy lepton collisions in Sec.~\ref{sec:conclusions_recs}. App.~\ref{app:madgraph} provides some instructions for reproducing our results and using (un)polarized EW boson PDFs in {\mgamc}. 
For completeness, a derivation of helicity-polarized EW boson PDFs at LO is given in App.~\ref{app:formalism}.


\section{The Effective Vector Boson Approximation for $\mu^+\mu^-$ collisions}\label{sec:formalism}

In this section, we summarize the EVA, i.e., the framework in which we work, and its use in evaluating scattering cross sections in many-TeV $\mu^+\mu^-$ collisions. While we focus on muons, the EVA is, in principle, applicable to any lepton-lepton, lepton-hadron, and hadron-hadron collider configuration. Extension to other colliders, however, may require substitutions of gauge coupling charges and/or convolutions with additional PDFs~\cite{Dawson:1984gx}. 
In Sec.~\ref{sec:formalism_theorem}, we state a scattering formula that will be the basis for all our numerical results and validation checks.
In Sec.~\ref{sec:formalism_pdfs}, we list the $q^2$ and $p_T^2$-evolved collinear PDFs that describe the density of EW bosons in muons at LO.
Finally, we document for the completeness in Sec.~\ref{sec:formalism_nu} the PDFs for SM neutrinos from muons.

\subsection{A scattering formula for $\mpmm$ collisions}\label{sec:formalism_theorem}

To described the fully differential production of an $n$-body, final state $\mathcal{F}$ with momenta $\{p_f\}$ via the high-energy VBF process $ V_{\lambda_A}V'_{\lambda_B} \to \mathcal{F}$, where $V_{\lambda_A}$ and $V'_{\lambda_B}$ are helicity-polarized EW gauge bosons, in $\mu^+\mu^-$ collisions at a center-of-mass (c.m.) energy of $\sqrt{s}$, we invoke the EVA. In practice, this means working from a scattering formula given by
\begin{align}
\sigma(\mu^+ \mu^- \to \mathcal{F} + X)
&=~
\tilde{f} \otimes \tilde{f} \otimes \hat{\sigma}
+ {\rm Power~and~Logarithmic~Corrections}
\\
&=~
\sum_{V_{\lambda_A},V'_{\lambda_B}}
\int_{\tau_0}^1 d\xi_1
\int_{\tau_0/\xi_1}^1 d\xi_2
\int dPS_{n} ~ 
\nonumber\\
& ~ \times ~
\tilde{f}_{V_{\lambda_A}/\mu^+}(\xi_1,\mu_f)
\tilde{f}_{V'_{\lambda_B}/\mu^-}(\xi_2,\mu_f)
\nonumber \\ 
& ~\times~ \frac{d\hat{\sigma}(V_{\lambda_A}V'_{\lambda_B} \to \mathcal{F})}{dPS_{n}}
\nonumber\\
&
~ + ~ \mathcal{O}\left(\frac{p_{T,l_k}^2}{M_{VV'}^2}\right)
+ \mathcal{O}\left(\frac{M_{V_k}^2}{M_{VV'}^2}\right)
+\mathcal{O}\left(\log  \frac{\mu_f^2}{M_{V_k}^2}\right)
.
\label{eq:factTheorem}
\end{align}

Here, $\sigma$ is the muon-level (beam-level) inclusive cross section for the production of $\mathcal{F}$ in association with an arbitrary state $X$. Explicitly, $X$ consists of at least two leptons $l$, where $l=\mu^\pm, \nu_\mu,$ or $\overline{\nu_\mu}$, in addition to particles originating from radiative corrections. The summation runs over all polarized EW boson $V_{\lambda} \in \{W^\pm_\lambda, Z_\lambda, \gamma_\lambda\}$,  with $\lambda\in\{0,\pm1\}$. Formally speaking, when the collection of states $\{V_{\lambda}\}$ is extended to left-handed (LH) and right-handed (RH) states $\nu_{\mu L}$ and $(\overline{\nu_{\mu}})_R$, the beam remnant $X$ includes weak bosons.

For beams $k=$1,2, the quantities $\tilde{f}_{V_\lambda/\mu^\pm}(\xi_k,\mu_f)$ are the bare PDFs that describe the likelihood that an unpolarized muon $\mu^\pm$ with energy $E_\mu = \sqrt{s}/2$ and momentum $p_{\mu} = E_\mu (1,0,0,\pm1)$ contains a ``parton'' $V$ with helicity $\lambda$, mass $M_V$, energy $E_V = \xi_k E_\mu$, and no transverse momentum $p_{T,V_\lambda}$. 
Following Ref.~\cite{Maltoni:2012pa}, we adopt the $\tilde{f}$ notation to stress that the PDFs in Eq.~\eqref{eq:factTheorem} are not resummed. 
The $\tilde{f}$ are related to resummed PDFs $f$ by
\begin{align}
f_{V_\lambda}(\xi,\mu_f)\ 
 =\ 
 \tilde{f}_{V_\lambda}(\xi,\mu_f)\ +\ \mathcal{O}\left((\alpha_W(\mu_f)\right)\ .
\end{align}
Generally, $E_V \neq E_{V'}$ in the frame of the $(VV')$-system since generally $M_V\neq M_V'$. 
In $\tilde{f}_{V_\lambda/\mu^\pm}$, the quantity $\mu_f$ is the collinear factorization scale and acts as the ultraviolet regulator of the bare PDF.
Physically, $\mu_f$ is the phase space upper bound on the norm of the space-like momentum transfer $q = (p_\mu - p_l)$ carried by $V_\lambda(q)$; alternatively, $\mu_f$ can be interpreted as the upper bound on the $p_T$ of lepton $l$ in $\mu^\pm \to V_\lambda + l$ splitting. The (phase space) integrals over the momentum fractions $\xi_k$ are bounded by the (dimensionless) kinematic threshold variable 
$\tau_0 = \min(M_{VV'}^2/s) = \min(M(\mathcal{F})/s)$. For $M_{VV'}< \sqrt{\tau_0 s}$, the $(VV')$-system has insufficient energy to produce the $n$-body state $\mathcal{F}$. 
The separately Lorentz-invariant phase space measure $dPS_{n}$ is given by the usual expression
\begin{equation}
 dPS_{n}(p_A+p_B; \{p_f\}) = (2\pi)^4\delta^4\left(p_A + p_B - \sum_{p_j\in\{p_f\}}^n p_j\right)\prod_{p_j\in\{p_f\}}^n \frac{d^3 p_j}{(2\pi)^3 2E_j}.
 \label{eq:phaseSpaceDef}
\end{equation}

In Eq.~\eqref{eq:factTheorem}, $d\hat{\sigma}/dPS_n$ is the totally differential, ``parton-level'' cross section for the hard-scattering process $ V_{\lambda_A}(p_A) ~ V'_{\lambda_B}(p_B) \to \mathcal{F}(\{p_f\})$,  which occurs at a hard scale
\begin{equation}
Q \equiv M_{VV'} =\sqrt{(p_A+p_B)^2} 
= \sqrt{\xi_1\xi_2 s} \geq \sqrt{\tau_0 s} \gg M_V.
 \end{equation}
Due to this equality, we use the terms ``hard-scattering system'' and ``$(VV')$-system'' interchangeably. The helicity-polarized cross sections can be computed from the formula~\cite{BuarqueFranzosi:2019boy}
\begin{equation}
 \frac{d\hat{\sigma}(V_{\lambda_A}V'_{\lambda_B} \to \mathcal{F})}{dPS_{n}} = 
 \frac{1}{2\lambda^{1/2}(Q^2,M_{V}^2,M_{V'}^2)}
 \sum_{\rm dof}\vert \mathcal{M}( V_{\lambda_A} V'_{\lambda_B} \to \mathcal{F}) \vert^2.
 \label{eq:partonXSec}
\end{equation}
Here, $\lambda(x,y,z)= (x-y-z)^2 - 4yz,$ is the usual K\"allen kinematic function that accounts for the masses of initial-state particles. Unlike traditional leading-twist approximations that neglect masses of initial-state partons, the ME $\mathcal{M}( V_{\lambda_A} V'_{\lambda_B} \to \mathcal{F})$ is evaluated with nonzero $M_V, M_{V'}$.
(In none of our results are weak boson masses set to zero.)
Moreover, unlike the scattering of unpolarized partons in unpolarized beams, no spin-averaging factor for initial-states $V_{\lambda_A} V'_{\lambda_B}$ is needed for helicity-polarized cross sections that are paired with PDFs for helicity-polarized partons. 
The summation in Eq.~\eqref{eq:partonXSec} runs over all discrete degrees of freedom (dof) related to $\mathcal{F}$, e.g., electric charge and color helicity multiplicities. Importantly, if the summations $\sum_{V_{\lambda_A},V'_{\lambda_B}}$ and $\sum_{\rm dof}$ do not run over all helicity polarizations for $VV'$ and $\mathcal{F}$, respectively, then the square of $\mathcal{M}$ is not Lorentz invariant. In such cases, an infrared-safe reference frame must be specified to define the helicities. For further details on evaluating helicity-polarized cross sections, particularly in relation to PDFs for polarized partons and polarized parton showers, see Ref.~\cite{BuarqueFranzosi:2019boy}.

Implicit in Eq.~\eqref{eq:factTheorem} is a restriction on the phase space integration measure $dPS_n$.
The purpose of this restriction is regulate $\mathcal{M}$ and render it meaningful.
For example: the ME for the process $\gamma\gamma \to q\overline{q}$, where $q$ is a massless quark, diverges without phase space cuts on $t$-channel momenta. 
Cuts should also ensure that $s$-, $t$-, and $u$-channel invariants in the $V_{\lambda_A} V'_{\lambda_B} \to \mathcal{F}$ hard process are comparable to one another and to the hard scale $M_{VV'}$. In principle, this means that logarithms $(L)$ of ratios of these invariants, which appear in hard-scattering cross sections, are never numerically large.
In practice, we vary phase space cuts to explore the growth of these logarithms in initial-state $\mu^\pm\to V_\lambda l$ splitting (see Sec.~\ref{sec:validity_evo}).  While the MEs for all the processes that we investigate are regulated, we set looser phase space cuts on final-state kinematics to balance computational demands. For some processes, logarithm can grow as large as $\mathcal{O}(5-10)$, and therefore remain within perturbative limits in the sense $\alpha_W\times L \ll 1$. We have checked (see Sec.~\ref{sec:validity_evo}) that tighter phase space cuts do not qualitatively change our findings.

When deriving Eq.~\eqref{eq:factTheorem}, a number of assumptions are made.
Two important ones are both related to enforcing large separations of scales in $ V_{\lambda_A} V'_{\lambda_B}$ scattering.
The first is that weak bosons are massive but that the invariant mass of the $(V_{\lambda_A} V'_{\lambda_B})$-system is much larger, i.e., $M_V \ll M_{VV'}$.
Nonzero $M_V$ for $V=W,Z$ ensure that their longitudinal polarization vectors, which scale as $\vareps^\mu(p_V,\lambda=0)\sim p_V^\mu/M_V + \mathcal{O}(M_V/E_V)$, remain non-vanishing when contracted with $\mu\to l$ currents. We reiterate that including initial-state parton masses here differs from typical treatments of QCD partons in hadron collisions, which are assumed massless in the absence of specialized schemes~\cite{Aivazis:1993kh,Aivazis:1993pi,Dawson:2014pea,Han:2014nja}. Outside this limit, Eq.~\eqref{eq:factTheorem} receives quasi-universal power corrections of the form $\delta\sigma\sim(M_V^{2k}/M_{VV'}^{2k+2})^k$ for $k>1$, the size of which are quantified in Sec.~\ref{sec:validity_mass}. The qualifier ``quasi-universal'' refers to the fact that such corrections originate from the derivation of $\tilde{f}_{V_0^\pm}$ PDFs, and therefore appear for any $V_{\lambda_A}V'_{\lambda_V}$ scattering process with at least one longitudinally polarized $W^\pm_0$ or $Z_0$.
(Specifically, they come from expanding the ME for $\mu\to l+V_0$ splitting.)
It is worth noting that the Goldstone Equivalence Theorem requires that these terms be small~\cite{Lee:1977yc,Lee:1977eg}; for further insights on relationship, see Refs.~\cite{Chen:2016wkt,Chen:2017ekt,Chen:2019dkx,Cuomo:2019siu}.

The second important assumption is the stipulation that EW bosons are emitted at shallow angles in $\mu\to l + V_\lambda$ splittings, i.e., $p_{T,l}\sim p_{T,V_\lambda} \ll  M_{VV'}$. This is a standard but necessary condition for collinear factorization in gauge theories~\cite{Peskin:1995ev,Collins:2011zzd}. As in QCD computations, universal power corrections of the form $\delta\sigma\sim(p_{T,l}^{2k}/M_{VV'}^{2k+2})$ for $k>1$ can be incorporated by higher-order perturbative computations, e.g., next-to-leading order (NLO) in $\alpha_W$ or $\alpha$, parton showers, or ME matching to higher leg multiplicities (see Sec.~\ref{sec:validity_matching}).
To be explicit, ``universal'' here refers to the fact that such corrections originate from the derivation of both $\tilde{f}_{V_0^\pm}$ and $\tilde{f}_{V_T^\pm}$, meaning that they are present for any $V_{\lambda_A}V'_{\lambda_V}$ scattering process.
(Specifically, they come from expanding the ME for $\mu\to l+V_\lambda$ splitting.)

In its present form, Eq.~\eqref{eq:factTheorem} is subject to universal and quasi-universal logarithmic corrections of the form $\delta\sigma/\sigma \sim \mathcal{O}[\log  (\mu_f^2/M_{V_k}^2)]$, where $g_W=\sqrt{4\pi\alpha_W}\approx0.65$ is the SM weak coupling constant and $\mu_f$ has the physical interpretation as described above Eq.~\eqref{eq:phaseSpaceDef}.
Na\"ively, one may argue that these corrections are sub-leading since they are coupling suppressed. However, $g_W$ is not a small number and collinear logarithms can compensate for this. For instance: taking $\mu_f=1-10\TeV$ implies corrections of $\delta\sigma\sim(g_W^2/4\pi)\log(\mu_f^2/M_W^2)$ that are $\mathcal{O}(20\%-30\%)$.
While we ultimately report in Sec.~\ref{sec:validity} a prescription for obtaining agreement between full and EWA-based calculations, the uncertainties associated with choosing $\mu_f$ reported there and in Sec.~\ref{sec:leptons} undercut our findings. 

Since Eq.~\eqref{eq:factTheorem} is only a LO expression, and therefore does not resum any logarithms, the only (quasi-)universal logarithms that we study are those coming from the $\mu^\pm \to V_T l$ splittings themselves.
For precision computations, an RG-improved version EVA with renormalized PDFs $f_{V_\lambda^\pm}$, running couplings, and an EW Sudakov form factor are necessary. 
Equation \eqref{eq:factTheorem} is written such that renormalized PDFs can be incorporated by the  replacement: $\tilde{f}_{V_\lambda^\pm}(\xi,\mu_f) \to f_{V_\lambda^\pm}(\xi,\mu_f').$
(Implicit in this replacement is that $\mu_f'$ in $f(\xi,\mu_f')$ is acting as a phase space cutoff and the RGE scale.)
 Were we to replace $\tilde{f}_{V_\lambda^\pm}$ with their renormalized versions, then the absence of a Sudakov factor still implies 
that the scattering formula is not scale invariant in an RG evolution sense. That is to say, the anomalous dimensions associated with $f_{V_{\lambda_A}/\mu^\pm}$ and $f_{V_{\lambda_B}/\mu^\mp}$ do not necessarily cancel those associated with a renormalized partonic cross section $\hat{\sigma}^R(V_{\lambda_A}V_{\lambda_B}\to \mathcal{F})$.
Sudakov factors can be incorporated following the classic treatment of Ref.~\cite{Contopanagos:1996nh} or modern treatments like Ref.~\cite{Pagani:2021vyk}. However, investigating and quantifying the impact of these improvements as well as those related to $\gamma_T/Z_T$ mixing~\cite{Chen:2016wkt,Bauer:2017isx,Fornal:2018znf,Bauer:2018xag,Bauer:2018arx,Han:2020uid,Han:2021kes}, which we also neglect, is left to future work.

\subsection{$q^2$ and $p_T^2$-evolved collinear EW PDFs}\label{sec:formalism_pdfs}

The expressions for EW boson PDFs $\tilde{f}_{V_{\lambda_A}/\mu^\pm}$ depend strongly on their precise formulation; compare for example Refs.~\cite{Dawson:1984gx,Kane:1984bb,Kunszt:1987tk,Altarelli:1987ue,Barger:1987nn,Kuss:1996ww,Bernreuther:2015llj}. As discussed in Sec.~\ref{sec:validity_evo}, seemingly innocuous conceptual differences can lead to substantial numerical differences in real computations. Therefore, we now summarize the PDFs used in this study.

\begin{table}[!t]
\begin{center}
\resizebox{\textwidth}{!}{
\renewcommand*{\arraystretch}{0.95}
\begin{tabular}{c|c|c|c|c|c}
\hline\hline
\multirow{2}{*}{Vertex} & Coupling  & \multirow{2}{*}{$g_R^f$}  & \multirow{2}{*}{$g_L^f$}  & \multirow{2}{*}{$g_V^f$}  & \multirow{2}{*}{$g_A^f$} \\
	& strength	&	& &  & \\
\hline\hline
$V-f-f'$	& $\tilde{g}$				& $(g_V^f+g_A^f)$		& $(g_V^f-g_A^f)$		& $\frac{(g_R+g_L)}{2}$ 	& $\frac{(g_R-g_L)}{2}$ \\
$\gamma-f-f$	& $e Q^f$				& $1$				& $1$				& $1$				& $0$\\
$Z-f-f$	    & $\frac{g}{\cos\theta_W}$	& $-Q^f\sin^2\theta_W$	& $(T_3^f)_L-Q^f\sin^2\theta_W$				& $\frac{1}{2}(T_3^f)_L-Q^f\sin^2\theta_W$ 			& $-\frac{1}{2}(T_3^f)_L$ \\
$W-f-f'$	& $\frac{g}{\sqrt{2}}$			& $0$				& $1$				& $\frac{1}{2}$ 			& $-\frac{1}{2}$ \\
\hline\hline
\end{tabular}
}
\caption{
EW chiral couplings and coupling strength normalizations used in the EVA for fermions $f,f'$ with weak isospin charge $(T_3^f)_L=\pm1/2$ and electric charge $Q^f$, with normalization $Q^\ell=-1$.
}
\label{tab:ewa_coup}
\end{center}
\end{table}

In PDFs for $W_\lambda, Z_\lambda, \gamma_\lambda$ bosons from high-energy muons, one has the freedom to parameterize the momentum transfer in $\mu(p_\mu)\to l(p_l) + V_\lambda(q)$ splittings either by the squared virtuality $q^2 = (p_\mu-p_l)^2 < 0$ propagated by $V_\lambda$, or by the squared transverse momentum $p_T^2$ carried away by $l$. While the two quantities are related by $q^2(1-\xi) = -p_T^2$, where $\xi = E_V / E_\mu$ is the fraction of $\mu$'s energy held by $V_\lambda$, 
the resulting PDF sets for transversely polarized $V_\lambda$ differ analytically. 
Consequentially, for fixed $z$, $\lambda$, and $\mu_f$, one can obtain large differences due a relative contribution that scale as $\delta \tilde{f}_{V_T/    \mu^\pm}\sim\log(1-\xi)$.
This logarithm diverges in the large-$\xi$ limit and corresponds to a nonzero $q^2$ but a vanishing $p_T^2$. Such differences have been sporadically discussed throughout the literature~\cite{Duncan:1985vj,Dawson:1986tc,Willenbrock:1986cr,Dicus:1987ez,Altarelli:1987ue,Kuss:1995yv,Kuss:1996ww,Borel:2012by} but not systematically compared. In light of this, we investigate both sets of PDFs.

For the couplings in Table~\ref{tab:ewa_coup}, and assuming $q^2$ evolution, the LO PDFs for polarized $V_\lambda\in\{W_\lambda^\pm,Z_\lambda\}$ from LH $(\tilde{f}_L)$ and RH $(\tilde{f}_R)$ fermions in the hard scattering frame are 
\begin{subequations}
\label{eq:formalism_pdfs_q2}
\begin{align}
\tilde{f}_{V_+/f_L}(\xi,\mu_f^2) 	&= \frac{g_V^2}{4\pi^2} \frac{g_L^2(1-\xi)^2}{2\xi} \log \left[\frac{\mu_f^2}{M_V^2}\right], \\
\tilde{f}_{V_-/f_L}(\xi,\mu_f^2) 	&= \frac{g_V^2}{4\pi^2} \frac{g_L^2}{2\xi} \log \left[\frac{\mu_f^2}{M_V^2}\right], \\
\tilde{f}_{V_0 / f_L}(\xi,\mu_f^2) 	&= \frac{g_V^2}{4\pi^2} \frac{g_L^2(1-\xi)}{\xi},\\  
\tilde{f}_{V_+/f_R}(\xi,\mu_f^2) 	&= \left(\frac{g_R}{g_L}\right)^2 \times \tilde{f}_{V_-/f_L}(\xi,\mu_f^2), \\
\tilde{f}_{V_-/f_R}(\xi,\mu_f^2) 	&= \left(\frac{g_R}{g_L}\right)^2 \times \tilde{f}_{V_+/f_L}(\xi,\mu_f^2), \\
\tilde{f}_{V_0/f_R}(\xi,\mu_f^2) 	&= \left(\frac{g_R}{g_L}\right)^2 \times \tilde{f}_{V_0/f_L}(\xi,\mu_f^2). 
\end{align}
\end{subequations}

Choosing instead to integrate over $p_T^2$ leads analogously to the following PDFs for $V_\lambda$:
\begin{subequations}
\label{eq:formalism_pdfs_pT}
\begin{align}
\tilde{h}_{V_+/f_L}(\xi,\mu_f^2) 	&= \frac{g_V^2}{4\pi^2} \frac{g_L^2(1-\xi)^2}{2\xi} \log \left[\frac{\mu_f^2}{(1-\xi)M_V^2}\right],  \\
\tilde{h}_{V_-/f_L}(\xi,\mu_f^2) 	&= \frac{g_V^2}{4\pi^2} \frac{g_L^2}{2\xi} \log \left[\frac{\mu_f^2}{(1-\xi)M_V^2}\right], \\
\tilde{h}_{V_0 / f_L}(\xi,\mu_f^2) 	&= \frac{g_V^2}{4\pi^2} \frac{g_L^2(1-\xi)}{\xi},\\  
\tilde{h}_{V_+/f_R}(\xi,\mu_f^2) 	&= \left(\frac{g_R}{g_L}\right)^2 \times \tilde{h}_{V_-/f_L}(\xi,\mu_f^2), \\
\tilde{h}_{V_-/f_R}(\xi,\mu_f^2) 	&= \left(\frac{g_R}{g_L}\right)^2 \times \tilde{h}_{V_+/f_L}(\xi,\mu_f^2), \\
\tilde{h}_{V_0/f_R}(\xi,\mu_f^2) 	&= \left(\frac{g_R}{g_L}\right)^2 \times \tilde{h}_{V_0/f_L}(\xi,\mu_f^2). 
\end{align}
\end{subequations}
To obtain the LO PDF for $\gamma_\lambda$ from polarized muons in either evolution scheme, one must make the replacement $M_{V}\to m_\mu$ in the $\tilde{f}_{V_T}$ PDFs and neglect the $\tilde{f}_{V_0}$ PDF.
Given a scheme, we construct polarized EW boson PDFs for \textit{unpolarized} muon beams, denoted by $\tilde{f}_{V_\lambda/\mu^\pm}$, from those PDFs for \textit{polarized} muons, denote by $\tilde{f}_{V_\lambda/\mu^\pm_\lambda}$, through the relation
\begin{equation}
 \tilde{f}_{V_\lambda/\mu^\pm}(\xi,\mu_f) =
 \cfrac{\tilde{f}_{V_\lambda/\mu^\pm_L}(\xi,\mu_f) + \tilde{f}_{V_\lambda/\mu^\pm_R}(\xi,\mu_f)}{2}.
\label{eq:pdfDef_unpolarizedMuon}
\end{equation}
As a technical note, both schemes are available in {\mgamc} (see App.~\ref{app:madgraph} for details) but stress that RG evolution of EW boson PDFs from leptons is not yet supported.

Differences between the two sets of PDFs appear only in the collinear logarithms for transversely polarized $V_\lambda$.
In this sense, the impact of $\log(1-\xi)$ corrections is process dependent and thus is labeled ``quasi-universal.''
The absence of scale evolution in PDFs for longitudinally polarized $V_\lambda$ is well-known and implies that traditional means of estimating scale uncertainty in pQCD, e.g., three-point scale variation, are not applicable to longitudinally polarized weak boson PDFs. 
In principal, one can obtain $\tilde{f}_{V_T/f_{L/R}}$ from $\tilde{h}_{V_T/f_{L/R}}$, or vice versa, with appropriately chosen $\mu_f$. To further highlight the parallels with pQCD, we note that absorbing factors of $(1-\xi)$ into factorization scales is common practice in Soft-Collinear Effective Field Theory (SCET)~\cite{Becher:2007ty,Becher:2014oda}. We reiterate that the PDFs here are only accurate to LO. This means that charge-flipping splittings such as $\mu^- \to \gamma^* \to \mu^+$ and $\mu^- \to \gamma^* \to W^+$, which appear first at NLO, are neglected.

\subsection{Collinear PDFs for SM neutrinos}\label{sec:formalism_nu}

We briefly note that the derivation of $W^\pm_\lambda$ PDFs in $\mu\to l + V_\lambda$ splitting also implies the existence of neutrino PDFs. As we are working in the SM, only massless, LH neutrinos (and RH antineutrinos) exist. Therefore, by probability conservation, the $\mu^-_L \to \nu_{\mu L}$ PDF at leading order accuracy when evolved by $q^2$ and $p_T^2$ are 
\begin{align}
 \tilde{f}_{\nu_{\mu L}/\mu^-_L}(\xi,\mu_f^2) 	&= \tilde{f}_{W_{\lambda=+}^-/\mu^-_L}\left((1-\xi),\mu_f^2\right) +  \tilde{f}_{W_{\lambda=-}^-/\mu^-_L}\left((1-\xi),\mu_f^2\right) 
 \\
 & = \frac{g^2}{16\pi^2} \left( \frac{1+\xi^2}{1-\xi} \right)\log \left[\frac{\mu_f^2}{M_V^2}\right],
\\ 
 \tilde{h}_{\nu_{\mu L}/\mu^-_L}(\xi,\mu_f^2) 	&= \tilde{h}_{W_{\lambda=+}^-/\mu^-_L}\left((1-\xi),\mu_f^2\right) +  \tilde{h}_{W_{\lambda=-}^-/\mu^-_L}\left((1-\xi),\mu_f^2\right) 
 \\
 &= \frac{g^2}{16\pi^2} \left( \frac{1+\xi^2}{1-\xi} \right)\log \left[\frac{\mu_f^2}{\xi M_V^2}\right].
\end{align}
As we are interested in VBF at $\mpmm$ colliders, we do consider further the role of neutrino PDFs from muon beams; for recent discussion on these PDFs, see Ref.~\cite{Han:2020uid}.
Moreover, while we have also implemented these PDFs into the public release of {\mgamc}, access to them is temporarily restricted due to the unregulated divergence at $\xi\to1$. Likewise, throughout this study, we neglect the importance of $\mu\to \mu$ PDFs due to the complication of soft/collinear photon emissions, which necessitates resummation~\cite{Greco:2016izi}; we refer readers to studies by Refs.~\cite{Greco:2016izi,Bertone:2019hks,Frixione:2019lga,Han:2021kes,Frixione:2021wzh,Frixione:2021zdp}.

\section{Computational Setup}\label{sec:setup}

In this section we summarize the computational framework used in this study. Here, we only document the Monte Carlo (MC) tool chain and its tuning. Details on the EVA itself and usage in {\mgamc} are documented in Sec.~\ref{sec:formalism}  and App.~\ref{app:madgraph}.

To simulate high-$p_T$ muon collisions, we employ a development release of version 3.3.0 of \mgFull~\cite{Stelzer:1994ta,Alwall:2014hca}. In this software suite, fully differential events are obtained from tree-level ME that are constructed~\cite{deAquino:2011ub} and evaluated~\cite{Maltoni:2002qb} using helicity amplitudes defined in the \texttt{HELAS} basis~\cite{Murayama:1992gi}, with QCD color algebra decomposed according to color flow~\cite{Maltoni:2002mq}. Helicity-polarized ME are obtained by truncating spin-averaging over initial-state states and/or spin-summing over final-state states~\cite{BuarqueFranzosi:2019boy}. Analysis of parton-level events is handled by \texttt{MadAnalysis5}~\cite{Conte:2012fm,Conte:2018vmg}.

\subsection*{Standard Model Inputs}
For all ME and PDFs, we take the following EW inputs and masses~\cite{ParticleDataGroup:2020ssz}
\begin{align}
M_W &= 80.419\GeV, \quad M_Z = 91.188\GeV, \quad G_F = 1.16639\times10^{-5}{\GeV}^2,
\\
m_H &= 125\GeV,     \quad m_t = 173\GeV, \quad  m_b = 4.7\GeV. 
\end{align}
This implies a QED coupling of $\alpha_{\rm QED}^{-1}(\mu_r=M_Z) \approx 132.507$. 
While we consistently modify EW couplings EW inputs are varied but we do not RG-evolve them. Importantly, we have structured {\mgamc} such that EW couplings and masses present in EW boson PDFs are set to those values stipulated in the \texttt{param\_card.dat} configuration file. Changes to EW inputs in this file are automatically propagated into EW boson PDFs. We reiterate that all ME and PDFs assume non-zero $W$ and $Z$ boson masses.
We use the light lepton masses 
\begin{align}
 m_e = 510.9989461\times10^{-6}{\GeV} \quad\text{and}\quad m_\mu = 105.6583745\times10^{-3}{\GeV}
 \label{eq:leptonMasses}
\end{align}
for the collinear logarithms contained in the $\gamma_\lambda$ PDFs. These masses are hard-coded into the $\gamma_\lambda$ PDFs and are independent of  \texttt{param\_card.dat}. While it is technically possible use massless $e^\pm/\mu^\pm$ in MEs, in this paper we choose to use massive leptons.


\section{EVA at high energies}\label{sec:validity}

A chief goal of factorization is to simplify in a systematic manner complicated, multi-scale MEs that describe many-body processes into a set of simpler, 1-to-2-scale MEs. In practice, this divide-and-conquer approach improves the efficiency and stability of numerical computations. Importantly, the formal  perturbative accuracy of factorized calculations  can also be improved through quasi-universal RGE methods, e.g., Sudakov resummation and DGLAP evolution. For the specific case of VBF in multi-TeV $\mpmm$ collisions, factorizing collinear $\mu\to V_\lambda l$ splittings into weak boson PDFs enables one to reorganize computations of an inherent 3-scale,  $2\to (n+2)$ scattering process  (the three scales being $M_V$, $p_T^l$ and $M_{VV'}$) into the product of two 2-scale computations ($M_{V}$ with $\mu_f\sim p_T^l$, and $M_V$ with $M_{VV'}$) involving process-independent PDFs and process-dependent $2\to n$ MEs.

As described in Sec.~\ref{sec:formalism_theorem}, 
the EWA is accurate up to universal and quasi-universal power corrections of the order  $\mathcal{O}(p_T^{l2}/M_{VV'}^2)$ and $\mathcal{O}(M_V^2/M_{VV'}^2)$, which originate from expanding the ME for transversely and longitudinally polarized weak bosons in $\mu\to V_\lambda l$ splittings,  as well as universal and quasi-universal logarithmic corrections of the order $\mathcal{O}[\log(\mu_f^2/M_{V}^2)]$, which stem from working at LO in the EW theory. In principle, both classes of corrections can be reduced via standard techniques, e.g., higher-order perturbative calculations and Sudakov resummation. In the absence of such improvements, however, there exist theoretical uncertainties in the formulation of weak PDFs that we now explore.
In Sec.~\ref{sec:validity_setup} we describe our common setup to study power-law and logarithmic corrections. 
In Sec.~\ref{sec:validity_scale} we describe how we quantify uncertainties associated with the cutoff scale $\mu_f$.
We then study $\mathcal{O}(M_V^2/M_{VV'}^2)$ corrections in Sec.~\ref{sec:validity_mass}, and the dependence on collider energy in Sec.~\ref{sec:validity_collider}. A subclass of $\mathcal{O}[\log(\mu_f^2/M_{V}^2)]$ corrections are then investigated in Sec.~\ref{sec:validity_evo}. Finally, in Sec.~\ref{sec:validity_matching}, we give a proof-of-principle demonstration of matrix element matching with transversely polarized weak boson PDFs and explore $\mathcal{O}(p_T^{l2}/M_{VV'}^2)$ corrections.


\subsection{Process choice and polarization decomposition}\label{sec:validity_setup}

To quantify uncertainties that stem from factorizing polarized EW bosons from initial-state $\mu^\pm \to V_\lambda l$ emissions into PDFs, we chose the two benchmark processes 
\begin{align}
e^+\mu^- \to H H \;  \bar\nu_e \nu_\mu\,
\quad\text{and}\quad
e^+\mu^- \to \gamma\gamma\gamma \;  \bar\nu_e \nu_\mu\,.
\label{eq:validity_setup_fullProc}
\end{align}
Following Ref.~\cite{Costantini:2020stv}, we work with $e^+\mu^-$ collisions in order to remove $s$-channel, $\mpmm$ annihilation diagrams in a gauge-invariant manner. As such channels have sizable contributions to inclusive cross sections, their removal helps isolate the VBF sub-processes. Under the EWA, these beam-level processes correspond to the partonic processes
\begin{align}
\sum_{\lambda_A,\lambda_B\in\{0,\pm1\}}  W_{\lambda_A}^+ W_{\lambda_B}^-    \to HH\,
\quad\text{and}\quad
\sum_{\lambda_A,\lambda_B\in\{0,\pm1\}}  W_{\lambda_A}^+ W_{\lambda_B}^-    \to \gamma\gamma\gamma\,.
\end{align}
In practice, we restrict ourselves  throughout this section to the EWA helicity configurations
\begin{align}
  W_{0}^+ W_{0}^-    \to HH\,
  \quad\text{and}\quad
\sum_{\lambda_A,\lambda_B\in\{\pm1\}}  W_{\lambda_A}^+ W_{\lambda_B}^-    \to \gamma\gamma\gamma\,.
\label{eq:validity_setup_evaProc}
\end{align}

\begin{table}[t!]
\begin{center}
\resizebox{\textwidth}{!}{ 
\renewcommand*{\arraystretch}{1.3}\setlength{\tabcolsep}{10pt}
	\begin{tabular}{r|c c c c c c}
		\hline \hline
		&\multicolumn{6}{c}{$\sigma_{\rm EWA}$ [fb] (Polarization Fraction)}\\
		&\multicolumn{2}{c}{$\sqrt{s}=4$ TeV}&\multicolumn{2}{c}{$\sqrt{s}=14$ TeV}&\multicolumn{2}{c}{$\sqrt{s}=30$ TeV}\\
        \hline
        $\sum_{\lambda_A,\lambda_B}W_{\lambda_A}^+ W_{\lambda_B}^-\to H H$  
                                    & $2.08~^{+4\%}_{-2\%}$ & $(-)$ 
                                    & $6.01~^{+2\%}_{-1\%}$ & $(-)$ 
                                    & $9.48~^{+2\%}_{-1\%}$ & $(-)$\\
        $W_0^+W_0^-\to H H$         & $2.03$ & $(97\%)$   
                                    & $5.91$ & $(98\%)$   
                                    & $9.16$ & $(97\%)$ \\
        $W_0^\pm W_T^\mp\to H H$    & $560\times10^{-6}~^{+75\%}_{-73\%}$   &  $(<0.5\%)$ 
                                    & $1.10\times10^{-3}~^{+72\%}_{-70\%}$  & $(<0.5\%)$ 
                                    & $1.44\times10^{-3}~^{+71\%}_{-70\%}$  & $(<0.5\%)$ \\
        $W_T^+W_T^-\to H H$         & $51.5\times10^{-3}~^{+140\%}_{-75\%}$ & $(2\%)$ 
                                    & $113\times10^{-3}~^{+130\%}_{-72\%}$  & $(2\%)$ 
                                    & $156\times10^{-3}~^{+130\%}_{-769\%}$ & $(2\%)$ \\
        \hline
		$\sum_{\lambda_A,\lambda_B}W_{\lambda_A}^+ W_{\lambda_B}^-\to\gamma\gamma\gamma$  
		                                            & $146\times10^{-3}~^{+93\%}_{-60\%}$ & $(-)$  
		                                            & $396\times10^{-3}~^{+76\%}_{-52\%}$  & $(-)$ 
		                                            & $519\times10^{-3}~^{+71\%}_{-50\%}$ & $(-)$ \\
        $W_0^+W_0^-\to \gamma\gamma\gamma$          & $894\times10^{-6}$ & $(0.6\%)$  
                                                    & $1.50\times10^{-3}$ & $(<0.5\%)$ 
                                                    & $1.70\times10^{-3}$ & $(<0.5\%)$ \\
        $W_0^\pm W_T^\mp\to \gamma\gamma\gamma$     & $3.56\times10^{-3}~^{+72\%}_{-63\%}$ & $(2\%)$ 
                                                    & $5.88\times10^{-3}~^{+64\%}_{-58\%}$ & $(2\%)$ 
                                                    & $6.55\times10^{-3}~^{+63\%}_{-57\%}$ & $(1\%)$ \\
        $W_T W_T\to\gamma\gamma\gamma$              & $141\times10^{-3}~^{+94\%}_{-60\%}$ & $(97\%)$ 
                                                    & $389\times10^{-3}~^{+76\%}_{-52\%}$ & $(98\%)$ 
                                                    & $510\times10^{-3}~^{+71\%}_{-50\%}$ & $(98\%)$ \\
		\hline
				&\multicolumn{6}{c}{$\sigma_{\rm Full}$ [fb]}\\
        \hline
		$e^+\mu^-\to HH\;\bar\nu_e\nu_\mu$                  & $1.26$              & 
		                                                    & $4.43$  &  
		                                                    & $9.60$ & \\
		$e^+\mu^-\to \gamma\gamma\gamma\;\bar\nu_e\nu_\mu$  & $248\cdot10^{-3}$   & 
		                                                    & $558\cdot10^{-3}$ & 
		                                                    & $4.04\cdot10^{-1}$ & \\
		\hline\hline
	\end{tabular}
	}
	\caption{
	Upper: Unpolarized and polarized, EWA-level cross sections [fb] for the process $W_{\lambda_A}^+ W_{\lambda_B}^-\to HH$, with scale uncertainties $[\%]$ and polarization fractions $[\%]$, in $e^+\mu^-$ collisions at $\sqrt{s}=4$, 14, and 30\TeV.
	Middle: The same but for the process $W_{\lambda_A}^+ W_{\lambda_B}^-\to\gamma\gamma\gamma$, assuming the phase space cuts of Eq.~\eqref{eq:validity_cuts}. 
	Lower: The cross sections for the analogous processes using the full $2\to4$ or $2\to5$ ME.
	No restrictions are applied to the invariant masses $M(HH)$ and $M(\gamma\gamma\gamma)$.}
	\label{tab:polarDecomp}
\end{center}
\end{table}

We consider these specific processes and configurations due to the high purity of helicity polarizations that drive them. For polarizations $\lambda_A$ and $\lambda_B$ defined in the $(W^+ W^-)$ frame, we find by explicit calculation~\cite{BuarqueFranzosi:2019boy} that \confirm{$97\%-99\%$} of $HH$ production in the EWA is dominated by longitudinally polarized $W^+W^-$ scattering, i.e., $(\lambda_A,\lambda_B)=(0,0)$,  for $\sqrt{s}=4-30\TeV$. In contrast, $\gamma\gamma\gamma$ production  is driven at the \confirm{$97\%-99\%$} level, albeit with a large scale uncertainty, by transversely polarized $W^+W^-$ scattering, i.e., $(\lambda_A,\lambda_B)=(T,T')$, where $T,T'=\pm1$, when assuming the following fiducial phase space cuts on photons
\begin{equation}
 p_T^\gamma>50\GeV,\quad \vert\eta^\gamma\vert<3, \quad\text{and}\quad \Delta R(\gamma,\gamma)>0.4.
 \label{eq:validity_cuts}
\end{equation}
In making this distinction between $(0,0)$ and $(T,T')$ configurations, we can showcase possible differences of the EWA as applied to longitudinal and transverse polarizations.
Many other processes, such as heavy Higgs production and top quark pair production, receive comparable contributions from multiple polarization configurations, which we believe can lead to ambiguities in interpreting the following comparisons.

\subsection{Defining scale uncertainties for unrenormalized $W_T/Z_T$ PDFs}\label{sec:validity_scale}

A key difference between the bare, LO PDFs in Sec.~\ref{sec:formalism_pdfs} and their renormalized variants is the definition of $\mu_f$. For renormalized PDFs, $\mu_f$ is the RGE scale generated through dimensional regularization; varying $\mu_f$ is a standard procedure for quantifying perturbative uncertainties in QCD predictions. 
In the present case of $\mu \to V_{\lambda}l$ splitting at LO, $\mu_f$ is literally a boundary on a phase space integral over either the virtuality $\sqrt{\vert q^2\vert}$ of $V_\lambda$,
if one uses Eq.~\eqref{eq:formalism_pdfs_q2},
or the transverse momentum $p_T^l$ of $l$, 
if one uses Eq.~\eqref{eq:formalism_pdfs_pT}.

For the PDFs of Sec.~\ref{sec:formalism_pdfs}, setting $\mu_f$ proportional to the $(VV')$ scattering scale $M_{VV'}$ is a natural choice as this attempts to captures the whole phase space in $VV'$ scattering~\cite{Dawson:1984gx,Barger:1987nn}.
However, much smaller choices are also favored. Integrating up to $p_T^l\sim \mu_f \propto M_{VV'}$ suggests a potential breakdown of the collinear approximation since 
one assumes $p_T^l = p_T^V \ll M_{VV'}$.
As discussed in Sec.~\ref{sec:validity_matching}, it is the wide-angle contribution of $\mu \to V_{\lambda}l$ splitting that coincides with the regime $p_T^l = p_T^V \sim M_{VV'}$.
Therefore, there is an ambiguity, or uncertainty, in the choice of $\mu_f$, and increasing or lowering $\mu_f$ corresponds to conjecturing how much phase space is actually captured by collinear kinematics. 
It is not guaranteed that arguments used in  hadron collisions to fix $\mu_f$, e.g., Refs.~\cite{Maltoni:2007tc,Degrande:2016aje}, are applicable here.
Furthermore, this uncertainty is only one part of the possible uncertainties of the EWA, as illustrated in the factorization formula of Eq.~\eqref{eq:factTheorem}. 
Exploring how such ambiguities relate to disagreements between the EWA and full matrix element computations is a reason for this study.

For the $\gamma\gamma\gamma$ process, we focus on PDF evolution by virtuality $(q^2)$ and set the baseline collinear factorization scale  to be half the partonic c.m.~energy, given by
\begin{equation}
 \mu_f = \zeta \frac{\sqrt{\hat{s}}}{2} = \zeta  \frac{M_{VV'}}{2}, \quad\text{with}\quad \zeta=1.
 \label{eq:scaleDefault}
\end{equation}
Three-point scale uncertainties for $W^\pm_T$ PDFs are obtained by varying $\zeta$ discretely over the range $\zeta\in\{0.5,1.0,2.0\}$.  
While our inspiration to use this procedure draws from common practices in QCD,
we reiterate that the physical interpretation is not the same as for renormalized PDFs in QCD.
There are also alternative ways to quantify uncertainties in the EWA~\cite{Accomando:2006mc,Borel:2012by,Cuomo:2019siu}.
For representative collider energies $\sqrt{s}=4,~14$, and 30\TeV, the beam-level cross sections under the EWA $(\sigma_{\rm EWA})$, scale uncertainties [\%],  and polarization fractions [\%] for $W_{\lambda_A}^+ W_{\lambda_B}^-\to HH$ and $\gamma\gamma\gamma$ are summarized in the top two panels of Table~\ref{tab:polarDecomp}. For comparison, we show in the lower panel of Table~\ref{tab:polarDecomp} the corresponding cross sections $(\sigma_{\rm Full})$ using the full MEs, i.e., without the EWA. The sizable differences between $\sigma_{\rm EVA}$ and $\sigma_{\rm Full}$, as well as the large scale uncertainties of the EWA result, will now be discussed.

\subsection{Dependence on hard-scattering scale}\label{sec:validity_mass}

\begin{figure}[t!]
\centering
\subfigure[]{\includegraphics[width=.48\textwidth]{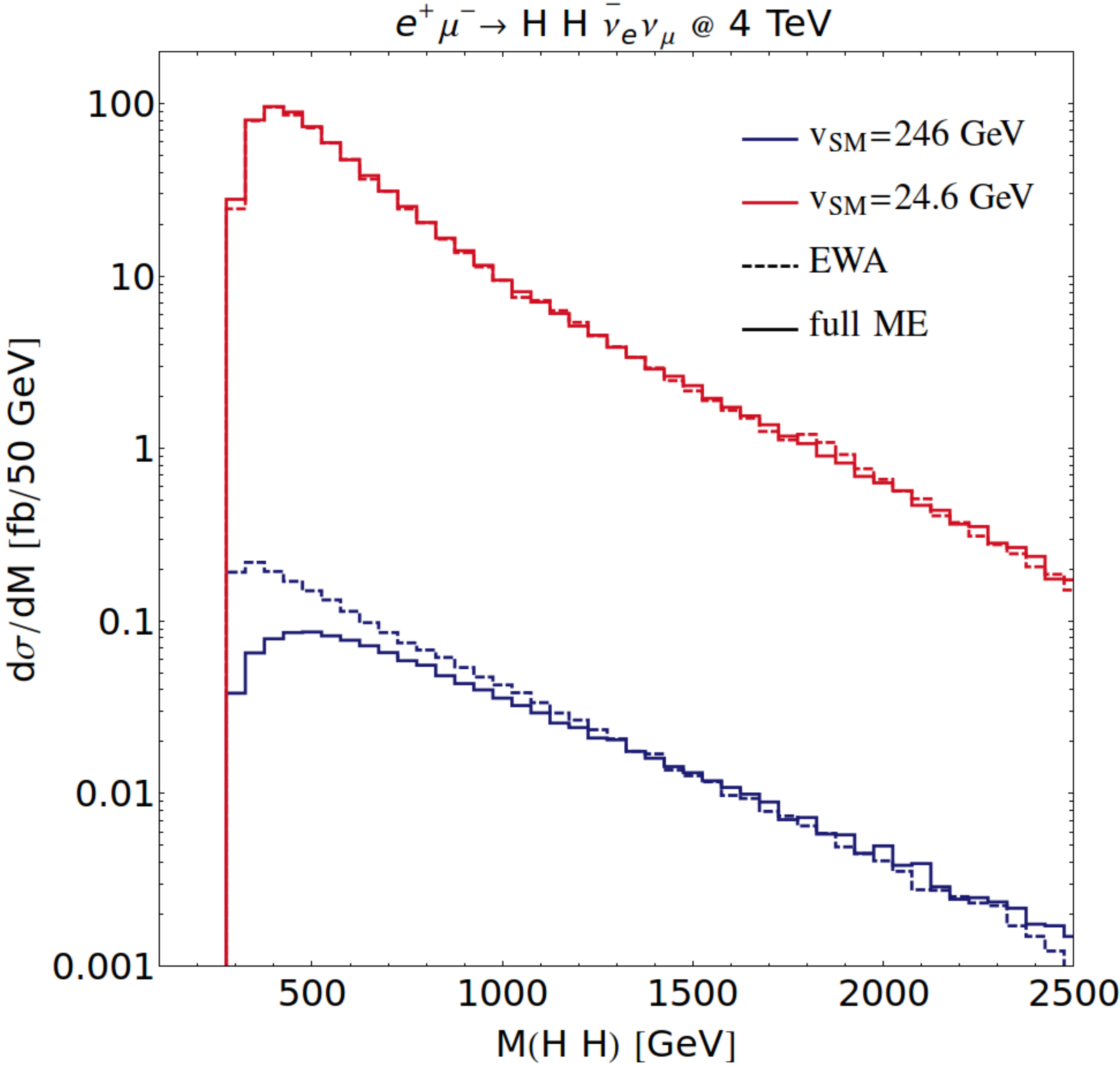} \label{fig:madEVA_hh_scale_dep}}
\subfigure[]{\includegraphics[width=.48\textwidth]{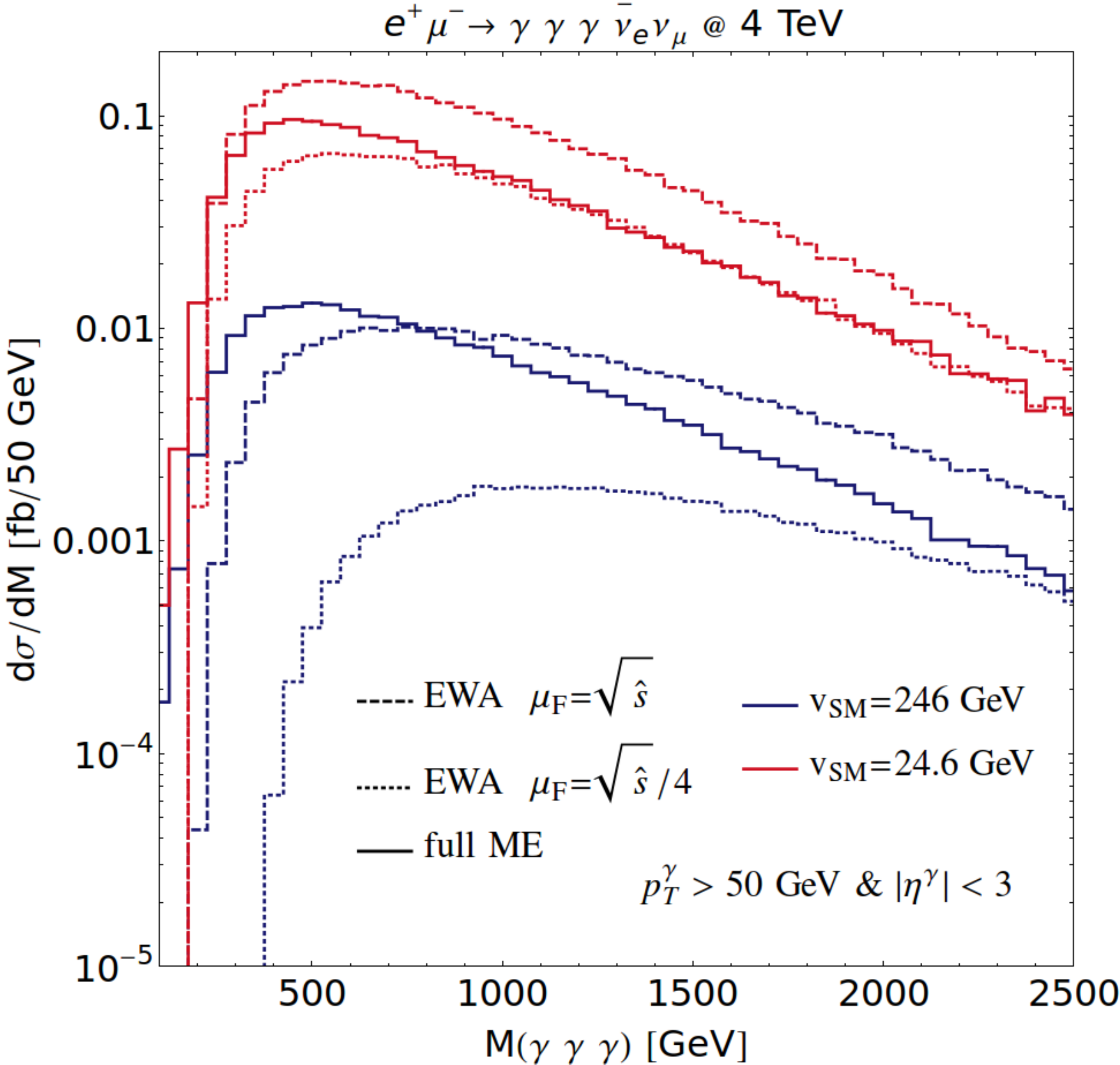} \label{fig:madEVA_aaa_scale_dep}}
\caption{
(a) The invariant mass distribution $M(HH)$ of the process $W^+W^-\to HH$, as predicted by the full $2\to4$ ME (solid) and the EWA $W^+_0W^-_0\to HH$ ME (dash) at $\sqrt{s}=4\TeV$, assuming the SM vev 
$\sqrt{2}\langle\Phi\rangle = v_{\rm SM}\approx246\GeV$ (darker, lower curves) and a scenario where $\sqrt{2}\langle\Phi\rangle = v_{\rm SM}/10$ (lighter, upper curves).
(b) Same but for the $M(\gamma\gamma\gamma)$ distribution of the process $W^+W^-\to \gamma\gamma\gamma$, assuming the fiducial cuts in Eq.~\eqref{eq:validity_cuts}, and for the $W^+_T W^-_T\to \gamma\gamma\gamma$ ME in the EWA with $\mu_f= M(\gamma\gamma\gamma)$ (dash) and $\mu_f= M(\gamma\gamma\gamma)/4$ (dots).}
\label{fig:evascale}
\end{figure}

We start our presentation on EWA uncertainties with what we find to be the most telling: that the accuracy of EWA cross sections for VBF depends crucially on the size of $(M_V^2/M_{VV'}^2)$ power corrections. To show this, we plot in Fig.~\ref{fig:evascale}(a) the invariant mass distributions at $\sqrt{s}=4\TeV$ of the $(HH)$-system using the full $2\to4$ ME (solid) and the EVA $2\to2$ ME (dashed). We assume two scenarios: one where the SM Higgs vev is its usual value $\sqrt{2}\langle\Phi\rangle = v_{\rm SM} \approx 246\GeV$ (dark, lower curves), and a hypothetical situation where the vev is reduced by a factor 10 (light, upper curves), i.e., where $\sqrt{2}\langle\Phi\rangle = v_{\rm SM}/10 \approx 24.6\GeV$. In the small-vev scenario, we keep $M_H$ and all EW gauge couplings to be their SM values in the Thomson limit. This implies $M_W\approx8.04\GeV$ and $M_Z\approx9.14\GeV$.

Focusing first on the SM case, we clearly see that the EWA and the full ME computations are in agreement for $M(HH)\gtrsim1\TeV$. Below this threshold, the EWA curve significantly overestimate the full ME. In the lowest bins, the differences between the curves reach approximately factors of $3-5$. This excess in the EVA prediction accounts for the differences in cross sections  reported in Table~\ref{tab:polarDecomp}.
Differences between the full ME and $f_{W_{0}^\pm}$ PDFs consist of corrections associated with expanding in powers of $(M_W^2/M^2(HH))$ and $(p_T^{\nu 2}/M^2(HH))\sim(M_W^2/M^2(HH))$. Importantly, we can rule out a meaningful dependence on $\mu_f$ since $W^+_{\lambda_A} W^-_{\lambda_B}\to HH$  is driven almost exclusively by $W^+_0W^-_0$ scattering. To check that these power corrections are driving the disparity between the full and approximated MEs, we turn to the reduced-vev case. Remarkably, if we reduce $M_W$ by a factor of 10, the disagreement between the EWA and the full ME disappears to within MC statistical uncertainties.

The same scenarios are presented for $\gamma\gamma\gamma$ production in Fig.~\ref{fig:evascale} (b). There, we plot the invariant mass distribution of the $(\gamma\gamma\gamma)$-system for the full (solid) and EWA MEs. As $W^+_{\lambda_A} W^-_{\lambda_B}\to \gamma\gamma\gamma$ is driven by $W^+_T W^-_T$ scattering, there is an ambiguity associated with our choice for $\mu_f$ in the $W_T^\pm$ PDFs. Therefore, we consider the envelop spanned by setting $\mu_f = \sqrt{\hat{s}}$ (dash) and $\mu_f = \sqrt{\hat{s}}/4$ (dot). In the SM case (light curves), the scale uncertainty envelope spans a huge gap that sandwiches the full ME for $M(\gamma\gamma\gamma)\gtrsim 750\GeV$. This large scale variation can be understood by considering the logarithms in the $W_T^\pm$ PDFs themselves. For a fixed $M(\gamma\gamma\gamma)$, the ratio of the two EVA distributions is given by
\begin{align}
 \cfrac{\frac{d\sigma_{\rm EVA}}{dM(\gamma\gamma\gamma)}\big\vert_{\mu_f=\sqrt{\hat{s}}}}{\frac{d\sigma_{\rm EVA}}{dM(\gamma\gamma\gamma)}\big\vert_{\mu_f=\sqrt{\hat{s}}/4}}
 \Bigg \vert_{\rm fixed~M(\gamma\gamma\gamma)}
 &= 
 \cfrac{\log^2\left(\frac{\hat{s}}{M_W^2}\right)}{\log^2\left(\frac{\hat{s}}{16M_W^2}\right)} \Bigg \vert_{\hat{s}=M^2(\gamma\gamma\gamma)}
 \\
 &= 
 \cfrac {\log^2\frac{M^2(\gamma\gamma\gamma)}{M_W^2}}
 {\log^2\frac{M^2(\gamma\gamma\gamma)}{M_W^2} + \log^2(16) - 2\log(16)\log\frac{M^2(\gamma\gamma\gamma)}{M_W^2}}
.
\label{eq:scaleUncRatio}
\end{align}

For the representative triphoton invariant masses $M(\gamma\gamma\gamma)\in\{0.5, 1, 2, 2.5\}\TeV$, we obtain roughly the respective ratios $\{17, 4.9, 3.1, 2.8\}$, in agreement with the distribution. For $M(\gamma\gamma\gamma)\lesssim750\GeV$, the full ME curve sits just above the EWA envelope. This is in contrast to $W_0^+W_0^-\to HH$, where the full ME distribution sits below the EWA rate. We attribute this to $W_T^+W_{T'}^-$ scattering having a weaker dependence on power corrections than $W_0^+W_0^-$ scattering.  
Differences between the full ME and $f_{W_T^\pm}$ PDFs are associated with expanding in powers of $(p_T^{\nu 2}/M^2(HH))\sim(M_W^2/M^2(HH))$. However, there is no second expansion in powers of $(M_W^2/M^2(\gamma\gamma\gamma))$ as one has for the $f_{W_0^\pm}$ PDFs. For $M(\gamma\gamma\gamma)>1.5\TeV$, the distribution of the full ME approaches the EWA curve for $\mu_f=\sqrt{\hat s}/4$, suggesting a preferred choice for setting $\mu_f$.

In the reduced-vev scenario (dark curves), we observe several noteworthy features. First is an improved agreement between the full and EWA distributions for $M(\gamma\gamma\gamma)\gtrsim 250\GeV$. For even lower invariant masses, the full ME is again higher than the EWA band. Second, we find that the  full ME converges to the EWA curve for $\mu_f=\sqrt{\hat s}/4$ when $M(\gamma\gamma\gamma)\gtrsim750\TeV$. Third is the appearance of a smaller scale uncertainty envelope, in accordance with Eq.~\eqref{eq:scaleUncRatio}. Numerically, this follows from the fact that for small variations of the argument $x$, the quantity $\log(x)$ varies less when $x$ is large than when $x$ is near unity. Physically,  this means that in the reduced-vev case, typical $M(\gamma\gamma\gamma)$ are further away from the $W$'s mass threshold, and therefore is less sensitive to $\mathcal{O}(\log(M^2(\gamma\gamma\gamma)/M_W^2))$ variations. Despite being smaller in this scenario, we stress that the scale uncertainty band remains sizable. For instance: using Eq.~\eqref{eq:scaleUncRatio} and our benchmark values for $M(\gamma\gamma\gamma)$, we obtain the ratios $\{2.3, 2.0, 1.8, 1.7\}$. This indicates that for realistic EW boson masses, one must go to asymptotically large $M(\gamma\gamma\gamma)$ in order to obtain $\mathcal{O}(10\%-20\%)$ uncertainties. From an alternative perspective, the large $\mu_f$ dependence is indicative of the need to extend the formula of Eq.~\eqref{eq:factTheorem} by an EW Sudakov form factor and/or RG evolution for weak boson PDFs, as studied in Refs.~\cite{Bauer:2016kkv,Chen:2016wkt,Bauer:2017bnh,Bauer:2017isx,Bauer:2018arx,Manohar:2018kfx,Han:2020uid,Han:2021kes,Pagani:2021vyk}.

\begin{table}[t!]
\begin{center}
\resizebox{\textwidth}{!}{ 
\renewcommand*{\arraystretch}{1.3}\setlength{\tabcolsep}{10pt}
	\begin{tabular}{r|c c c c c c}
		\hline \hline
		&\multicolumn{6}{c}{$\sigma_{\rm EVA}$ [fb] (Polarization Fraction)}\\
		&\multicolumn{2}{c}{$\sqrt{s}=4$ TeV}&\multicolumn{2}{c}{$\sqrt{s}=14$ TeV}&\multicolumn{2}{c}{$\sqrt{s}=30$ TeV}\\
        \hline
        $\sum_{\lambda_A,\lambda_B}W_{\lambda_A}^+ W_{\lambda_B}^-\to H H$  
                                    & $364\times10^{-3}~^{+2\%}_{-2\%}$ & $(-)$ 
                                    & $2.44~^{+1\%}_{-1\%}$ & $(-)$ 
                                    & \confirm{$4.63~^{+1\%}_{-1\%}$} & $(-)$   \\
        $W_0^+W_0^-\to H H$         & $353\times10^{-3}$ & $(97\%)$   
                                    & $2.41$ & $(99\%)$   
                                    & $4.59$ & $(99\%)$ \\
        $W_0^\pm W_T^\mp\to H H$    & $13.7\times10^{-6}~^{+34\%}_{-34\%}$ & $(<0.5\%)$   
                                    & $47.6\times10^{-6}~^{+34\%}_{-34\%}$ & $(<0.5\%)$   
                                    & $71.5\times10^{-6}~^{+33\%}_{-33\%}$ & $(<0.5\%)$ \\
        $W_T^+W_T^-\to H H$         & $10.8\times10^{-3}~^{+75\%}_{-54\%}$ & $(3\%)$   
                                    & $34.6\times10^{-3}~^{+71\%}_{-52\%}$ & $(1\%)$   
                                    & $53.1\times10^{-3}~^{+69\%}_{-50\%}$ & $(1\%)$ \\
        \hline
		$\sum_{\lambda_A,\lambda_B}W_{\lambda_A}^+ W_{\lambda_B}^-\to\gamma\gamma\gamma$  
		                                            & $87.9\times10^{-3}~^{+70\%}_{-51\%}$ & $(-)$ 
		                                            & $308\times10^{-3}~^{+62\%}_{-46\%}$ & $(-)$ 
		                                            & $423\times10^{-3}~^{+59\%}_{-45\%}$ & $(-)$    \\
        $W_0^+W_0^-\to \gamma\gamma\gamma$          & $36.8\times10^{-6}$ & $(<0.5\%)$   
                                                    & $166\times10^{-6}$ & $(<0.5\%)$   
                                                    & $240\times10^{-6}$ & $(<0.5\%)$ \\
        $W_0^\pm W_T^\mp\to \gamma\gamma\gamma$     & $572\times10^{-6}~^{+33\%}_{-33\%}$ & $(1\%)$   
                                                    & $1.53\times10^{-3}~^{+31\%}_{-31\%}$ & $(0.5\%)$   
                                                    & $1.86\times10^{-3}~^{+31\%}_{-31\%}$ & $(<0.5\%)$ \\
        $W_T W_T\to\gamma\gamma\gamma$              & $87.3\times10^{-3}~^{+71\%}_{-52\%}$ & $(99\%)$   
                                                    & $307\times10^{-3}~^{+62\%}_{-47\%}$  & $(99\%)$   
                                                    & $421\times10^{-3}~^{+59\%}_{-45\%}$  & $(99\%)$ \\
		\hline
				&\multicolumn{6}{c}{$\sigma_{\rm Full}$ [fb]}\\
        \hline
		$e^+\mu^-\to HH\;\bar\nu_e\nu_\mu$                  & $325\cdot10^{-3}$ & 
		                                                    & $2.31$  &  
		                                                    & $4.61$ & \\
		$e^+\mu^-\to \gamma\gamma\gamma\;\bar\nu_e\nu_\mu$  & $84.8\cdot10^{-3}$ & 
		                                                    & $309\cdot10^{-3}$  & 
		                                                    & $412\cdot10^{-3}$  & \\
		\hline\hline
	\end{tabular}
	}
	\caption{
	Same as Table~\ref{tab:polarDecomp} but requiring $M(HH)>1\TeV$ and $M(\gamma\gamma\gamma)>1\TeV$.
	}\label{tab:polarDecomp_massCut}
\end{center}
\end{table}

From these distributions, we can conclude that the EWA is acutely sensitive to power corrections of the form $(M_V^2/M_{VV'}^2)$. This is particularly true when scattering longitudinally polarized weak bosons. Distributions also suggest a weaker dependence on power corrections when scattering transversely polarized weak bosons. We attribute this difference to the different expansions needed to derive longitudinal and transverse weak boson PDFs:
$W_T$ PDFs require a single power expansion whereas $W_0$ PDFs require a double expansion. Altogether, this points to evidence of the EWA's success for both transverse and longitudinal $V_{\lambda_A} V_{\lambda_B}'$ scattering when $M_{VV'}>\mathcal{O}(1\TeV)$, or $(M_V^2/M_{VV'}^2)\lesssim0.01$. 
To further demonstrate this at the level of cross sections, we show in Table~\ref{tab:polarDecomp_massCut} the same quantities as in Table~\ref{tab:polarDecomp} but require also that $M(HH)>1\TeV$ and $M(\gamma\gamma\gamma)>1\TeV$. The improved agreement between the full and EWA computations is due to the cuts on $M(WW)$.

\subsection{Dependence on collider energy}\label{sec:validity_collider}

In light of the above, we consider now the impact collider energy on the EWA's accuracy. Increasing $\sqrt{s}$ has two prominent effects on $V_{\lambda_A} V_{\lambda_B}'$ scattering: (i) For fixed momentum fractions $\xi_1$ and $\xi_2$, more energetic $\mpmm$ collisions lead to more energetic $(VV')$-systems, with $M_{VV'}^2 = \xi_1 \xi_2 s$. The corresponding enhancement of collinear logarithms indicates an enlargement of collinear regions of phase space. (ii) For a fixed hard-scattering scale $M_{VV'}$, increasing the collider energy leads to probing smaller $\xi_1$ and $\xi_2$. The corresponding enhancement of soft logarithms similarly indicates an enlargement of soft regions of phase space. (Soft logarithms appear after integrating $\tilde{f}_{V_\lambda}(\xi_i)\sim 1/\xi_i$ over $\xi_i$; see, e.g., Ref.~\cite{Costantini:2020stv}.)

\begin{figure}[t!]
\centering
\subfigure[]{\includegraphics[width=.48\textwidth]{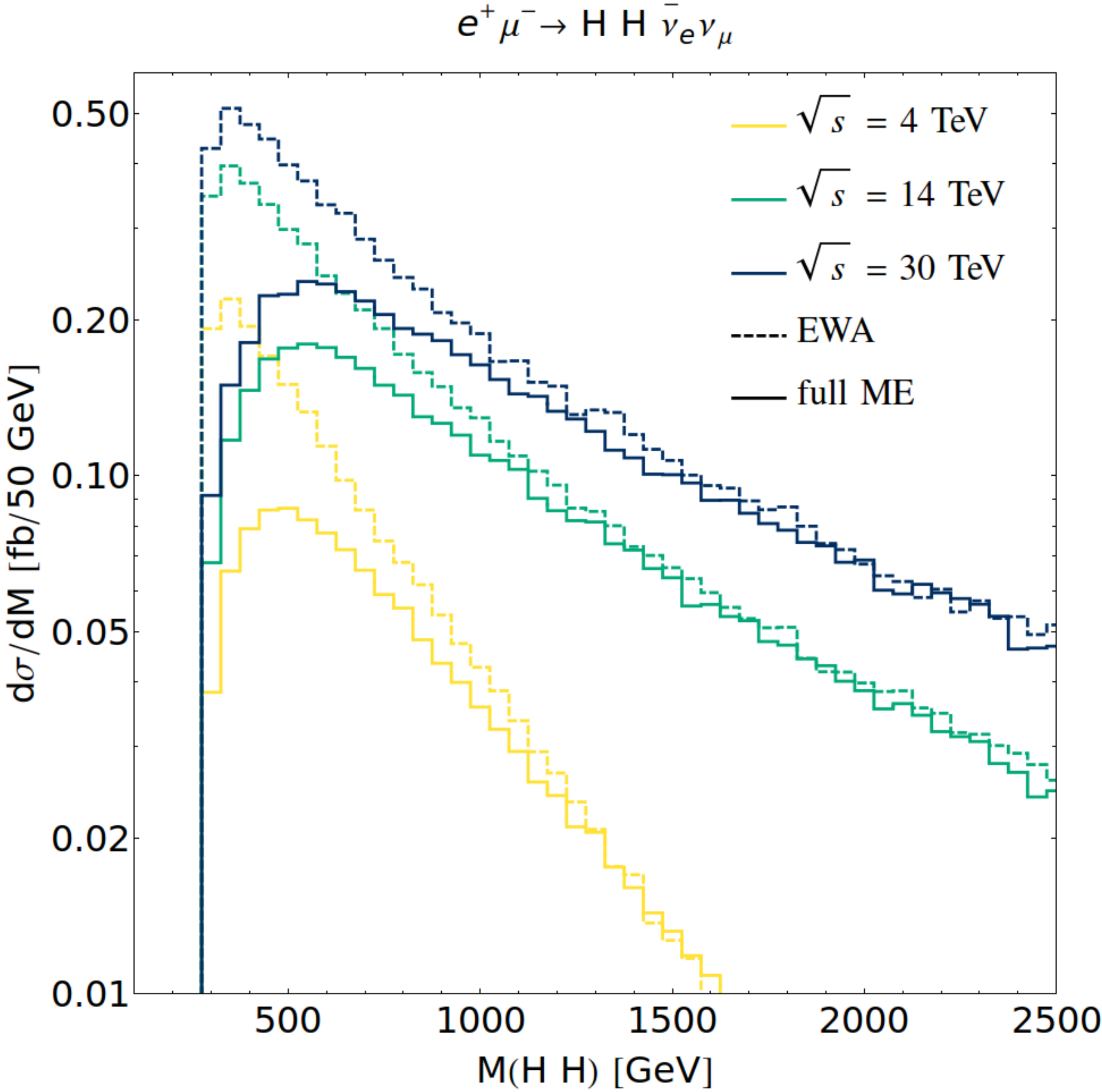} \label{fig:madEVA_hh_energy_dep}}
\subfigure[]{\includegraphics[width=.48\textwidth]{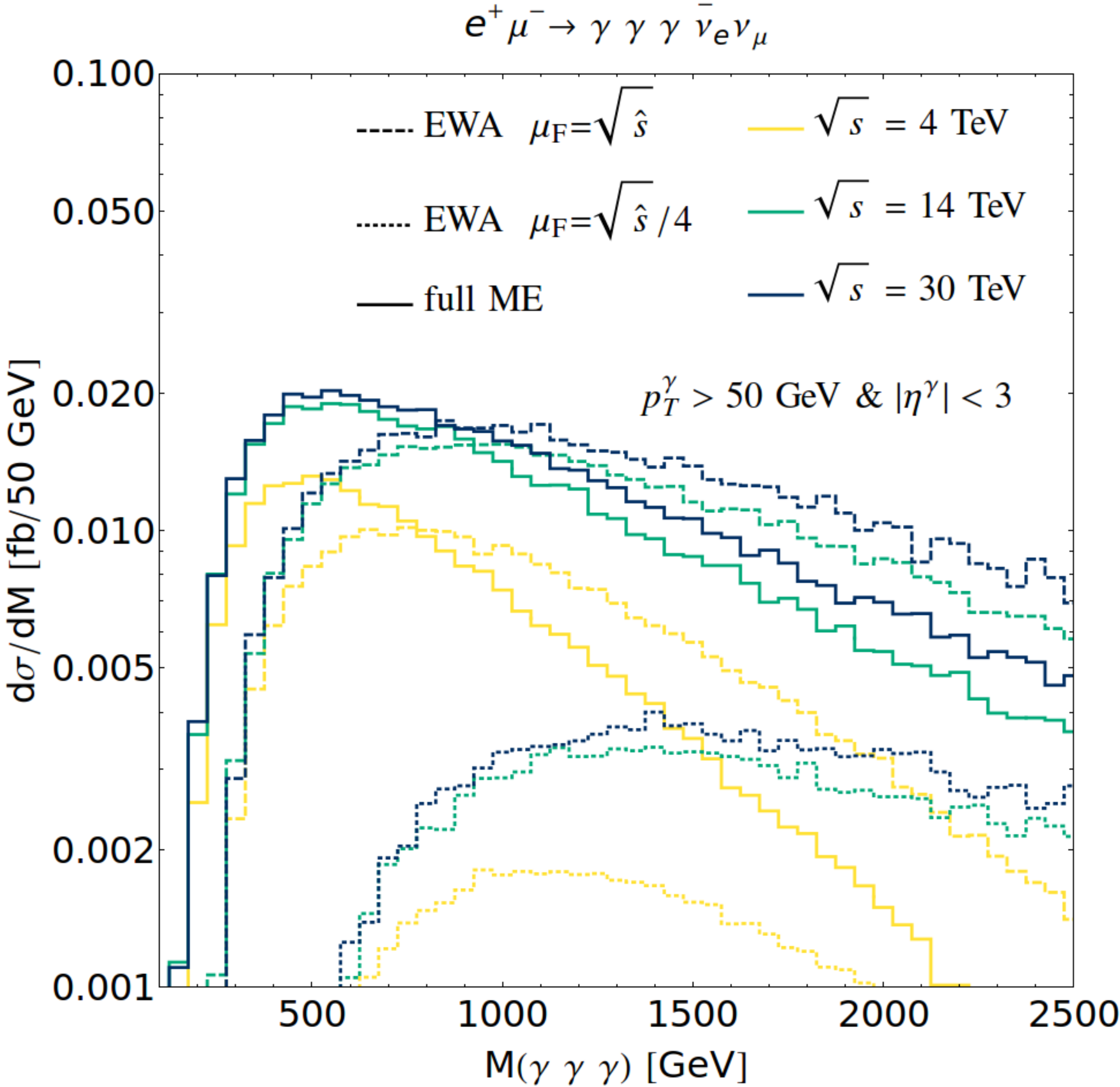} \label{fig:madEVA_aaa_energy_dep}}
\caption{(a) The invariant mass distribution $M(HH)$ of the process $W^+W^-\to HH$, as predicted by the full $2\to4$ ME (solid) and the EVA $W^+_0W^-_0\to HH$ ME (dash) at $\sqrt{s}=4\TeV$ (lower light curves) and $14\TeV$ (upper dark curves).
(b) Same but for the $M(\gamma\gamma\gamma)$ distribution of the process $W^+W^-\to \gamma\gamma\gamma$, assuming the fiducial cuts in Eq.~\eqref{eq:validity_cuts}, and for the $W^+_T W^-_T\to \gamma\gamma\gamma$ ME in the EWA with $\mu_f= M(\gamma\gamma\gamma)$ (dash) and $\mu_f= M(\gamma\gamma\gamma)/4$ (dots).}
\label{fig:evasqrts}
\end{figure}

To explore these effects, we present in Fig.~\ref{fig:evasqrts} the invariant mass distributions of (a) the $(HH)$-system and (b) the $(\gamma\gamma\gamma)$-system  at $\sqrt{s}=4\TeV$ (light), 14 TeV (darker), and 30 TeV (darkest), assuming the full MEs for the processes in Eq.~\eqref{eq:validity_setup_fullProc} (solid),
and the EWA MEs for the processes in Eq.~\eqref{eq:validity_setup_evaProc} (dash). For the $W_T^+ W_T^- \to \gamma\gamma\gamma$ process, we show the scale variation envelop obtained by setting $\mu_f=\sqrt{\hat{s}}$ (dash) and $\mu_f=\sqrt{\hat{s}}/4$ (dot). 

Focusing on the $M(HH)$ distribution in Fig.~\ref{fig:evasqrts}(a), several observations can be made. We start with the anticipated jump in cross section for increasing $\sqrt{s}$. For both the EWA and the full MEs, we find that increasing the collider energy by a factor of $3.5$ causes all total cross sections to increase by about a factor of $3$  (see Table~\ref{tab:polarDecomp}). Increases are much more dramatic at the differential level for $M(HH)\gtrsim1.5\TeV$ due to the significant opening of phase space. In this regime, we also find good agreement with the normalization and shape between the EWA and full MEs. At lower invariant masses, particularly for $M(HH)\lesssim500\GeV$, we find that the EWA overestimates the full ME in the same manner as observed in the previous section.  In this regime, the EWA distributions increase more quickly with rising $\sqrt{s}$ than the full ME distributions: in the lowest $M(HH)$ bins, the EWA ME overestimates the full ME by about a factor of $3-5$ at $\sqrt{s}=4\TeV$ and by about $3.4-4.7$ at $30\TeV$.
As longitudinal weak boson PDFs do not contain collinear logarithms, the enhancements in Fig.~\ref{fig:evasqrts}(a) are driven exclusively by soft logarithms. This implies that the EWA favors the production of relatively softer $W^\pm_0$, and hence lower $M(HH)$, a phenomenon that is sometimes~\cite{Chen:2016wkt} described as ``ultra collinear enhancements.'' Consequentially, increasing the collider energy reinforces the sensitivity to $(M_W^2/M^2(HH))$ power corrections, which must be negative. Despite this, the distributions show that regardless of $\sqrt{s}$ the EWA converges to the full ME computation for $M(H H)\gtrsim 1$ TeV.

Turning to the $M(\gamma\gamma\gamma)$ distribution in Fig.~\ref{fig:evasqrts}(b), we observe several of the same characteristics. Foremost we find that the full ME distribution consistently sits within the EVA scale uncertainty band for $M(\gamma\gamma\gamma)\gtrsim750\GeV-1\TeV$, for all $\sqrt{s}=4-30\TeV$. Though, for increasing $\sqrt{s}$ we find that the full ME expectation migrates away from the $\mu_f=\sqrt{s}/4$ boundary and towards the envelope's center. For a fixed $M(\gamma\gamma\gamma)$, we find that the thicknesses of the $\mu_f$ uncertainty bands remain about the same for increasing $\sqrt{s}$, with changes just outside MC statistical uncertainties. This is consistent with the ratio expression of Eq.~\eqref{eq:scaleUncRatio}, which does not obviously suggest an additional dependence on collider energy once $M(\gamma\gamma\gamma)$ is fixed. An important difference with respect to the $M(HH)$ case is the preference for larger values of $M(\gamma\gamma\gamma)$ with increasing $\sqrt{s}$. (For $W_0^+W_0^-\to HH$, smaller invariant masses are preferred at increasing collider energy.) As $W_T^+W_T^- \to \gamma\gamma\gamma$ is devoid of possible ME-level enhancements from longitudinal polarizations, we attribute these behaviors to the collinear logarithm contained in the $f_{W_T^\pm}$ PDF, which favors producing larger invariant masses. Notably, the collinear logarithms reinforce the accuracy of the EVA by favoring phase space regions where $(M_W^2/M^2(\gamma\gamma\gamma))$ is small.

\begin{figure}[t!]
\centering
\subfigure[]{\includegraphics[width=.48\textwidth]{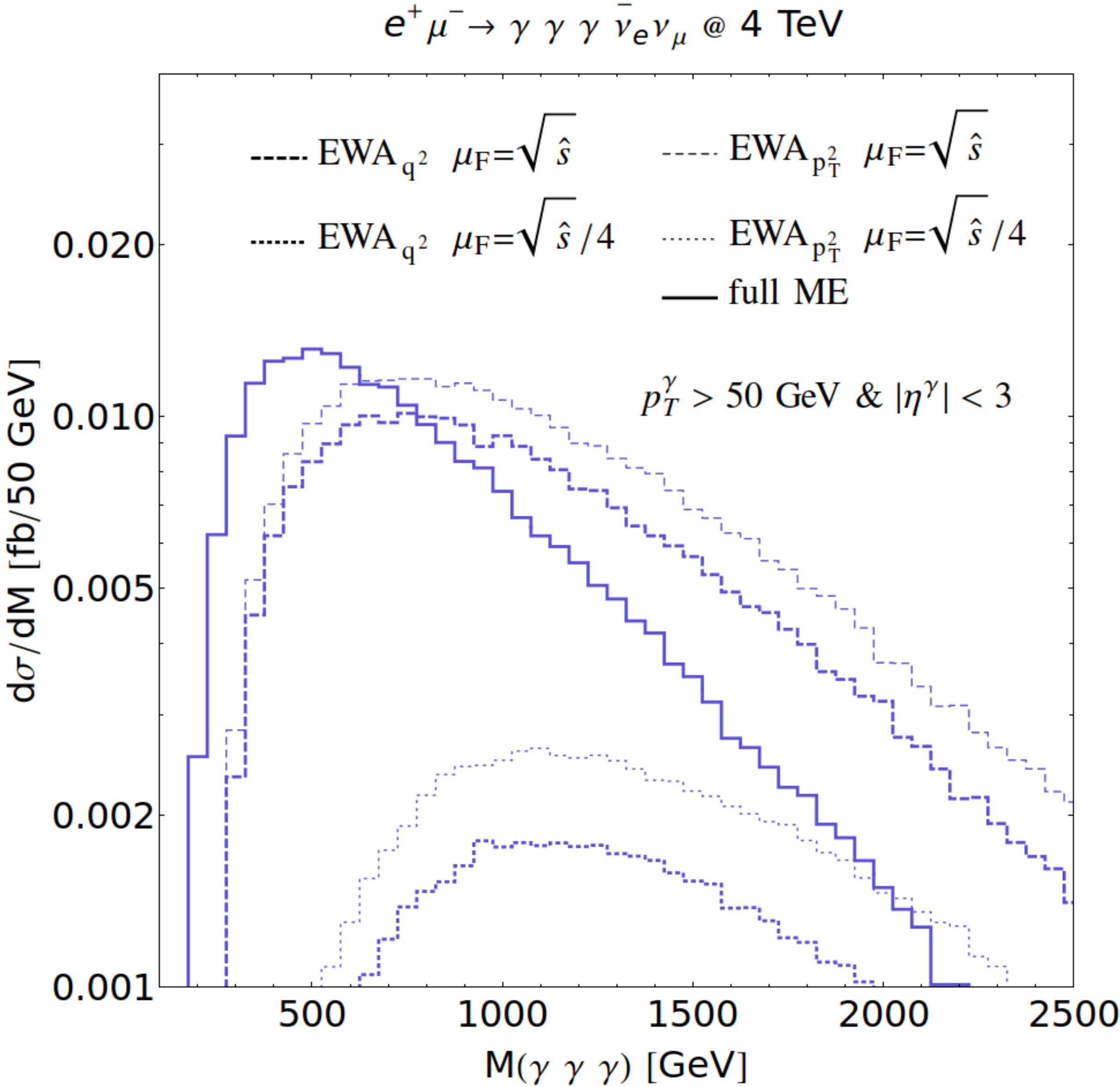}}
\subfigure[]{\includegraphics[width=.48\textwidth]{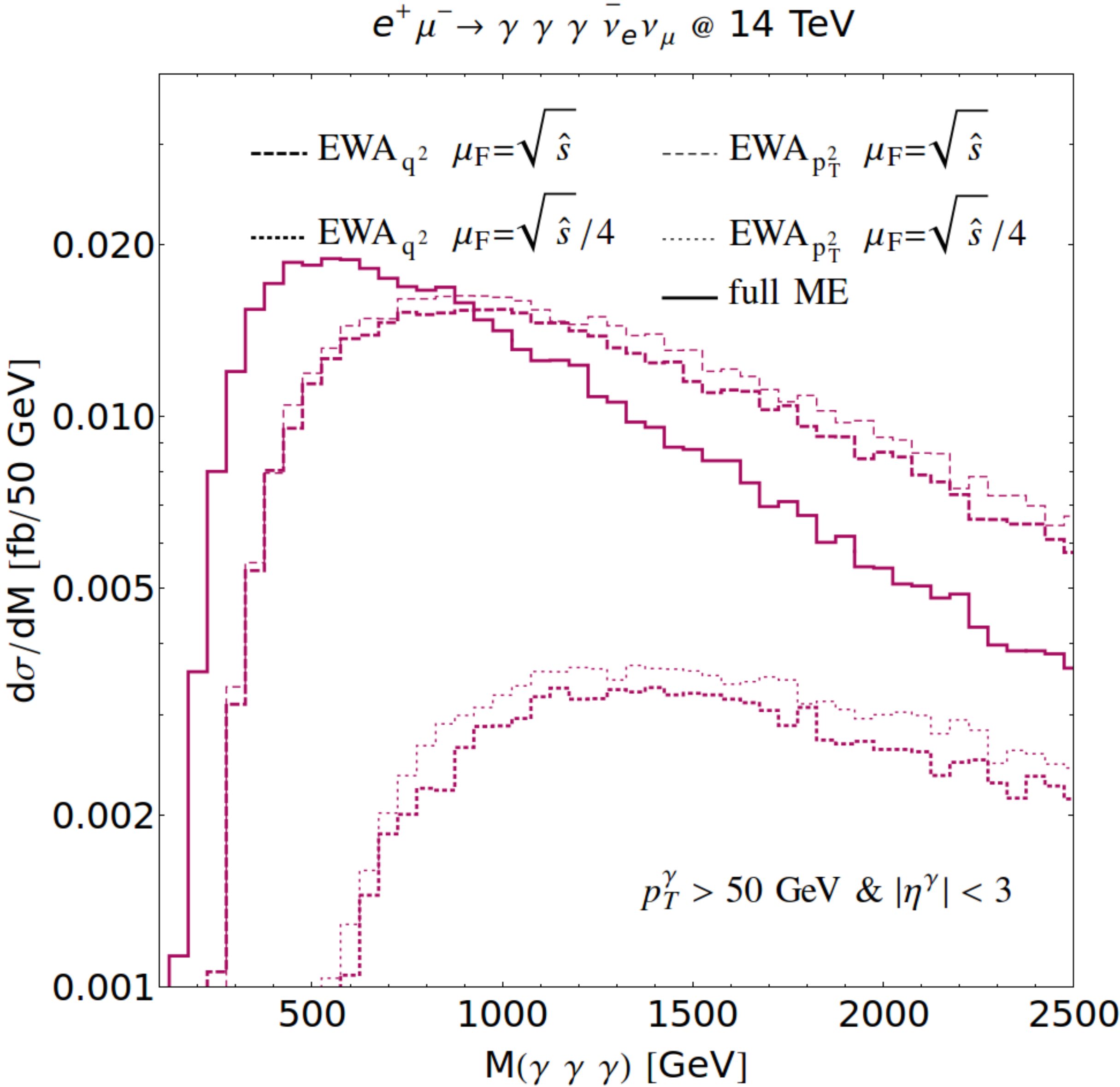}}
\subfigure[]{\includegraphics[width=.48\textwidth]{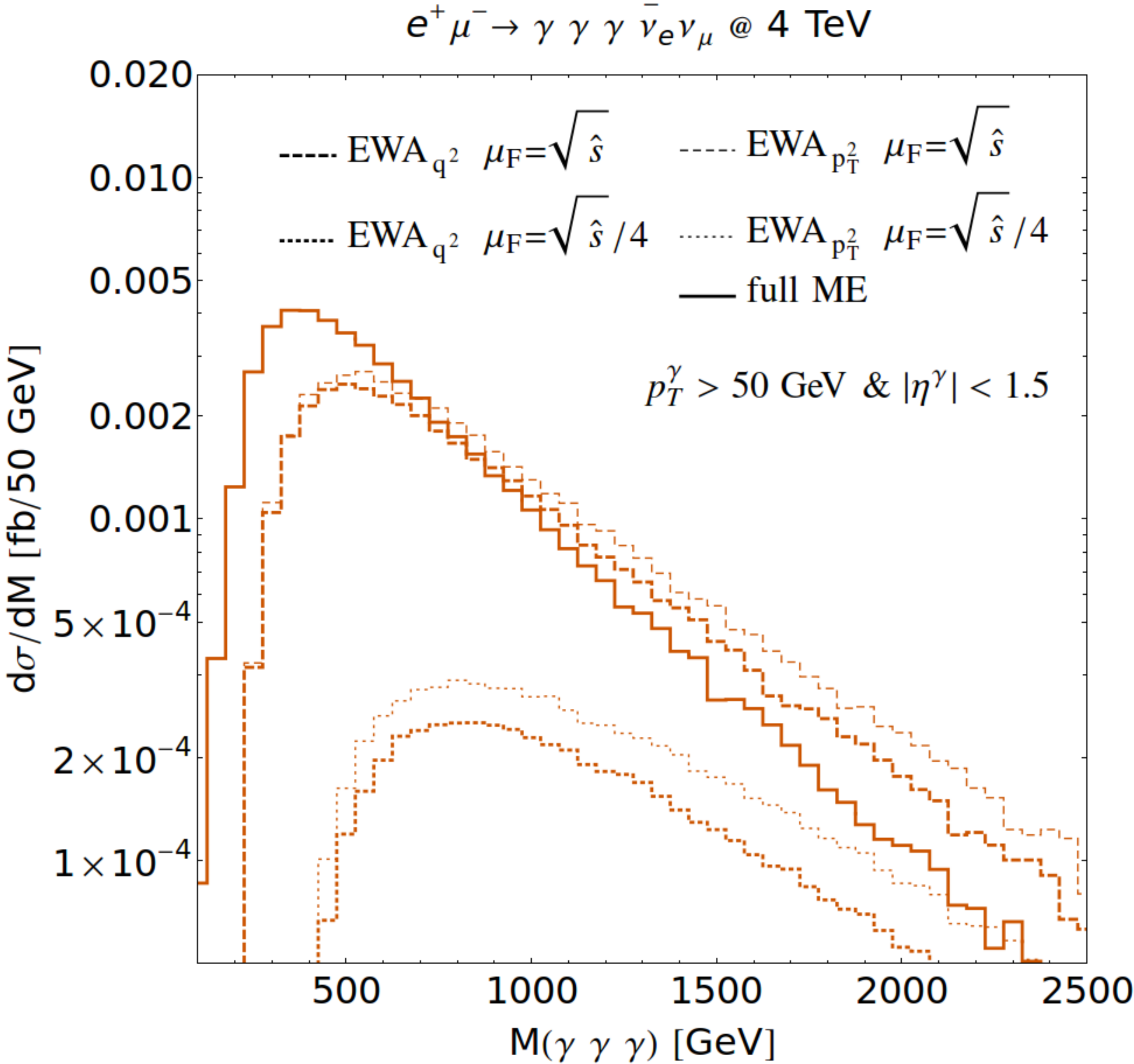}}
\subfigure[]{\includegraphics[width=.48\textwidth]{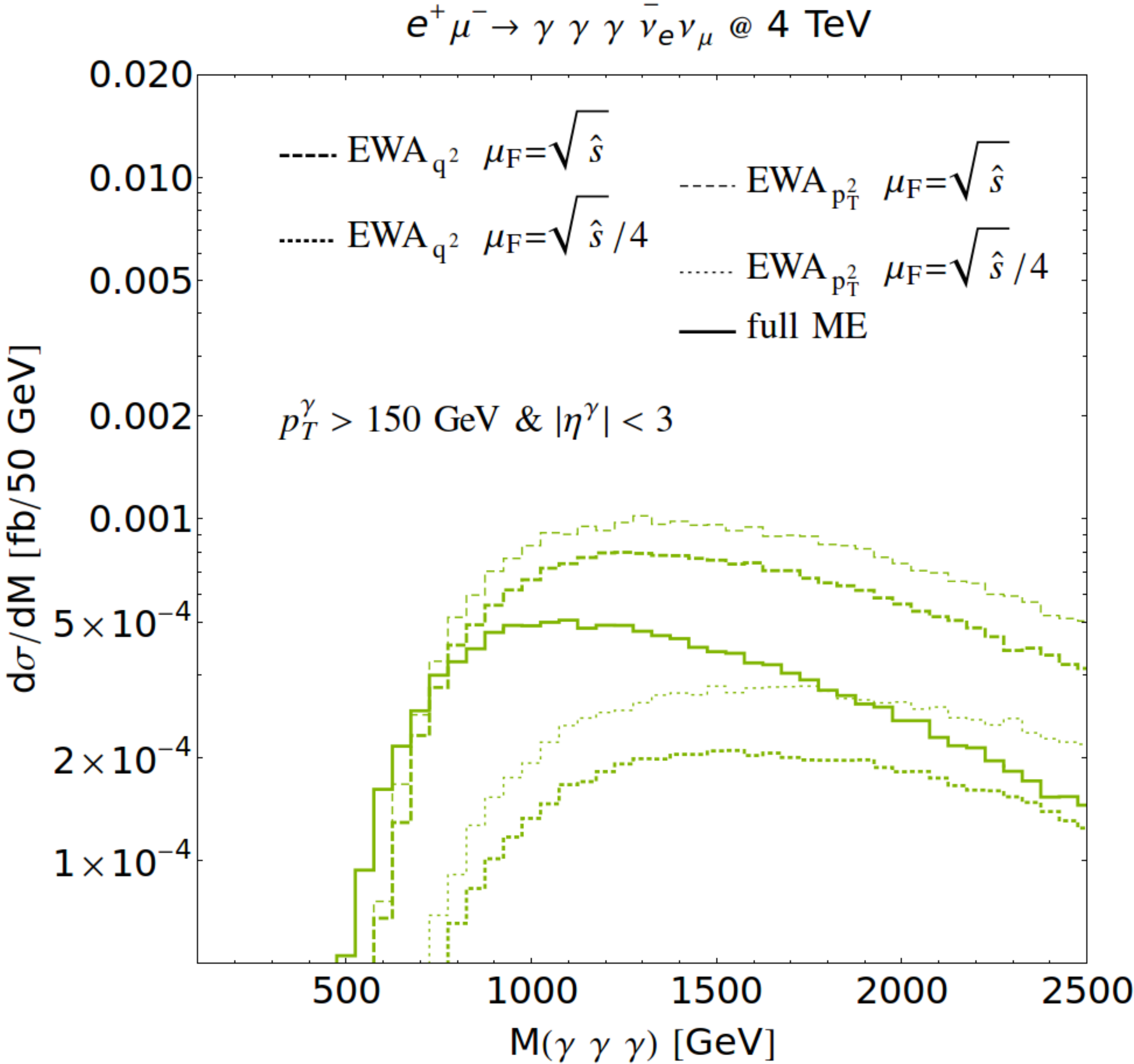}}
\caption{(a) The invariant $M(\gamma\gamma\gamma)$ distribution of the process $W^+W^-\to \gamma\gamma\gamma$, as predicted by the full $2\to5$ ME (solid) and the EVA $W^+_TW^-_T\to \gamma\gamma\gamma$ ME at $\sqrt{s}=4\TeV$, assuming $f_{W_T}$ PDF evolution by virtuality $q^2$ (thick lines) and transverse momentum $p_T^2$ (thin lines), 
with $\mu_f= M(\gamma\gamma\gamma)$ (dash) and $\mu_f= M(\gamma\gamma\gamma)/4$ (dots).
(b) The same as (a) but for $\sqrt{s}=14\TeV$.
(c) The same as (a) but for $\vert\eta^\gamma\vert<1.5$.
(c) The same as (a) but for $p_T^\gamma>150\GeV$.
}
\label{fig:aaascheme}
\end{figure}

\subsection{Dependence on evolution variable and phase space cuts}\label{sec:validity_evo}

As shown in Sec.~\ref{sec:formalism}, EW boson PDFs can be constructed using either the virtuality $q$ of $V_\lambda$ or transverse momentum $p_T$ of $l$ as the evolution variable in collinear $\mu\to V_\lambda l$ splitting. Given that the expressions for the $f_{W_T^\pm}$ PDFs differ, we investigate whether these schemes give appreciably different results. To explore this, we focus on the process $W_T^+W_{T'}^- \to \gamma\gamma\gamma$ since the $f_{W_0^\pm}$ PDFs are the same under both schemes.

In Fig.~\ref{fig:aaascheme}(a), we show the invariant mass distribution of the $(\gamma\gamma\gamma)$-system in $W_T^+W_{T'}^- \to \gamma\gamma\gamma$ at $\sqrt{s}=4\TeV$ using the full ME for the $2\to5$ scattering process (solid) and the analogous process under the EWA. For the EWA curves, we consider the scale variation envelop spanned by setting $\mu_f=\sqrt{\hat{s}}$ (dash) and $\mu_f=\sqrt{\hat{s}}/4$ (dot) for evolution by $q^2$ (thick curves) and evolution by $p_T^2$ (thin curves). We also impose the fiducial cuts in Eq.~\eqref{eq:validity_cuts}. As found in previous subsections, the distribution for the full ME sits in the EWA envelope for $M(\gamma\gamma\gamma)\gtrsim750\GeV$, when assuming evolution by $q^2$. In comparison,  the evolution-by-$p_T^2$ envelope is systematically shifted upward to a larger set of rates. Notably, this leads to the $\mu_f=\sqrt{\hat{s}}/4$ curve for evolution by $p_T^2$ to overestimate the full ME when $M(\gamma\gamma\gamma)\gtrsim2\TeV$. This shift is entirely due to the different arguments in the PDFs' collinear logarithms.

More specifically, for evolution by $q^2$, one has $\tilde{f}_{W_T^\pm}(\xi,\mu_f)\propto \log(\mu_f^2/M_W^2)$, whereas for evolution by $p_T^2$, one has $\tilde{h}_{W_T^\pm}(\xi,\mu_f)\propto \log(\mu_f^2/(1-\xi)M_W^2)$. This difference implies that for a fixed factorization scale, the logarithm in the evolution-by-$p_T^2$ PDF is relatively enhanced by a factor of $1/(1-\xi)$. This favors larger momentum fractions and therefore harder invariant mass distributions since $M^2(\gamma\gamma\gamma)=\xi_1\xi_2 s$. This enhancement and subsequent disagreement do not imply that evolution-by-$p_T^2$ scheme itself is incorrect. It only indicates that setting $\mu_f = \sqrt{\hat{s}}$ or $\mu_f = \sqrt{\hat{s}}/4$ are poor choices of factorization scale for this process and collider energy. Obviously, setting $\mu_f = \sqrt{(1-\xi)\hat{s}}$ or $\mu_f = \sqrt{(1-\xi)\hat{s}}/4$, which is a common practice SCET~\cite{Becher:2007ty,Becher:2014oda}, would recover the results for evolving by $q^2$.

In Fig.~\ref{fig:aaascheme}(b), we show the invariant mass distribution but for $\sqrt{s}=14\TeV$. Remarkably, we find that two the envelopes converge, and differences between the two evolution schemes are within MC statistical uncertainties. Importantly, the full ME distribution remains inside both envelopes for $M(\gamma\gamma\gamma)\gtrsim750\GeV$. To understand this improved agreement, recall that when the hard scattering scale $M(\gamma\gamma\gamma)$ is fixed, one probes smaller momentum fractions for increasing collider energies. This implies that as $\sqrt{s}$ increases, the $1/(1-\xi)$ enhancements cease and systematically approach unity, i.e., $1/(1-\xi)\to 1$.

In Fig.~\ref{fig:aaascheme}(c) and (d), we investigate the impact of tighter phase space cuts on the final-state system. In particular, we consider the cases when photons are 
(c)  more central with $\vert\eta^\gamma\vert < 1.5$, and 
(d) harder with $p_T^\gamma > 150\GeV$.
Aside from an obvious reduction in cross section, a few qualitative observations are worth reporting. 
In figure (c), we observe a shift in the distribution to smaller $M(\gamma\gamma\gamma)$, whereas in figure (d) the shift is to larger $M(\gamma\gamma\gamma)$. In both cases, the change is kinematical. Consider the invariant mass of the entire three-photon system in terms of two-photon systems, i.e.,
\begin{align}
    M^2(\gamma\gamma\gamma) &= 2M(\gamma_1\gamma_2) + 2M(\gamma_2\gamma_3) + 2M(\gamma_1\gamma_3), \quad\text{where}
    \\
    M^2(\gamma_i\gamma_j)&\sim p_T^i p_T^j[(y_i - y_j)^2 + (\phi_i - \phi_j)^2].
\end{align}
Here,  $\eta_i$ and $\phi_i$ are the pseudorapidity and azimuth of $\gamma_i$.
Requiring a smaller $\vert\eta^\gamma\vert$ window leads to smaller diphoton masses, and hence smaller triphoton masses. Likewise, requiring larger $p_T^\gamma$ leads to larger diphoton masses, and therefore larger triphoton masses.
Despite these different shifts, we do not find clear improvement or worsening of the agreement between either evolution scheme and the full ME. In both (c) and (d), the full ME distribution sits inside both envelopes for $M(\gamma\gamma\gamma)\gtrsim750\GeV$. However, as in the baseline case, the $p_T^2$-scheme again overestimates the full ME for $M(\gamma\gamma\gamma)\gtrsim2\TeV$ due to too large $\log(1-\xi)$ enhancements. This suggests that the specific dynamics of the hard scattering process may not have an appreciable impact the validity of the EWA, as one would hope.

\subsection{Matrix Element Matching with Collinear $W_T$ PDFs}\label{sec:validity_matching}

As a final check of our implementation of EW boson PDFs into {\mgamc} and as a proof-of-concept demonstration of the potential capabilities, we briefly explore matrix element matching (MEM) with transverse weak boson PDFs. The idea behind MEM is that one can divide computations for complicated, many-leg final states that are susceptible to numerical instabilities, e.g., $\mu^+\mu^- \to \overline{\nu_\mu} \nu_\mu  \gamma\gamma\gamma$, into two easier, more stable  parts:
(i) a ME with a smaller final-state multiplicity that represents a particular region of phase space of the original process, e.g., the process $\mu^+ W^- \to \overline{\nu_\mu} \gamma\gamma\gamma$ with a $\tilde{f}_{W^-/\mu^-}$ PDF, which describes the collinear $\mu^- \to W^- \nu_\mu$ splitting; and
(ii) the ME for the original process but where the phase space for (i) is excluded, e.g., $\mu^+\mu^- \to  \overline{\nu_\mu} \nu_\mu \gamma\gamma\gamma$ process with only wide-angle $\mu^- \to W^- \nu_\mu$ 
splittings. In principle, summing the two components should recover the full phase space for the original ME, up to power corrections that are formally small. The aim of this procedure is to efficiently describe regions of phase space that are otherwise difficult to model simultaneously due to instabilities associated with soft and/or collinear radiation. If MEM is successfully implemented, then the sum of (i) and (ii) should not only reproduce the original cross section, up to uncertainties, but also display an insensitivity to the (artificial) cutoff scale that divides the original process into regions (i) or (ii).

This subsection serves as a proof-of-concept check and exploration of the power-law-like corrections in the factorization formula of Eq.~\eqref{eq:factTheorem}. The discussion here also touches upon whether there is a natural or preferred choice for $\mu_f$, which might resolve the collinear/wide-angle ambiguity described in Sec.~\ref{sec:validity_scale}.
(It is not obvious that arguments for setting $\mu_f$ in hadron collisions, such as those given in Ref.~\cite{Maltoni:2007tc,Degrande:2016aje}, are applicable here.)
As discussed below, there are technical nuances at play that merit comprehensive exploration.
However, this is beyond the scope of our work. Future studies that expand on this section are therefore encouraged.

To sketch MEM conceptually for the case of matching collinear $W_T$ (or $Z_T$) PDFs with full ME, we focus on the scattering process $e^+\mu^- \to  \overline{\nu_e} \nu_\mu \gamma\gamma\gamma$. One can schematically divide the cross section of the process $(\sigma^{\rm Tot.})$ into three disjoint pieces that describe different modes of $\mu^- \to W^-\nu_\mu$ splitting:
(i) a collinear piece $(\sigma^{\rm C})$,
(ii) a quasi-collinear piece $(\sigma^{\rm QC})$, and 
(iii) a hard piece $(\sigma^{\rm H})$.
Taking $p_T^{{\nu_\mu}}$ as the evolution variable, one can write:
\begin{align}
 \sigma^{\rm Tot.}
 ~=~
 \underset{\sigma^{\rm C}}{\underbrace{\int_0^{\mu_f}dp_T^{{\nu_\mu}} \frac{d\sigma}{dp_T^{{\nu_\mu}}}}}
~+~
 \underset{\sigma^{\rm QC}}{\underbrace{\int_{\mu_f}^\Lambda dp_T^{{\nu_\mu}} \frac{d\sigma}{dp_T^{{\nu_\mu}}}}}
~+~
 \underset{\sigma^{\rm H}}{\underbrace{\int_\Lambda^{p_T^{\max}} dp_T^{{\nu_\mu}} \frac{d\sigma}{dp_T^{{\nu_\mu}}}}}.
\end{align}
Here, $\mu_f\gg M_W$ is a factorization scale separating collinear and wide-angle $\mu^- \to W^- \nu_\mu$ splittings. The second cutoff $\Lambda$, which satisfies $\mu_f\lesssim \Lambda \ll p_T^{\max}$, is some arbitrary scale such that collinear factorization remains a good approximation.  (The introduction of $\Lambda$ is simply for bookkeeping: it ensures $\sigma^{\rm QC}$ can be written in a convenient manner.) Finally, $p_T^{\max}$ is upper bound on $p_T^\nu$ allowed by momentum conservation and phase space cuts.

After integration over $p_T^\nu$, both $\sigma^{\rm C}$ and $\sigma^{\rm QC}$ will scale like a collinear logarithm and power corrections, which we neglect (retain) in the (quasi-)collinear expression:
\begin{align}
 \sigma^{\rm C} &= \int_0^{\mu_f} dp_T^{{\nu_\mu}} \frac{d\sigma}{dp_T^{{\nu_\mu}}} \sim \log\frac{\mu_f^2}{M_W^2}
 +
 \underset{\rm neglect}{
 \underbrace{
 \mathcal{O}\left(\frac{\mu_f^2}{M_{WW}^2}\right)
 }},
\\
 \sigma^{\rm QC} &= \int_{\mu_f}^\Lambda dp_T^{{\nu_\mu}} \frac{d\sigma}{dp_T^{{\nu_\mu}}} \sim \log\frac{\Lambda^2}{\mu_f^2}
  +
 \underset{\rm retain}{
 \underbrace{  
 \mathcal{O}\left(\frac{\Lambda^2}{M_{WW}^2}\right)
 }}.
\end{align}
Combining the quasi-collinear and hard terms, one obtains the cross section $(\sigma^{\rm W}\equiv\sigma^{\rm QC} +  \sigma^{\rm H})$ for wide-angle $\mu^- \to W^-\nu_\mu$ splitting that is independent of $\Lambda$ (since all $\Lambda$-dependent terms are kept). Moreover, were one to combine the leading logarithmic term of $\sigma^{\rm C}$ with $\sigma^{\rm W}$, then the logarithmic dependence on $\mu_f$ would vanish identically. One would also recover the total cross section, up to the neglected power corrections. Explicitly, one finds
\begin{equation}
\sigma^{\rm Sum} = 
\sigma^{\rm C} +  \sigma^{\rm W} 
\sim
\log\frac{\mu_f^2}{M_W^2}
+ 
 \log\frac{\Lambda^2}{\mu_f^2}
 +
 \underset{\rm retained}{
 \underbrace{  
 \mathcal{O}\left(\frac{\Lambda^2}{M_{WW}^2}\right)
 }}
+  \sigma^{\rm H}
+ 
\underset{\rm neglected}{
 \underbrace{
 \mathcal{O}\left(\frac{\mu_f^2}{M_{WW}^2}\right)
 }}
.
\end{equation}
This indicates that $\mu_f$ can be interpreted in MEM also as the ``matching scale'' that matching collinear and wide-angle regions of phase space in an inclusive calculation.

To demonstrate that MEM is possible with transverse weak boson PDFs, we focus on $W^+W^-_T\to \gamma\gamma\gamma$ in $e^+ \mu^-$ collisions and define the following disjoint regions of phase space:
\begin{align}
\text{Collinear Region}  &\quad:\quad e^+ W_T^- \to \overline{\nu_e} \gamma \gamma \gamma,~
\quad\text{for}\quad p_T^{\nu_\mu} \leq \mu_f \quad \text{or} \quad \sqrt{\vert q^2_{\mu\nu_\mu}\vert} \leq \mu_f,
\label{eq:validity_eWT_scatt}
\\
\text{Wide-Angle Region}  &\quad:\quad e^+ \mu^- \to \overline{\nu_e}\nu_\mu \gamma \gamma \gamma, \quad\text{for}\quad p_T^{\nu_\mu} > \mu_f \quad \text{or} \quad \sqrt{\vert q^2_{\mu\nu_\mu}\vert} > \mu_f.
\label{eq:validity_emu_scatt}
\end{align}
When mediated by the EWA, we can identify $e^+W_T^-$ scattering in Eq.~\eqref{eq:validity_eWT_scatt} as the collinear component of the inclusive $e^+ \mu^- \to \overline{\nu_e}\nu_\mu \gamma \gamma \gamma $ process. Analogously, we can identify Eq.~\eqref{eq:validity_emu_scatt} as the wide-angle component of the inclusive $e^+ \mu^- \to \overline{\nu_e}\nu_\mu \gamma \gamma \gamma $ process, when the appropriate phase space cut is applied to $p_T^{\nu_\mu}$ or the $\mu^- \to \nu_\mu$ momentum transfer $q_{\mu\nu_\mu}$.

To regulate poles associated with final-state photons, to avoid instabilities associated with collinear $e^+\to W^+\overline{\nu_e}$ splittings, and to minimize the power corrections described in Sec.~\ref{sec:validity_mass}, we impose the following phase space restrictions on both Eqs.~\eqref{eq:validity_eWT_scatt} and \eqref{eq:validity_emu_scatt}:
\begin{equation}
    p_T^{\overline{\nu_e}}>50\GeV, \quad \vert\eta^{\overline{\nu_e}}\vert<5, \quad
p_T^\gamma>50\GeV, \quad \vert\eta^\gamma\vert<3,
\quad
 M(\gamma\gamma\gamma) > 1~\text{or}~3\TeV.
 \label{eq:validity_matching_cutsBaseline}
\end{equation}
We treat the initial-state $W_T^-$ using the appropriate PDF and consider when the PDF is defined in terms of (i) $q^2$ as given in Eq.~\eqref{eq:formalism_pdfs_q2}, and (ii) $p_T^2$ as given in Eq.~\eqref{eq:formalism_pdfs_pT}.  Assuming that $\tilde{f}_{W_T^-}$ has been evaluated at a factorization scale $\mu_f$, then for the case of evolution by $q^2$, we remove collinear and shallow-angle splittings in Eq.~\eqref{eq:validity_emu_scatt} by requiring that the norm of the $\mu^- \to \nu_\mu$ momentum transfer is larger than $\mu_f$. Symbolically, this is given by
\begin{equation}
\vert q_{\mu\nu_\mu}^2 \vert > \mu_f^2, \quad\text{where}\quad q_{\mu\nu_\mu}^2 \equiv (p_\mu - p_{\nu_\mu})^2.
\label{eq:validity_momTransfer}
\end{equation}
This cut is implemented into {\mgamc} through the \texttt{dummy\_cuts} function  (file \texttt{dummy\_fct.f}). For the case of evolution by $p_T^2$, we require that $\nu_\mu$ satisfies the following restriction
\begin{equation}
 p_T^{\nu_\mu} > \mu_f.
\end{equation}

\begin{figure}[t!]
\centering
\subfigure[]{\includegraphics[width=.48\textwidth]{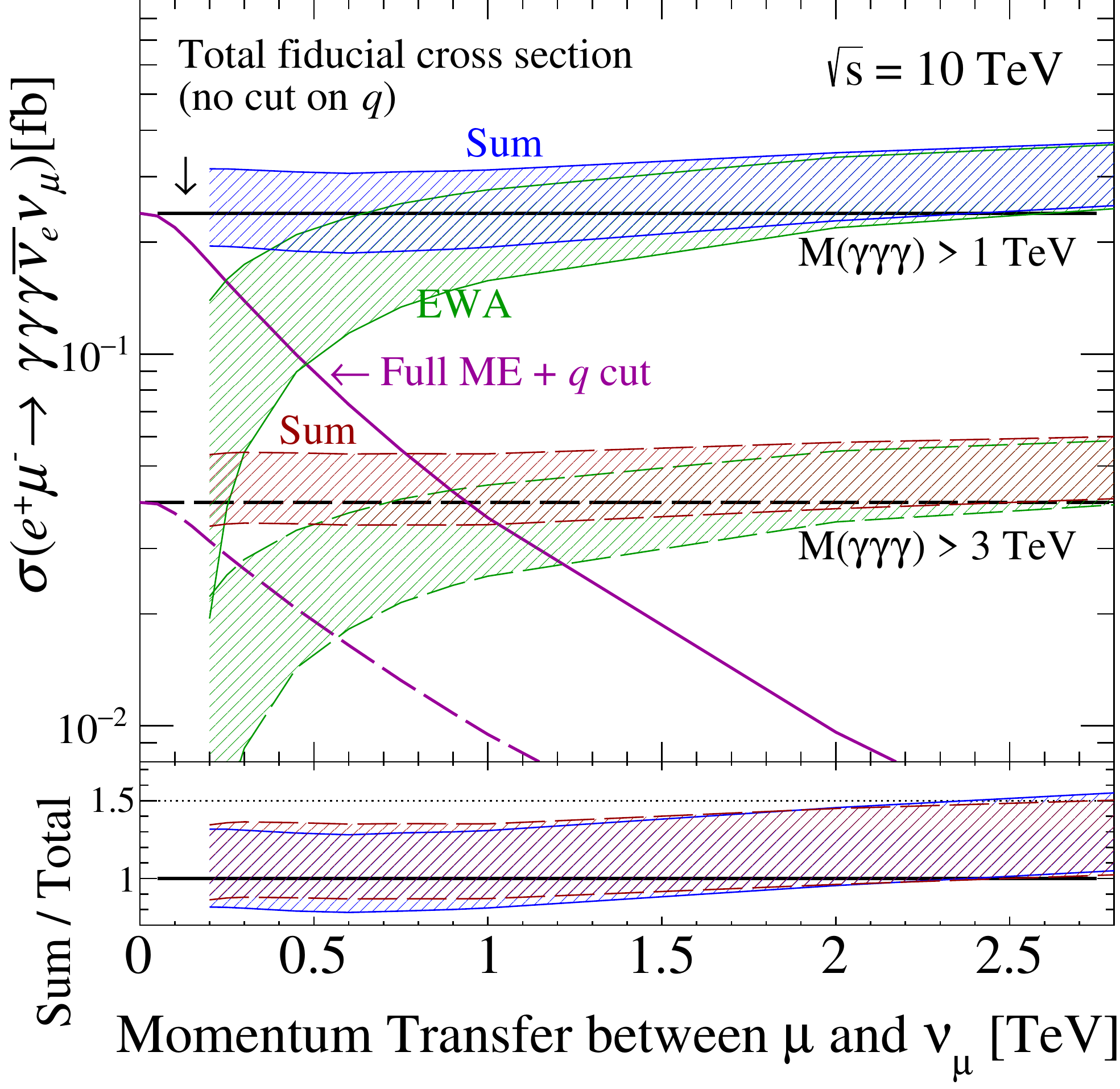} \label{fig:evaMatch_WW_AAA_ByQX_MuCo10TeV}}
\subfigure[]{\includegraphics[width=.48\textwidth]{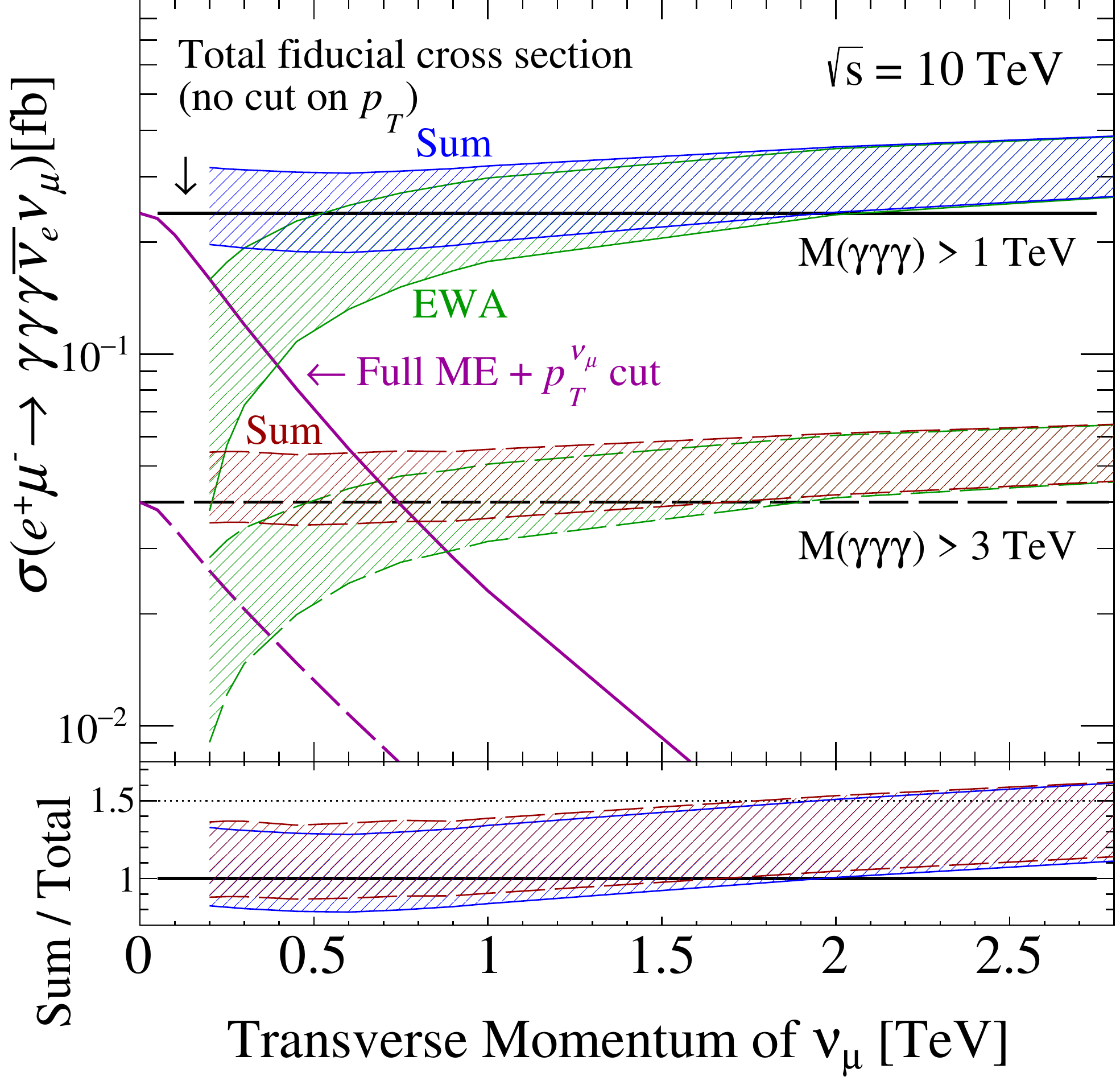} \label{fig:evaMatch_WW_AAA_ByPT_MuCo10TeV}}
\subfigure[]{\includegraphics[width=.48\textwidth]{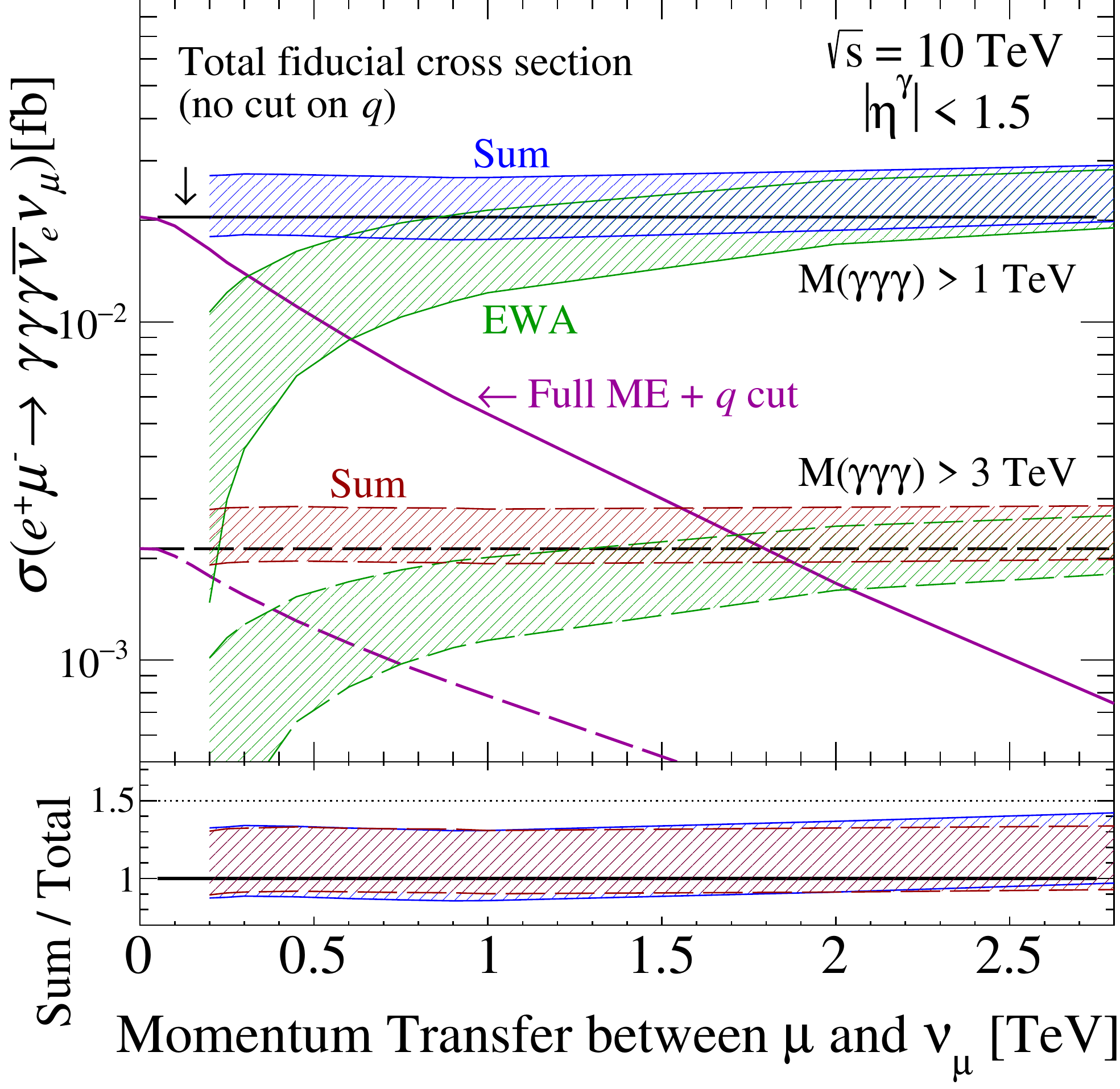} \label{fig:evaMatch_WW_AAA_ByQX_MuCo10TeV_TightETA}}
\subfigure[]{\includegraphics[width=.48\textwidth]{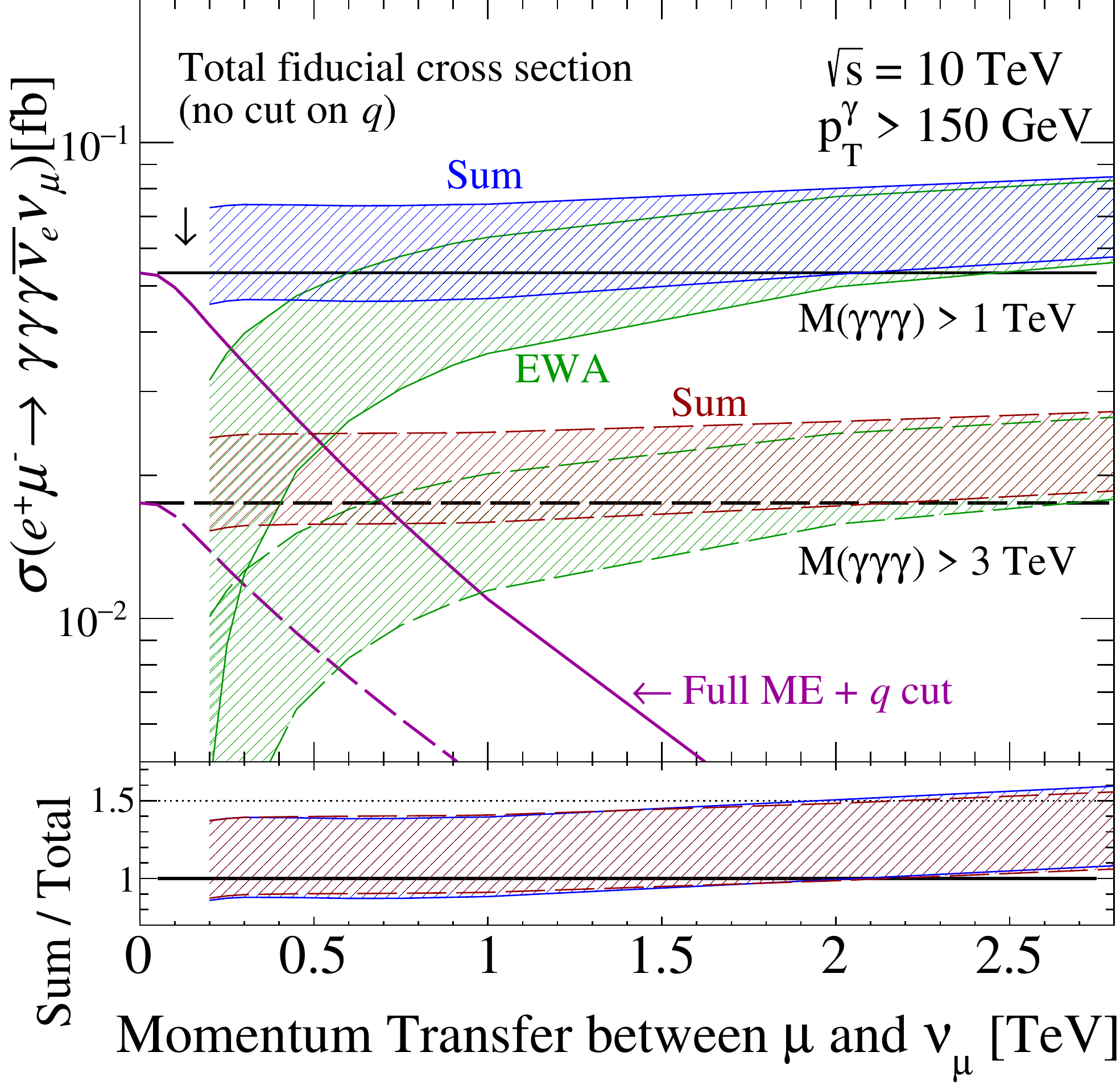} \label{fig:evaMatch_WW_AAA_ByQX_MuCo10TeV_TightPTX}}
\caption{Matrix element matching in EVA for $e^+\mu^- \to \gamma\gamma\gamma \overline{\nu_e}\nu_\mu$ using the (a) $q$-scheme and (b) $p_T$-scheme, at $\sqrt{s}=10\TeV$ and assuming cuts of Eq.~\eqref{eq:validity_matching_cutsBaseline}.
(c,d) Same as (a) but for $\vert\eta^\gamma\vert < 1.5$ and $p_T^\gamma>160\GeV$, respectively.
}
\label{fig:evaMatch_WW_AAA}
\end{figure}

In Fig.~\ref{fig:evaMatch_WW_AAA}, we show as a function of the factorization scale at $\sqrt{s}=10\TeV$:
(i) the total fiducial cross section for $e^+ \mu^- \to \overline{\nu_e}\nu_\mu \gamma \gamma \gamma$ without restrictions on the $\mu^- \to W^-$ splitting (flat black curve labeled ``Total''); 
(ii) the same but with restrictions on the $\mu^- \to W^-$ splitting (light purple curve labeled ``Full ME + $q$ cut'' or ``Full ME + $p_T$ cut'');
(iii) the fiducial cross section with its scale uncertainty band for $e^+ W^-_T \to \overline{\nu_e} \gamma \gamma \gamma$ (light green band labeled ``EWA'');
and (iv) the sum of the restricted cross section and EWA with scale uncertainty (dark blue or red bands labeled ``Sum'').
More specifically, we show the dependence when $\mu_f=\sqrt{\vert q_{\mu\nu_\mu}^2\vert}$ in Fig.~\ref{fig:evaMatch_WW_AAA}(a)  and when $\mu_f=p_T^{\nu_\mu}$ in Fig.~\ref{fig:evaMatch_WW_AAA}(b), at $M(\gamma\gamma\gamma)>1\TeV$ (solid lines) and
 $M(\gamma\gamma\gamma)>3\TeV$ (dashed lines).
For both $M(\gamma\gamma\gamma)$ cuts we also show in the lower panel of each plot the ratio of the summed result, with its uncertainty band, to the total fiducial cross section without collinear restrictions. The $f_{W_T}$ PDF and the restriction on $\mu^- \to W^-$ splittings are evaluated at the same value of $      \mu_f$. 
Uncertainty bands are obtained from three-point variation of $\mu_f$ in the EWA computations.

To establish a baseline, we start with Fig.~\ref{fig:evaMatch_WW_AAA_ByQX_MuCo10TeV} for evolution by $ q_{\mu\nu_\mu}$ at $M(\gamma\gamma\gamma)>1\TeV$. As one can anticipate, the fiducial cross sections with and without restrictions on $\mu^- \to W^-$ splittings converge to the same rate, about \confirm{$\sigma\sim0.3\fb$,} when effectively no cut is placed on the momentum transfer variable $\vert q_{\mu\nu_\mu}^2 \vert$, i.e.,  when $\mu_f\to0{\TeV}$. For increasing $\mu_f$, we observe a logarithmic-like dependence for both the full ME with a $\vert q_{\mu\nu_\mu}^2 \vert$ cut as well as the EWA result. In the former (latter), the rate decreases (increases) with increasing $\mu_f$. We observe that the restricted rate and the EWA rate are comparable over the approximate range $\mu_f\in(200\GeV,500\GeV)$. For \confirm{$\mu_f \gtrsim 1.5\TeV-2\TeV$,} we find that the full ME with a $\vert q_{\mu\nu_\mu}^2 \vert$ cut  becomes negligible, whereas the EWA result remains comparable to the total cross section. Over the range of $\mu_f$ investigated, we find that the EWA uncertainty band spans \confirm{from about $\pm75\%$ at $\mu_f\sim200\GeV$ to about $\pm20\%$ at $\mu_f\gtrsim2.5\TeV$}; we caution, however, that the large change to the uncertainty band reflects the extreme range of $\mu_f$ we investigate. (Varying $\mu_f$ downward to $\mu_f=M_W$, for instance, would induce a -100\% change since $\tilde{f}_{W_T}\sim \log\mu_f^2/M_W^2$.) Importantly, when adding the restricted and EWA cross sections, we find that the sum reproduces the total fiducial cross section to within uncertainties and shows a strong insensitivity to the matching scale for $\mu_f < 2\TeV$. This indicates that MEM was achieved. While not shown, we report that the central value for the summed result consistently overestimates the total result \confirm{by about $3\%-7\%$ for $\mu_f=200\GeV-1\TeV$, and up to $20\%-30\%$ for $\mu_f=2\TeV-3\TeV$.} For the case of $M(\gamma\gamma\gamma)>3\TeV$, we observe much of the same qualitative and quantitative behavior. The two notable differences are: (i) the obvious reduction in cross sections due to a more restrictive phase space cut, and (ii) a more stable summed result that features a central value (curve not shown) sitting nearly uniformly at \confirm{about $10\%$ above the total result for $\mu_f$ up to $\mu_f = 1\TeV$.}

\begin{figure}[t!]
\centering
\subfigure[]{\includegraphics[width=.48\textwidth]{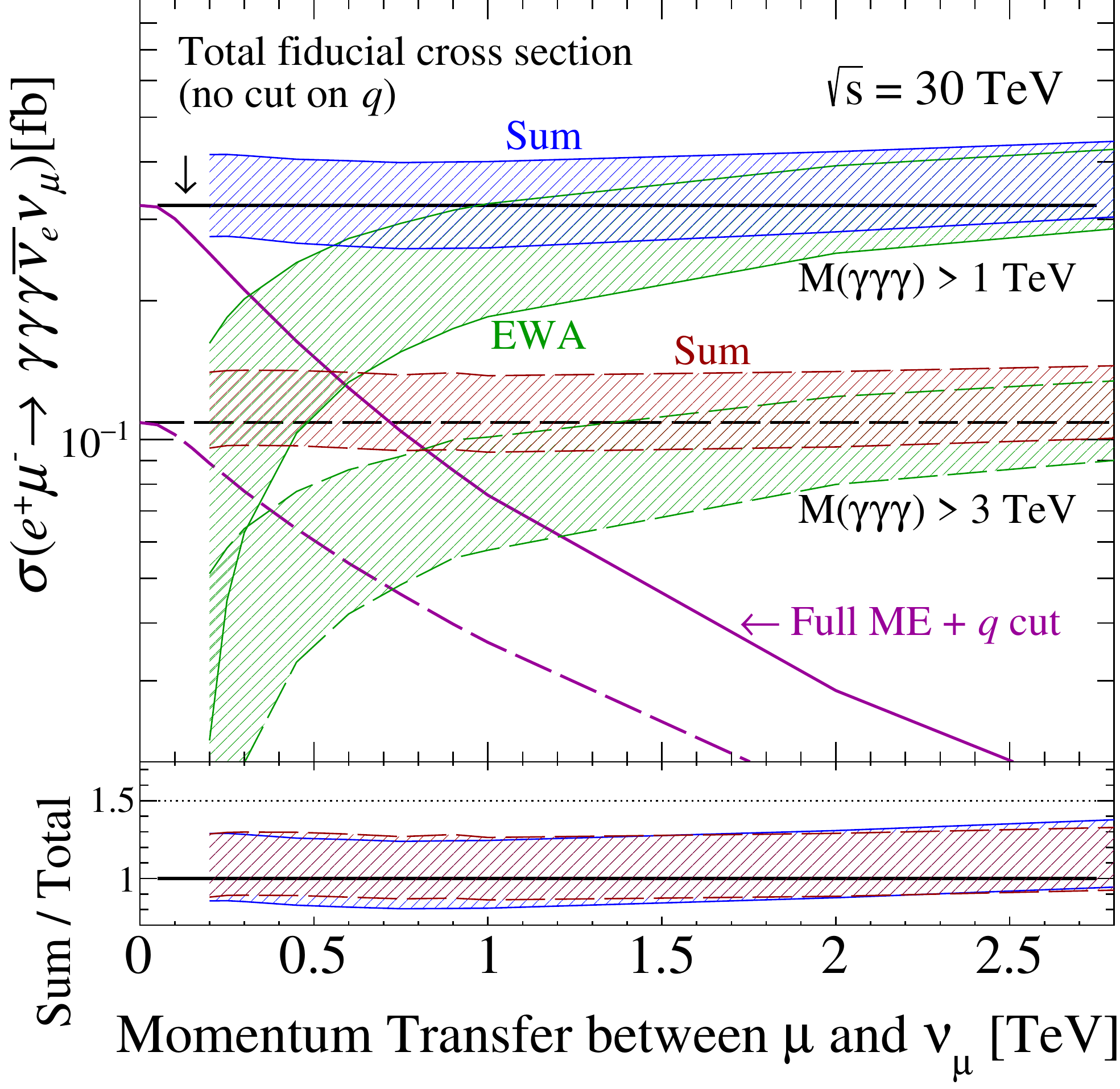} \label{fig:evaMatch_WW_AAA_ByQX_MuCo30TeV}}
\subfigure[]{\includegraphics[width=.48\textwidth]{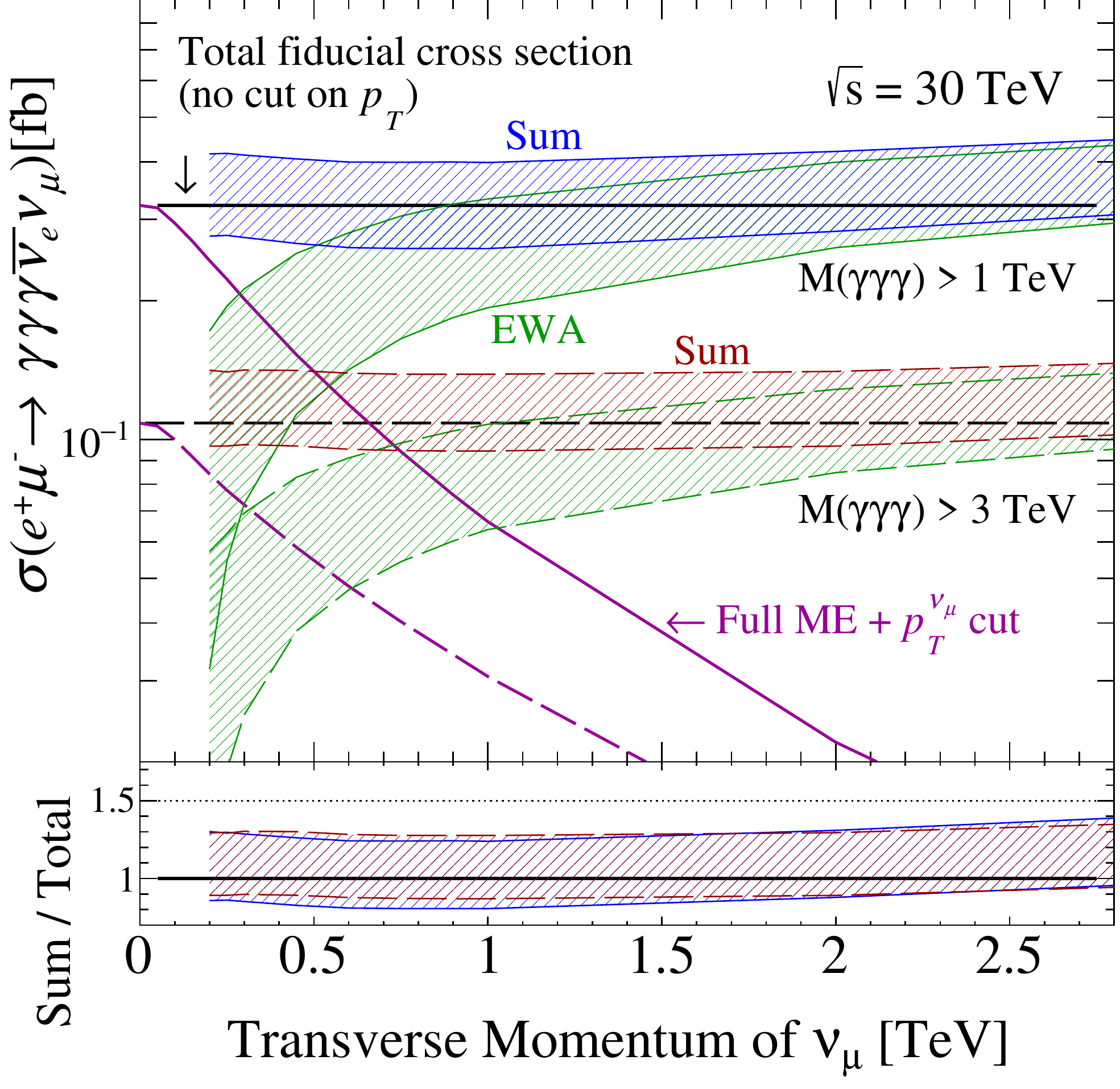} \label{fig:evaMatch_WW_AAA_ByPT_MuCo30TeV}}
\caption{Same as Fig.~\ref{fig:evaMatch_WW_AAA} (a) and (b) but for $\sqrt{s}=30\TeV$.}
\label{fig:evaMatch_WW_AAA_MuCo30TeV}
\end{figure}

Much can be learned from this exercise. First is that the power corrections $(\Delta \sigma^{\rm PC})$ that distinguish the total fiducial cross section without $t$-channel cuts $(\sigma^{\rm Total})$ from the summed result $(\sigma^{\rm Sum})$, and which scale as $\Delta\sigma^{\rm PC}=\sigma^{\rm Total} - \sigma^{\rm Sum}  \sim \mathcal{O}(\mu_f^2 / M_{WW}^2)$, are negative. (This is clear from the fact that the summed rate exceeds the total rate.) 
Second is that the $W_T$ PDFs, and by extension the $Z_T$ PDFs, work best when $\mu_f$ is set to lower values, e.g., $\mu_f\lesssim1\TeV$ when $\sqrt{s}=10\TeV$. Na\"ively, requiring relatively small $\mu_f$ may appear at odds with standard practices in pQCD, where $\mu_f$ are typically very high. However, in pQCD, one nearly always uses QCD PDFs that have been DGLAP-evolved; this has the effect of reducing the size of PDFs due to the RG running of $\alpha_s(\mu_r)$, thereby compensating for large collinear logarithms. The EW PDFs throughout this work are not evolved with EW-DGLAP equations and necessitates smaller $\mu_f$.
The third observation is that for sufficiently large values of $\mu_f$, the EWA rate begins to overestimate the total cross section. With little doubt, this can be attributed to a breakdown of the collinear approximation, which requires \textit{collinear} initial-state splittings, i.e., $\vert q_{\mu\nu_\mu}^2\vert\sim p_T^{\nu 2} \ll M_{WW}^2$. 
Fourth is that the relative independence of the summed result on a matching scale indicates that the logarithmic dependence on $\mu_f$ in each component effectively cancel, in accordance with expectations. This serves as a highly non-trivial check of MEM with $W_T/Z_T$ PDFs but also demonstrates the potential to support it in MC event generators.  
Finally, we note that the very similar and consistent size of the scale uncertainty bands across all channels can be attributed to the fact that we are working with fixed $\mu_f$. For example: in (a), the scale variations are given by the ratio $\sigma(\mu_f = \zeta \mu_0)/\sigma(\mu_f = \mu_0) = \log(\zeta^2 \mu_0^2/M_W^2)/\log(\mu_0^2/M_W^2)$ with $\zeta\in\{0.5,1,2\}$.

Focusing now on Fig.~\ref{fig:evaMatch_WW_AAA_ByPT_MuCo10TeV}, we show the same quantities as in Fig.~\ref{fig:evaMatch_WW_AAA_ByQX_MuCo10TeV} but for evolution by $p_T^{\nu_\mu}$. Qualitatively, we see many similarities to the previous case, including that the summed result reproduces the total fiducial cross section to within uncertainties for $\mu\lesssim1.5\TeV$. For larger $\mu_f$, the difference between the summed and total results exceeds the uncertainty band of the summed result. We attribute this breakdown of MEM to a breakdown of the collinear approximation. The breakdown is more explicit in this case since one is varying $p_T^{\nu_\mu}$ and is influenced by the large-$\xi$ enhancement discussed in Sec.~\ref{sec:validity_evo}.

To further explore the dependence on the kinematics of the hard $W^+W^-_T\to \gamma\gamma\gamma$ scattering process, we show in Figs.~\ref{fig:evaMatch_WW_AAA_ByQX_MuCo10TeV_TightETA} and \ref{fig:evaMatch_WW_AAA_ByQX_MuCo10TeV_TightPTX} the same quantities as in Figs.~\ref{fig:evaMatch_WW_AAA_ByQX_MuCo10TeV}, but consider the more restrictive phase space cuts (c) $\vert\eta\vert<1.5$ and (d) $p_T^\gamma>150\GeV$. We again observe many of the same qualitative features, indicating some degree of independence from the hard scattering process. (Fewer changes would suggest more universal-like behavior.) One notable difference is that the tighter $\eta^\gamma$ restriction helps extend the agreement between the summed and total results out to about $\mu_f\sim3\TeV$. Slightly better agreement is also observed for the tighter $p_T^\gamma$ requirement, but only until $\mu_f\sim2.25\TeV$.

Finally in Fig.~\ref{fig:evaMatch_WW_AAA_MuCo30TeV}(a) and (b), we show the same quantities as in Fig.~\ref{fig:evaMatch_WW_AAA}(a) and (b) but for $\sqrt{s}=30\TeV$. As the qualitative and quantitative findings are highly comparable, little needs to be said. One noteworthy difference, however, is that higher collider energy further alleviates differences between the total and summed results, and extends the agreement to $\mu_f\gtrsim3\TeV$. For all cases, the ``summed'' rate remains bigger than the ``total'' rate.


\section{Polarized Vector Boson Scattering at Muon Colliders}\label{sec:leptons}

A scenario in which the EVA promises to be highly relevant is that of a multi-TeV $\mpmm$ collider. It is worth reiterating that when considering a generic multi-particle state $\mathcal{F}$, the VBF mechanism becomes an increasingly important, if not dominant, production vehicle in lepton collisions as the energy increases~\cite{Costantini:2020stv}. Given this, it is natural to consider muon colliders as effective weak boson colliders and take full advantage of the EVA in order to systematically organize and  simplify the precision of scattering computations.

In this section, we explore the production of SM states generically parameterized by
\begin{equation}\label{eq:genEVA}
\sum_{V_\lambda\in\{\gamma_\lambda,\,Z_\lambda,\,W^\pm_\lambda\}} V_{\lambda_A}\, V' _{\lambda_B} \to \mathcal{F}, 
\end{equation}
where $\mathcal{F}$ contains up to $n_{\mathcal{F}}=4$ unpolarized states from the collection $\{H, t, \overline{t}, W^+, W^-, Z, \gamma \}$.
All helicity polarizations are defined in the hard-scattering frame, i.e., the rest frame of $\mathcal{F}$. We consider collider configurations over the range {$\sqrt{s}=2-30\TeV$}, and require final-state particles to obey the following kinematic and fiducial cuts:
\begin{equation}
M(\mathcal{F}) > 1\TeV, \quad p_T^I>50\GeV, \quad \vert y^I\vert <3, \quad\text{for}\quad \;I\in\{\mathcal{F}\}.
\label{eq:eva_cuts}
\end{equation}
The invariant mass cut of $1\TeV$ on the system $\mathcal{F}$ is needed to ensure that power corrections of the form $(M_V^2/M^2(\mathcal{F}))$ are negligible, in accordance with findings of Sec.~\ref{sec:validity}. We require moderate rapidities $y$ to avoid $t$- and $u$-channel singularities and instabilities associated with final-state particles, as advocated by Ref.~\cite{Han:2020uid}; in the massless limit, the pseudorapidity value of $y\to\eta=3$ corresponds to a polar angle of $\theta\approx5.7^\circ$. For all calculations involving $V_{T}V'_{T}$ or $V_{T}V'_{0}$ scattering, we set the central collinear factorization scale according to Eq.~\eqref{eq:scaleDefault} and display three-point scale uncertainties. Scale uncertainties are unavailable for $V_{0}V'_{0}$ scattering as the $\tilde{f}_{V_0}$ PDFs do not depend on $\mu_f$.

Our survey is organized in the following manner:
We start in Sec.~\ref{sec:leptons_higgs} with associated and multi-Higgs production. In Sec.~\ref{sec:leptons_top}, top quark and associated top quark production are discussed, followed by   diboson and triboson production in Secs.~\ref{sec:leptons_diboson} and \ref{sec:leptons_triboson}, respectively.

\subsection{Higgs production}\label{sec:leptons_higgs}

\begin{figure*}[!t]
\begin{center} 
    \subfigure[]{\includegraphics[width=0.48\textwidth]{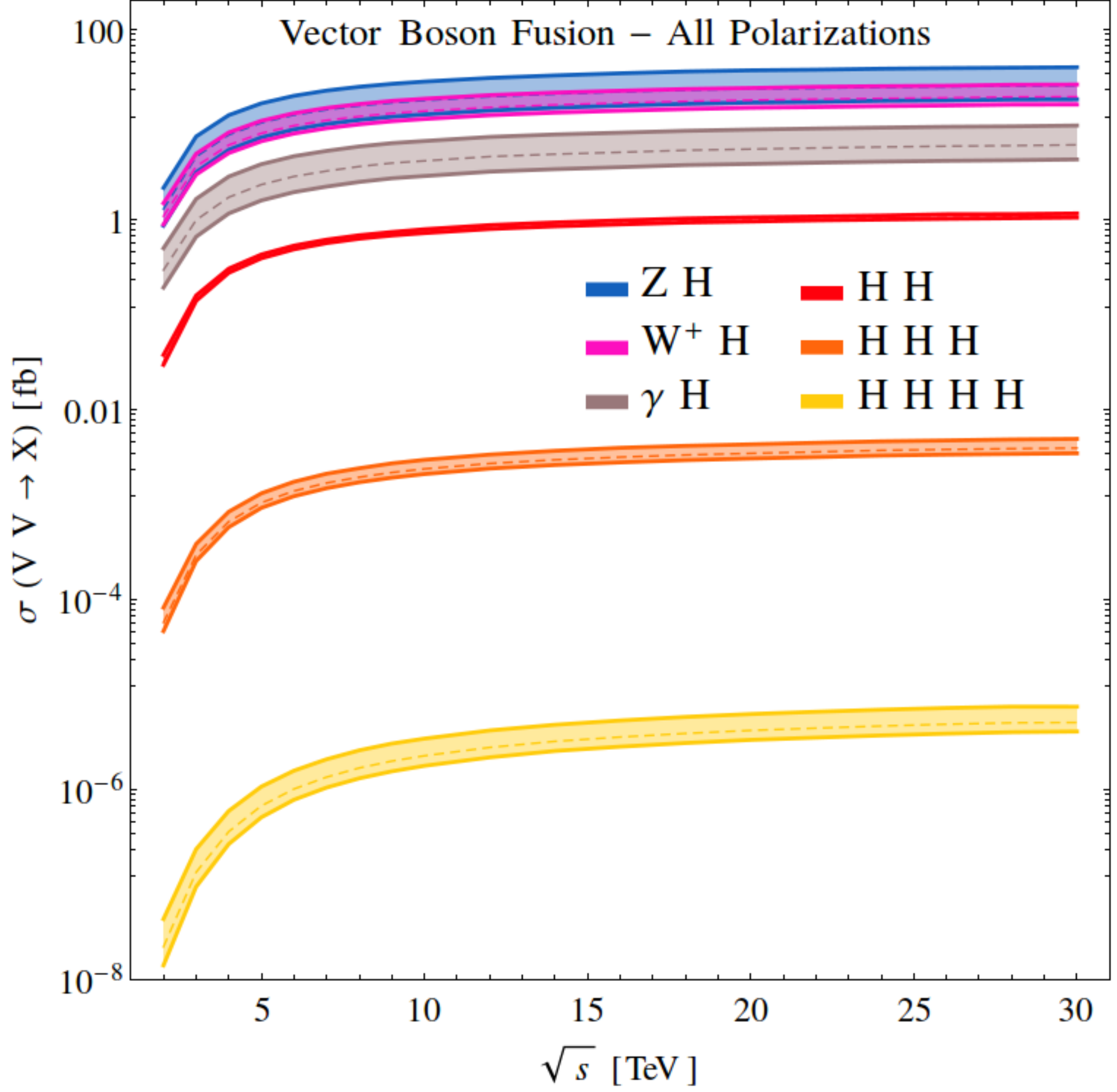}}
    \subfigure[]{\includegraphics[width=0.48\textwidth]{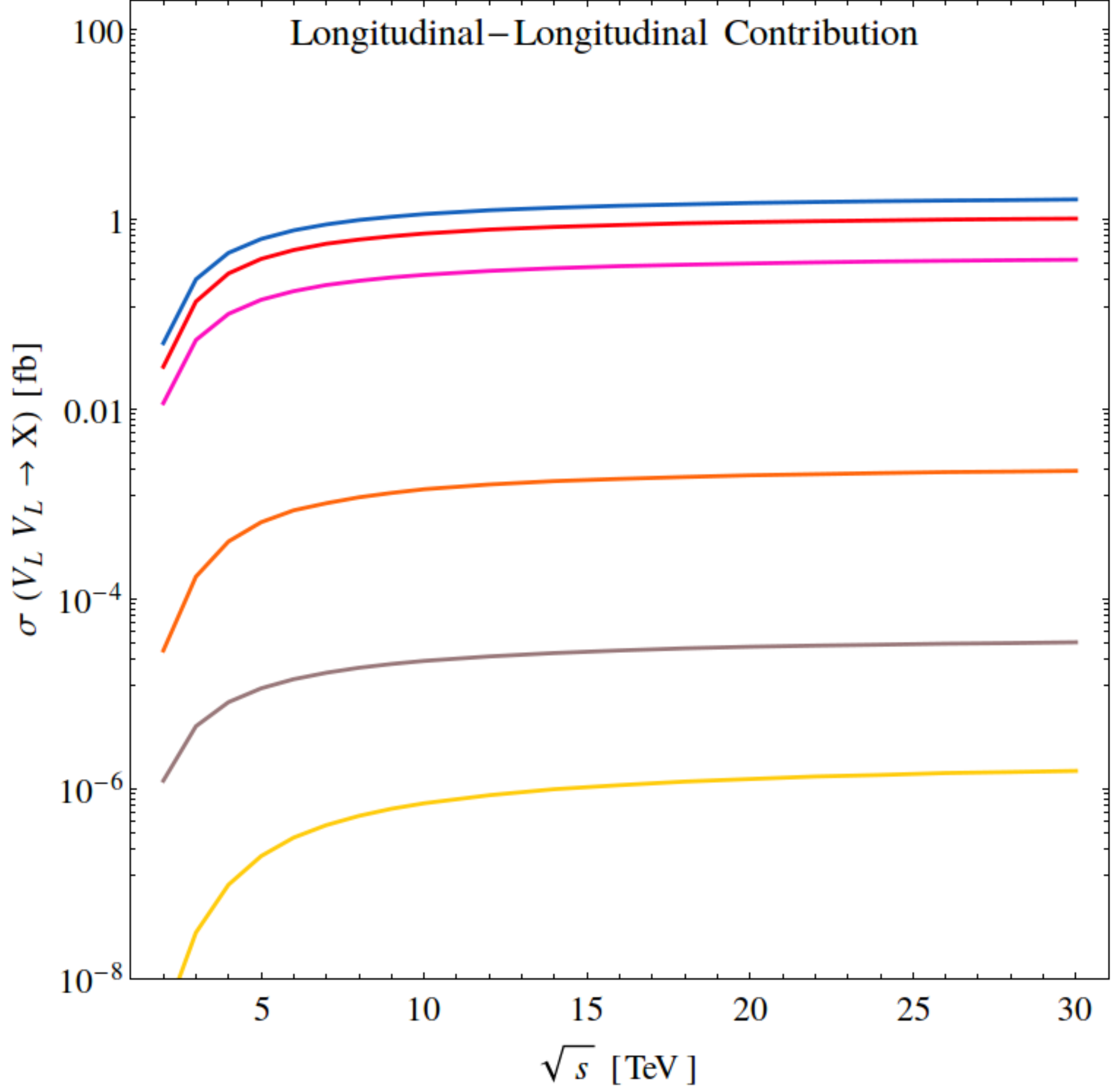}}
    \\
    \subfigure[]{\includegraphics[width=0.48\textwidth]{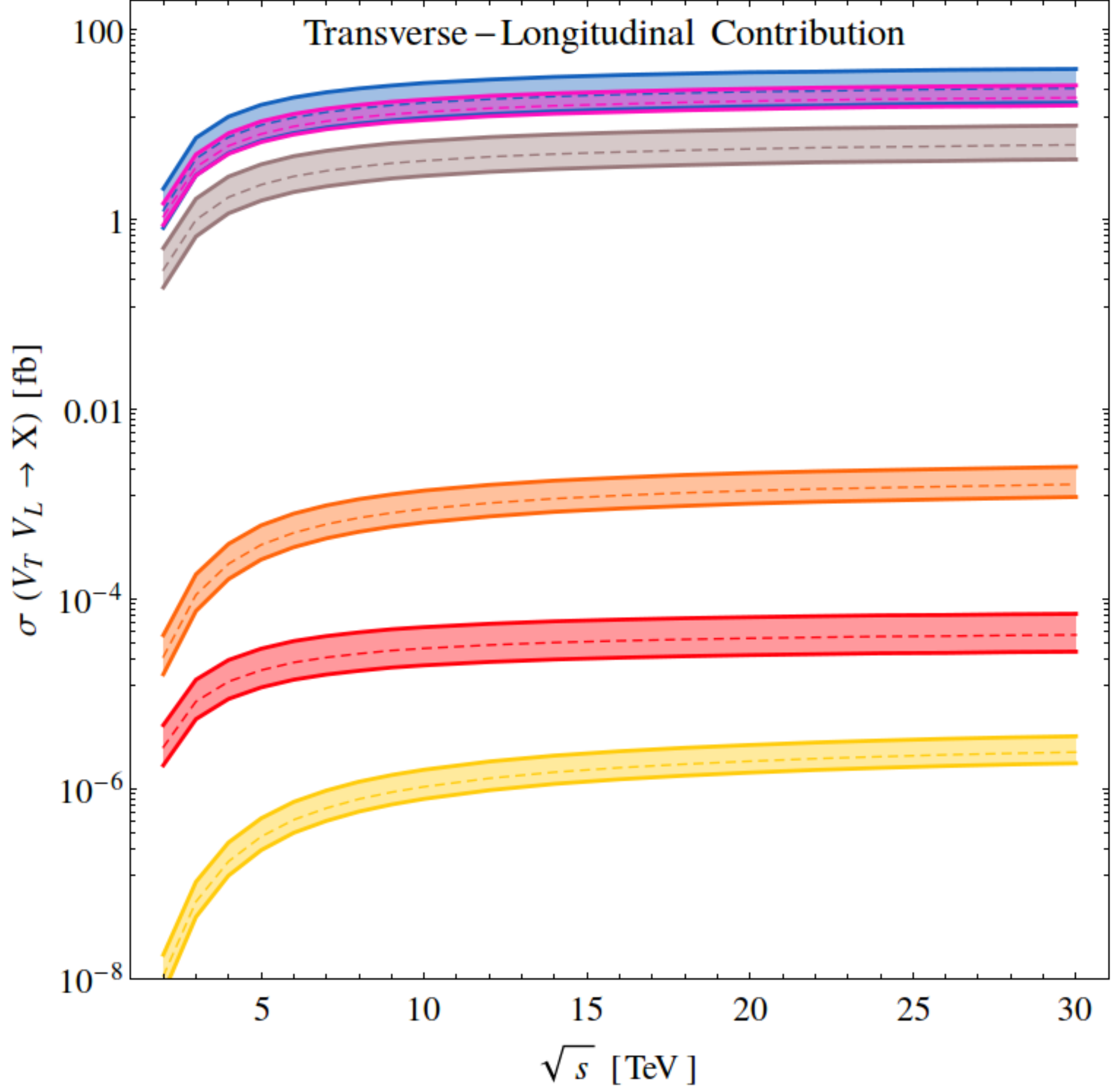}}   
    \subfigure[]{\includegraphics[width=0.48\textwidth]{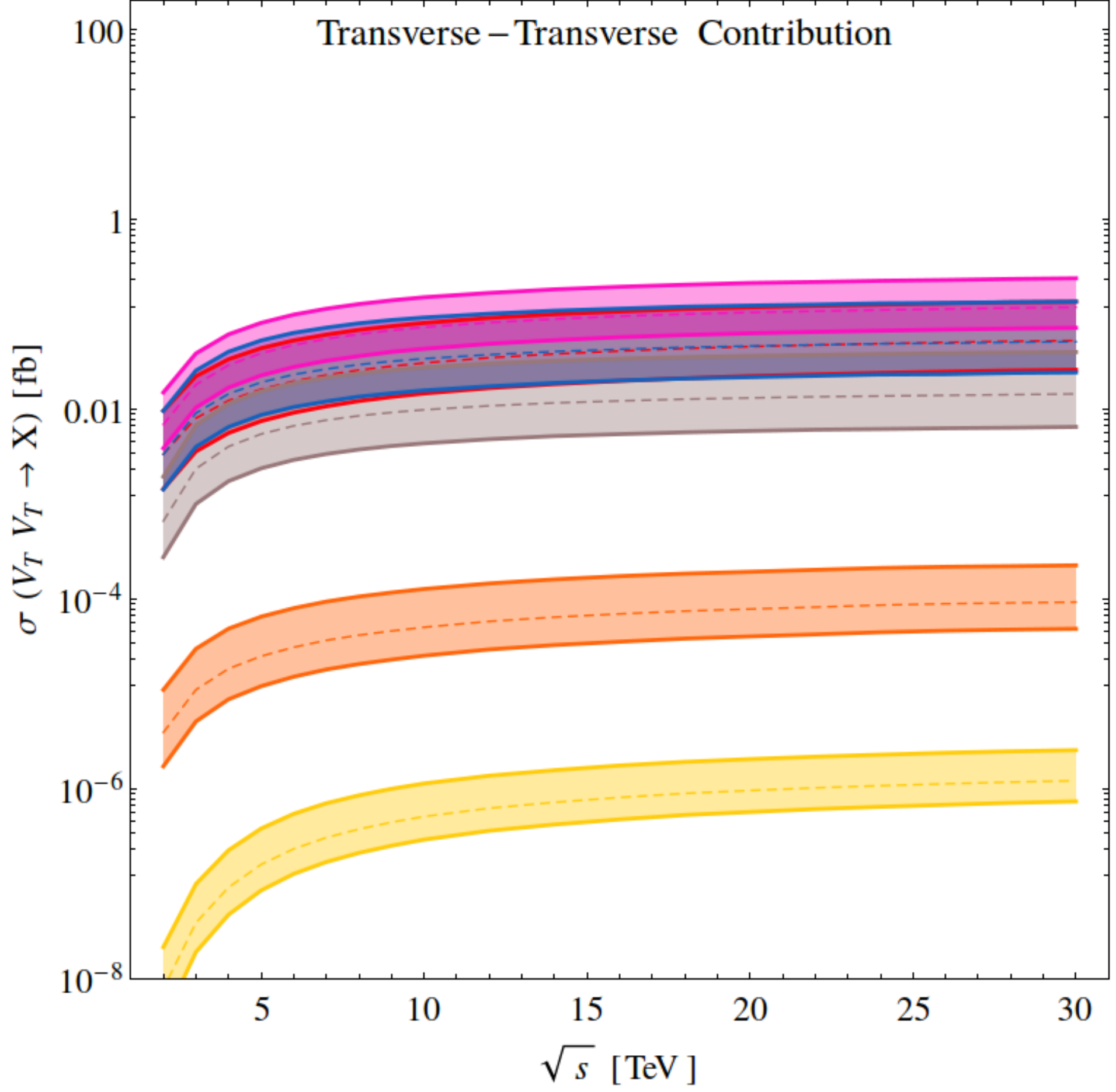}}
\end{center}
        \caption{As a function of collider energy,
        the fiducial cross section and scale uncertainties under the EVA for associated Higgs and multi-Higgs production in $\mpmm$ collisions from
        (a) all initial-state EW boson polarizations, 
        (b) longitudinal-longitudinal scattering, 
        (c) transverse-longitudinal scattering, and
        (d) transverse-transverse scattering.}
    \label{fig:eva_nH}
\end{figure*}

We first consider Higgs production in $\mpmm$ collisions and focus on the channels    
\begin{equation}
V_{\lambda_A} V'_{\lambda_B} \quad\to\quad 
ZH, \quad W^+H, \quad \gamma H,\quad HH, \quad HHH, \quad\text{and}\quad HHHH.
\label{eq:eva_nH}
\end{equation}
In Fig.~\ref{fig:eva_nH}, we show as a function of collider energy [TeV] the fiducial cross section [fb],  along with the associated three-point scale uncertainty band (from left to right in Eq.~\eqref{eq:eva_nH} corresponds the darkest to lightest color band), as mediated by
(a) all initial-state EW boson polarizations, 
(b) longitudinal-longitudinal scattering, 
(c) transverse-longitudinal scattering, and
(d) transverse-transverse scattering.

At the unpolarized level in Fig.~\ref{fig:eva_nH}(a), we observe a strong hierarchy between associated and multi-Higgs production, with $VH$ production rates being more than an order of magnitude larger than $HH$ production. More specifically, the $ZH$ and $W^+H$ rates span roughly $\sigma\sim5-25\fb$ for $\sqrt{s}\sim5-30\TeV$, while $HH$ production reaches about $\sigma\sim0.5-1\fb$ over the same range. By $C$-symmetry, the $W^-H$ production rate is the same as $W^+H$, and therefore is not shown.
The production of $\gamma H$ is universally smaller than $ZH$ and $WH$ by about a factor of $2-3$.
Since $ZH$ and $\gamma H$ are both mediated by $W^+W^-$ scattering, the difference can be attributed to the difference in $WWZ$ and $WW\gamma$ gauge couplings, where $\sigma_{ZH}/\sigma_{\gamma H} \sim (g_{ZWW}^2/g_{\gamma WWW}^2)= (g\cos\theta_W/g\sin\theta_W)^2\sim3$.
As the $HH$ channel is also driven by $W^+W^-$ fusion (recall that the $Z\mu\mu$ coupling is smaller than the $W\mu\nu_\mu$ coupling), it is tempting to also attribute the relative size of the $VH$ and $HH$ production rates to the coupling ratio $(g/\lambda)^2\sim 27$. However, as shown in Fig.~\ref{fig:eva_nH}(b) and discussed in the next paragraph, this is not actually the case.  All $VH$ and $HH$ channels are about 2-to-3 orders of magnitude larger than $HHH$ production, which is yet another 2-to-3 orders of magnitude larger than $HHHH$ production. The (relatively) tiny triple and quadruple Higgs cross sections follow from the compound effect of a small Higgs self-coupling and phase space suppression. For $\sqrt{s}\lesssim5\TeV$, all rates are sensitive to small changes in collider energy due to threshold effects; above this scale, the energy dependence becomes milder.

\begin{table}[t!]
\begin{center}
\resizebox{\textwidth}{!}{ 
\renewcommand*{\arraystretch}{1.1}\setlength{\tabcolsep}{10pt}
	\begin{tabular}{r|l l l l}
		\hline \hline
		& 
		&\multicolumn{3}{c}{$\sigma$ [fb]}\\
		& {\mgamc} syntax
		&$\sqrt{s}=3$ TeV&$\sqrt{s}=14$ TeV&$\sqrt{s}=30$ TeV\\
		\hline
		$\sum V_{\lambda_A} V'_{\lambda_B}\to Z\, H$&\texttt{vxp vxm > z h}&$4.5\cdot10^{0}$ $^{+63 \%}_{-31 \%}$&$2.1\cdot10^{1}$ $^{+57 \%}_{-28 \%}$&$2.5\cdot10^{1}$ $^{+56 \%}_{-28 \%}$\\
		$V_T V'_T\to Z\,H$&\texttt{vxp\{T\} vxm\{T\} > z h}&$9.0\cdot10^{-3}$ $^{+181 \%}_{-56 \%}$&$4.0\cdot10^{-2}$ $^{+169 \%}_{-53 \%}$&$5.1\cdot10^{-2}$ $^{+165 \%}_{-52 \%}$\\
		$V_0 V'_T\to Z\,H$&\texttt{vxp\{0\} vxm\{T\} > z h}&$4.3\cdot10^{0}$ $^{+66 \%}_{-33 \%}$&$1.9\cdot10^{1}$ $^{+61 \%}_{-30 \%}$&$2.4\cdot10^{1}$ $^{+59 \%}_{-30 \%}$\\
		$V_0 V'_0\to Z\,H$&\texttt{vxp\{0\} vxm\{0\} > z h}&$2.3\cdot10^{-1}$&$1.3\cdot10^{0}$&$1.6\cdot10^{0}$\\
		\hline
		$\sum V_{\lambda_A} V'_{\lambda_B}\to W^+ H$&\texttt{vxp vxm > w+ h}&$3.6\cdot10^{0}$ $^{+36 \%}_{-18 \%}$&$1.6\cdot10^{1}$ $^{+34 \%}_{-17 \%}$&$1.9\cdot10^{1}$ $^{+34 \%}_{-17 \%}$\\
		$V_T V'_T\to W^+ H$&\texttt{vxp\{T\} vxm\{T\} > w+ h}&$1.8\cdot10^{-2}$ $^{+113 \%}_{-43 \%}$&$8.8\cdot10^{-2}$ $^{+105 \%}_{-40 \%}$&$1.2\cdot10^{-1}$ $^{+102 \%}_{-39 \%}$\\
		$V_0 V'_T\to W^+ H$&\texttt{vxp\{0\} vxm\{T\} > w+ h}&$3.5\cdot10^{0}$ $^{+36 \%}_{-18 \%}$&$1.5\cdot10^{1}$ $^{+35 \%}_{-17 \%}$&$1.9\cdot10^{1}$ $^{+34 \%}_{-17 \%}$\\
		$V_0 V'_0\to W^+ H$&\texttt{vxp\{0\} vxm\{0\} > w+ h}&$5.3\cdot10^{-2}$ &$3.0\cdot10^{-1}$ &$3.7\cdot10^{-1}$\\
		\hline
		$\sum V_{\lambda_A} V'_{\lambda_B}\to \gamma\,H$&\texttt{vxp vxm > a h}&$9.8\cdot10^{-1}$ $^{+67 \%}_{-33 \%}$&$4.8\cdot10^{0}$ $^{+61 \%}_{-30 \%}$&$6.0\cdot10^{0}$ $^{+60 \%}_{-30 \%}$\\
		$V_T V'_T\to \gamma\,H$&\texttt{vxp\{T\} vxm\{T\} > a h}&$2.3\cdot10^{-3}$ $^{+188 \%}_{-57 \%}$&$1.1\cdot10^{-2}$ $^{+178 \%}_{-55 \%}$&$1.4\cdot10^{-2}$ $^{+176 \%}_{-55 \%}$\\
		$V_0 V'_T\to \gamma\,H$&\texttt{vxp\{0\} vxm\{T\} > a h}&$9.8\cdot10^{-1}$ $^{+67 \%}_{-33 \%}$&$4.8\cdot10^{0}$ $^{+61 \%}_{-30 \%}$&$6.0\cdot10^{0}$ $^{+59 \%}_{-30 \%}$\\
		$V_0 V'_0\to \gamma\,H$&\texttt{vxp\{0\} vxm\{0\} > a h}&$4.5\cdot10^{-6}$ &$2.7\cdot10^{-5}$ &$3.5\cdot10^{-5}$\\
		\hline
		$\sum V_{\lambda_A} V'_{\lambda_B}\to HH$&\texttt{vxp vxm >h h}&$1.4\cdot10^{-1}$ $^{+10 \%}_{-3 \%}$&$8.5\cdot10^{-1}$ $^{+7 \%}_{-2 \%}$&$1.1\cdot10^{0}$ $^{+8 \%}_{-2 \%}$\\
		$V_T V'_T\to HH$&\texttt{vxp\{T\} vxm\{T\} > h h}&$7.9\cdot10^{-3}$ $^{+177 \%}_{-55 \%}$&$3.8\cdot10^{-2}$ $^{+162 \%}_{-52 \%}$&$5.2\cdot10^{-2}$ $^{+157 \%}_{-51 \%}$\\
		$V_0 V'_T\to HH$&\texttt{vxp\{0\} vxm\{T\} > h h}&$8.2\cdot10^{-6}$ $^{+69 \%}_{-35 \%}$&$3.4\cdot10^{-5}$ $^{+67 \%}_{-38 \%}$&$4.1\cdot10^{-5}$ $^{+67 \%}_{-33 \%}$\\
		$V_0 V'_0\to HH$&\texttt{vxp\{0\} vxm\{0\} > h h}&$1.3\cdot10^{-1}$ &$8.2\cdot10^{-1}$ &$1.0\cdot10^{0}$\\
		\hline
		$\sum V_{\lambda_A} V'_{\lambda_B}\to HHH$&\texttt{vxp vxm > h h h}&$2.9\cdot10^{-4}$ $^{+30 \%}_{-14 \%}$&$2.9\cdot10^{-3}$ $^{+24 \%}_{-12 \%}$&$3.9\cdot10^{-3}$ $^{+25 \%}_{-12 \%}$\\
		$V_T V'_T\to HHH$&\texttt{vxp\{T\} vxm\{T\} > h h h}&$1.1\cdot10^{-5}$ $^{+170 \%}_{-54 \%}$&$6.4\cdot10^{-5}$ $^{+150 \%}_{-49 \%}$&$9.2\cdot10^{-5}$ $^{+144 \%}_{-48 \%}$\\
		$V_0 V'_T\to HHH$&\texttt{vxp\{0\} vxm\{T\} > h h h}&$1.1\cdot10^{-4}$ $^{+65 \%}_{-32 \%}$&$1.1\cdot10^{-3}$ $^{+55 \%}_{-27 \%}$&$1.6\cdot10^{-3}$ $^{+53 \%}_{-26 \%}$\\
		$V_0 V'_0\to HHH$&\texttt{vxp\{0\} vxm\{0\} > h h h}&$1.7\cdot10^{-4}$ &$1.7\cdot10^{-3}$ &$2.3\cdot10^{-3}$\\
		\hline
		$\sum V_{\lambda_A} V'_{\lambda_B}\to HHHH$&\texttt{vxp vxm > h h h h}&$1.3\cdot10^{-7}$ $^{+75 \%}_{-30 \%}$&$3.2\cdot10^{-6}$ $^{+49 \%}_{-21 \%}$&$5.0\cdot10^{-6}$ $^{+46 \%}_{-20 \%}$\\
		$V_T V'_T\to HHHH$&\texttt{vxp\{T\} vxm\{T\} > h h h h}&$3.8\cdot10^{-8}$ $^{+156 \%}_{-51 \%}$&$7.1\cdot10^{-7}$ $^{+118 \%}_{-42 \%}$&$1.2\cdot10^{-6}$ $^{+110 \%}_{-39 \%}$\\
		$V_0 V'_T\to HHHH$&\texttt{vxp\{0\} vxm\{T\} > h h h h}&$6.4\cdot10^{-8}$ $^{+62 \%}_{-31 \%}$&$1.5\cdot10^{-6}$ $^{+49 \%}_{-25 \%}$&$2.4\cdot10^{-6}$ $^{+47 \%}_{-24 \%}$\\
		$V_0 V'_0\to HHHH$&\texttt{vxp\{0\} vxm\{0\} > h h h h}&$3.0\cdot10^{-8}$ &$9.8\cdot10^{-7}$ &$1.5\cdot10^{-6}$\\
		\hline
		\hline
	\end{tabular}
	} 
	\caption{
	Fiducial cross sections with scale uncertainties for associated Higgs  and multi-Higgs production in $\mpmm$ collisions
		through $V_{\lambda_A} V'_{\lambda_B}$ scattering, where  $V_\lambda\in\{\gamma_\lambda,Z_\lambda,W_\lambda^\pm\}$,
	under the EVA for polarization-summed and polarized initial states, at $\sqrt{s}=3,~14,$ and 30\TeV.
	All helicity polarizations are defined in the hard-scattering frame.
	Phase space cuts are summarized in Eq.~\eqref{eq:eva_cuts}.
	Also shown is the {\mgamc} syntax for modeling the process assuming the multi-particle definitions 
	``\texttt{vxp=z,w+,a}'' and ``\texttt{vxm=z,w-,a}.''
	The configuration $(0,T)$ implies a sum over both $(0,T)$ and $(T,0)$, and uses the syntax
	\texttt{generate vxp\{0\} vxm\{T\} > ...; add process vxp\{T\} vxm\{0\} > ...}
	}  
	\label{tab:eva_nH}
\end{center}
\end{table}

As a function of polarization, one sees from Figs.~\ref{fig:eva_nH}(b-d) several notable characteristics.
For instance: 
The $VH$ channels are driven almost exclusively by $V_{0}V_{T}'$ scattering, with a sub-leading component of $V_{0}V_{0}'$ scattering. $HH$ is dominated by $V_{0}V_{0}'$ scattering, which accounts for the smaller dependence on factorization scales  when summing over all polarizations, but also contains a sub-leading $V_{T}V_{T}'$ contribution. Notably, $ZH$, $W^+H$, and $HH$ all have comparable $V_{0}V_{0}'$ scattering rates, which is in line with the Goldstone Equivalence Theorem. This indicates that the $\sigma_{VH}$-$\sigma_{HH}$ hierarchy observed in Figs.~\ref{fig:eva_nH}(a) is actually due to the compound effect of logarithmic enhancements in $\tilde{f}_{V_T}$ PDFs and helicity configurations allowed by angular momentum conservation, e.g., $V_T V'_0\to HH$ is helicity suppressed.
We report that $HHH$ production has significant and comparable contributions from $V_{0}V_{0}'$ and $V_{T}V_{0}'$ scattering, but only a marginal contribution from $V_{T}V_{T}'$.
Interestingly, $HHHH$ receives comparable contributions from all polarization configurations.
As one can expect, production from $V_{T}V_{T}$ scattering exhibits larger scale uncertainties than in $V_{T}V_{0}$ scattering. For each unpolarized and polarized scattering configuration, we document in Table~\ref{tab:eva_nH} the relevant {\mgamc} process syntax that enables our computation and the fiducial cross section [fb] with its scale uncertainty [\%], for representative collider energies.

\subsection{Top and associated top production}\label{sec:leptons_top}

\begin{figure*}[!t]
\begin{center} 
    \subfigure[]{\includegraphics[width=0.48\textwidth]{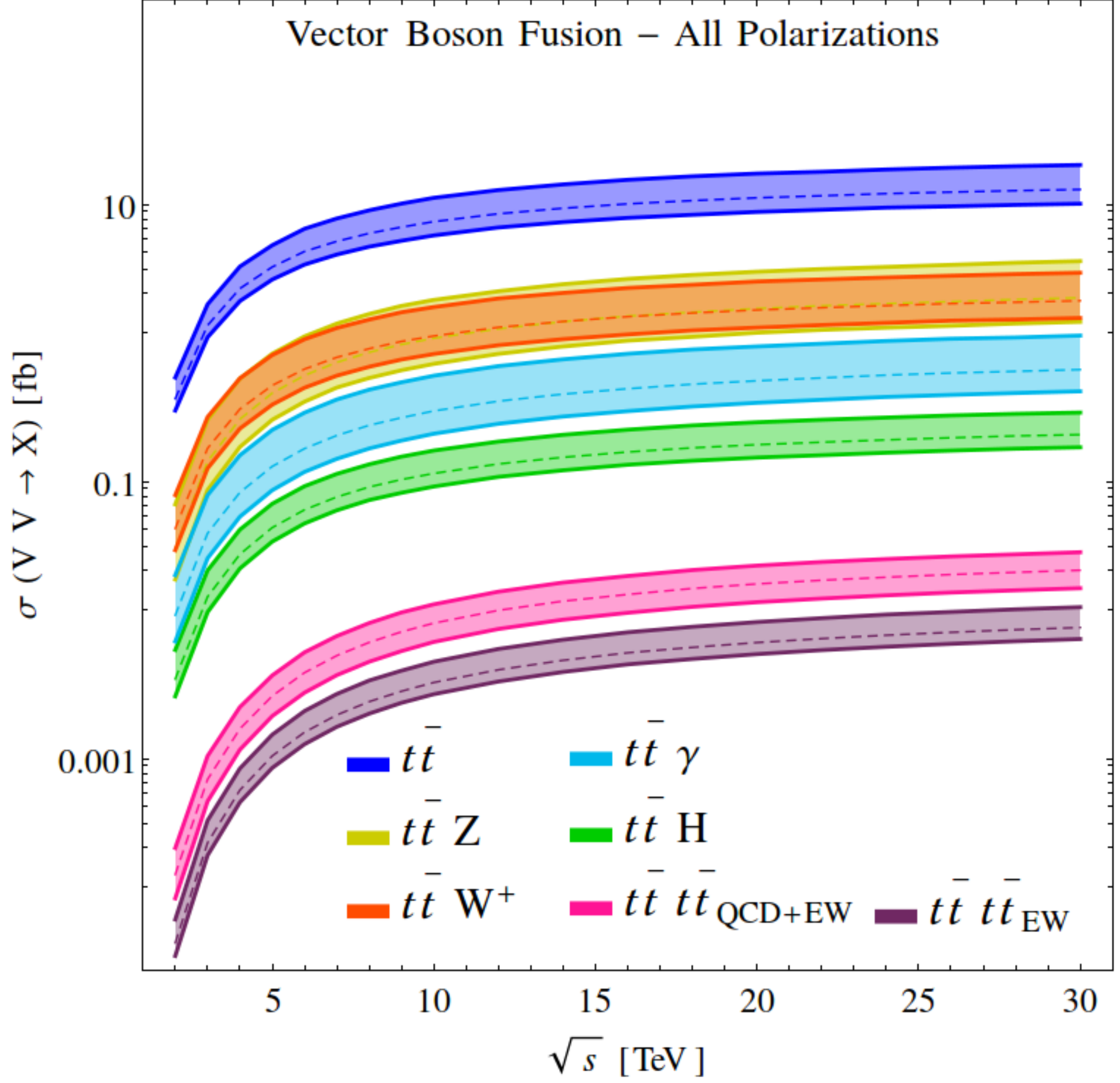}}
    \subfigure[]{\includegraphics[width=0.48\textwidth]{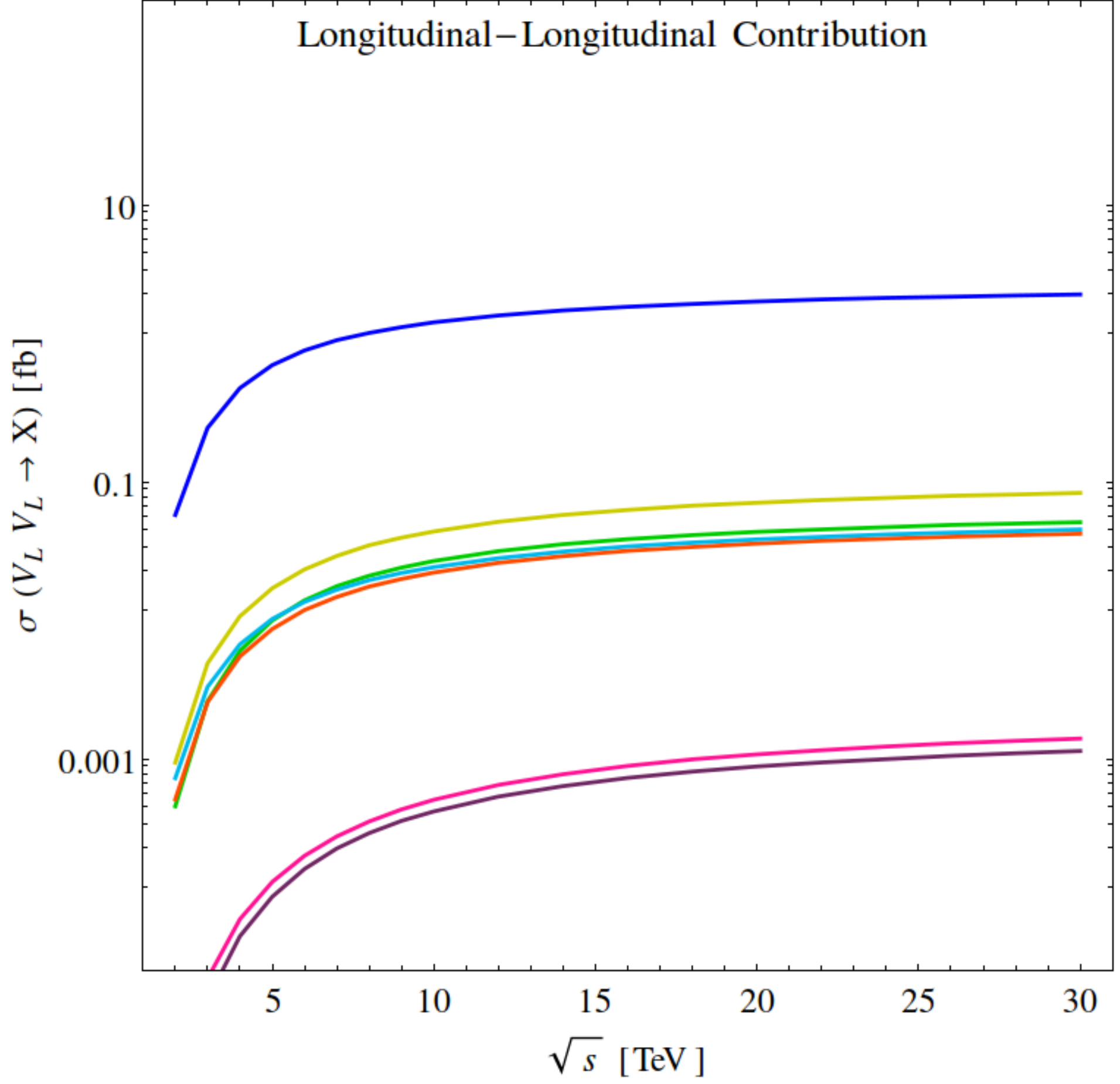}}
    \\
    \subfigure[]{\includegraphics[width=0.48\textwidth]{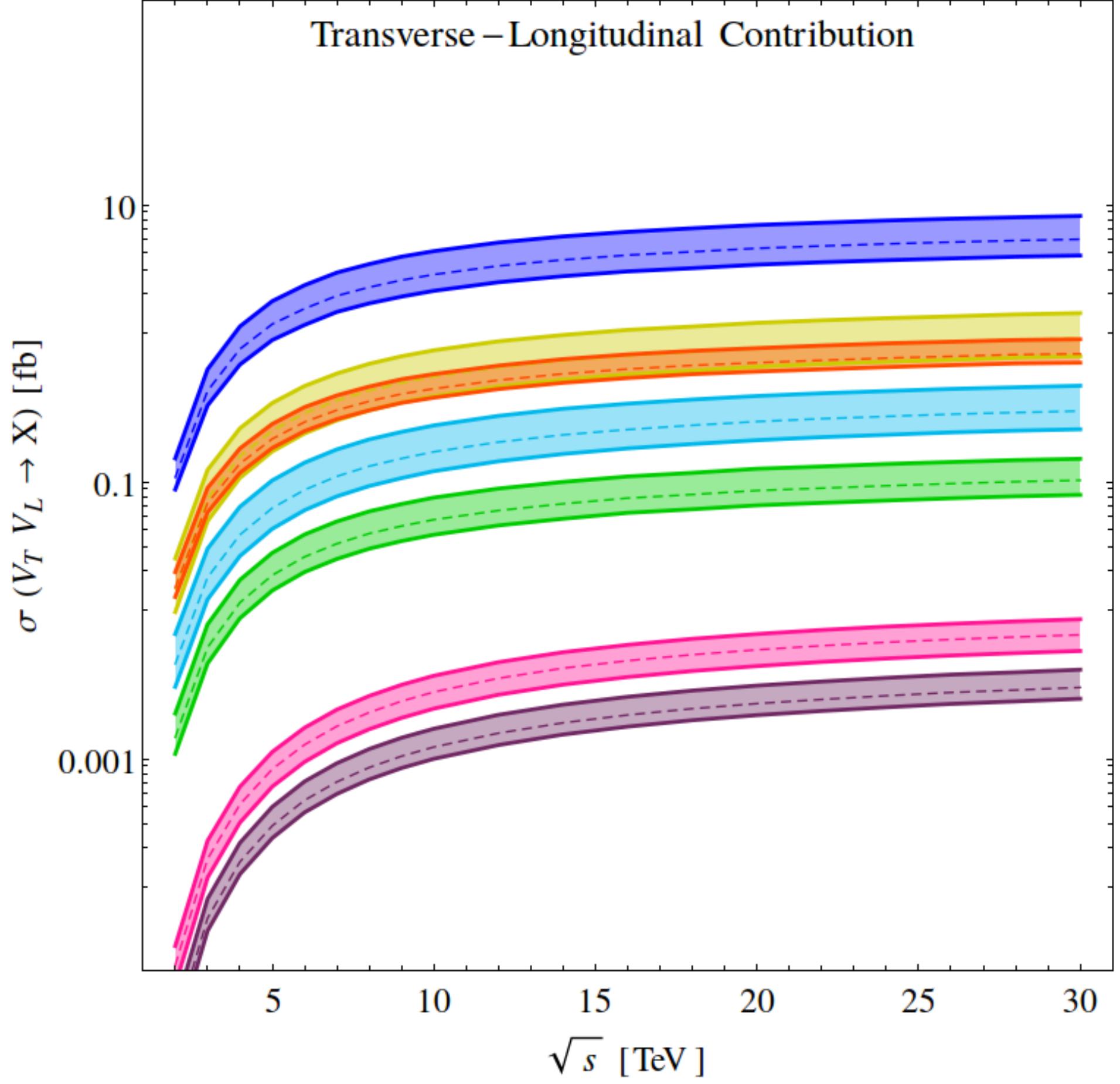}}   
    \subfigure[]{\includegraphics[width=0.48\textwidth]{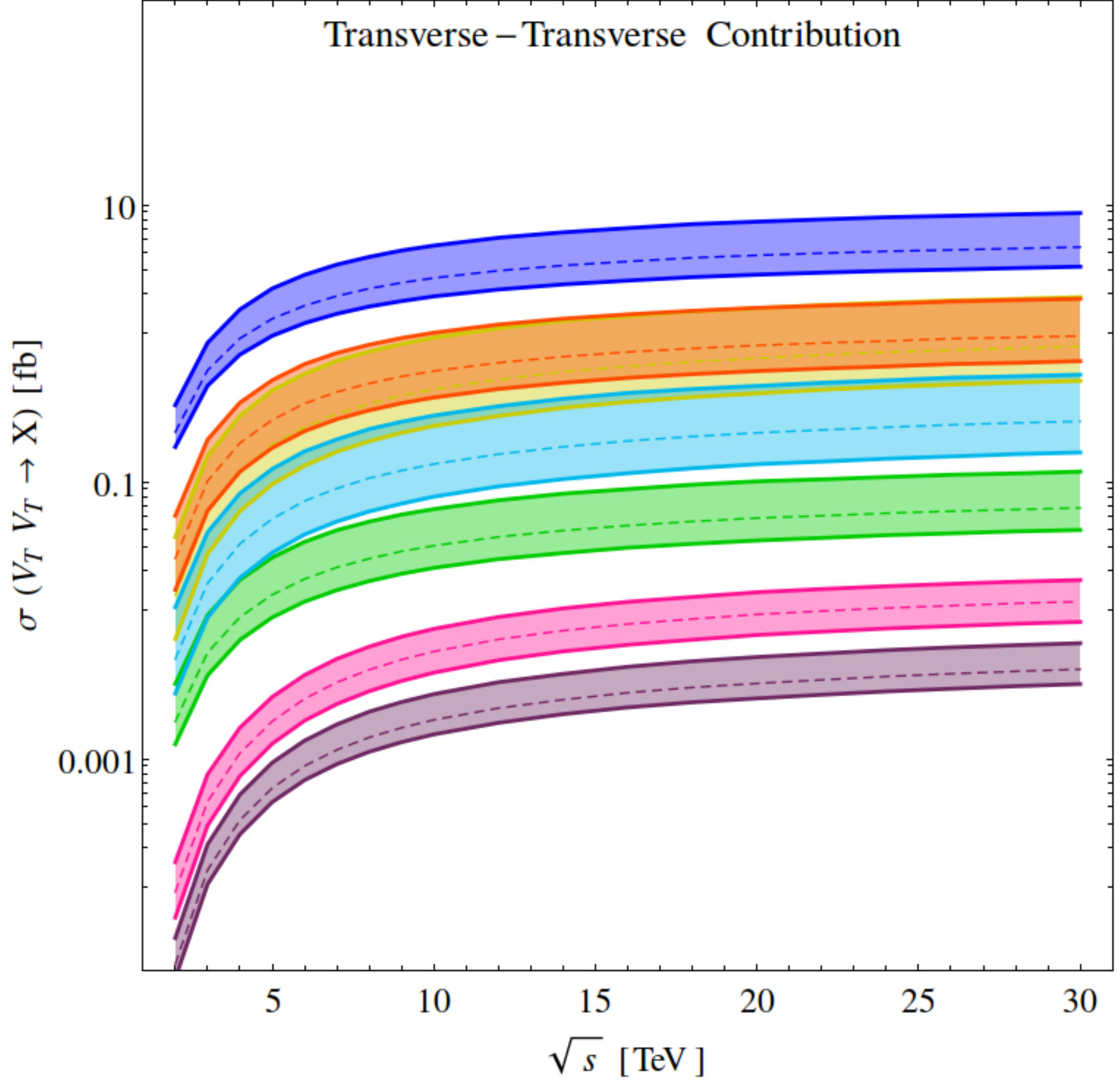}}
\end{center}
        \caption{Same as Fig.~\ref{fig:eva_nH} but for $t\overline{t}$ pair production and $t\overline{t}+X$ associated production.}
    \label{fig:eva_TTX}
\end{figure*}

Next, we address $t\overline{t}$ pair and associated $t\overline{t}$ production, focusing on the channels    
\begin{equation}
V_{\lambda_A} V'_{\lambda_B} \quad\to\quad 
t\overline{t}, \quad t\overline{t}Z, \quad t\overline{t}W^+, \quad t\overline{t}\gamma, \quad t\overline{t}H, 
\quad\text{and}\quad t\overline{t} t\overline{t}.
\label{eq:eva_top}
\end{equation}
For the $t\overline{t} t\overline{t}$ process, we consider both the production though QCD and EW couplings $(t\overline{t} t\overline{t}_{\rm QCD+EW})$ as well as purely through EW couplings $(t\overline{t} t\overline{t}_{\rm EW})$.
In Fig.~\ref{fig:eva_TTX} we present the unpolarized and helicity-polarized cross sections as a function of collider energy  in the same manner as in Fig.~\ref{fig:eva_nH}. An immediate observation is that for all processes each of the polarization combinations of initial-state EW bosons $V_{\lambda_A} V_{\lambda_B}'$ give a comparable contribution to the unpolarized process. This is in contrast to associated and multi-Higgs production in Sec.~\ref{sec:leptons_higgs}, where typically one particular $V_{\lambda_A} V'_{\lambda_B}$ configuration drives the total process.  For all polarized and unpolarized cases, $t\overline{t}$ production exhibits the largest cross sections, whereas pure EW production of $t\overline{t}t\overline{t}$ exhibits the lowest rates. The difference between the channels is  about three orders of magnitude. Mixed QCD+EW production of $t\overline{t} t\overline{t}$ sits just above the pure EW rate; notably, the difference between the two $t\overline{t} t\overline{t}$ processes is larger than their uncertainty bands. All $t\overline{t}V$ and $t\overline{t}H$ cross sections are sandwiched between the two processes and exhibit a rate hierarchy that is in line with na\"ive EW coupling enhancement/suppression.  While the hierarchy is independent of collider energy, some reordering can be observed when individual helicity configurations are considered.

\begin{table}[!t]
\begin{center}
\resizebox{\textwidth}{!}{ 
\renewcommand*{\arraystretch}{1.1}\setlength{\tabcolsep}{10pt}
	\begin{tabular}{r|l l l l}
		\hline \hline
		& 
		&\multicolumn{3}{c}{$\sigma$ [fb]}\\
		& {\mgamc} syntax
		&$\sqrt{s}=3$ TeV&$\sqrt{s}=14$ TeV&$\sqrt{s}=30$ TeV\\
		\hline
		$\sum V_{\lambda_A} V'_{\lambda_B}\to t\overline{t}$&\texttt{vxp vxm > t t\~}&$1.3\cdot10^{0}$ $^{+42 \%}_{-18 \%}$&$9.4\cdot10^{0}$ $^{+49 \%}_{-21 \%}$&$1.3\cdot10^{1}$ $^{+50 \%}_{-21 \%}$\\
        $V_T V'_T\to t\bar t $&\texttt{vxp\{T\} vxm\{T\} > t t\~}&$6.4\cdot10^{-1}$ $^{+58 \%}_{-23 \%}$&$3.7\cdot10^{0}$ $^{+73 \%}_{-27 \%}$&$5.0\cdot10^{0}$ $^{+76 \%}_{-28 \%}$\\
        $V_0 V'_T\to t\bar t$&\texttt{vxp\{0\} vxm\{T\} > t t\~}&$4.6\cdot10^{-1}$ $^{+43 \%}_{-22 \%}$&$4.0\cdot10^{0}$ $^{+48 \%}_{-24 \%}$&$5.7\cdot10^{0}$ $^{+47 \%}_{-24 \%}$\\
        $V_0 V'_0\to t\bar t$&\texttt{vxp\{0\} vxm\{0\} > t t\~}&$2.5\cdot10^{-1}$ &$1.7\cdot10^{0}$ &$2.3\cdot10^{0}$\\
		\hline
		$\sum V_{\lambda_A} V'_{\lambda_B}\to t\bar t\; Z$&\texttt{vxp vxm > t t\~\ z}&$1.4\cdot10^{-1}$ $^{+99 \%}_{-38 \%}$&$1.4\cdot10^{0}$ $^{+86 \%}_{-34 \%}$&$2.1\cdot10^{0}$ $^{+84 \%}_{-33 \%}$\\
		$V_T V'_T\to t\bar t\; Z$&\texttt{vxp\{T\} vxm\{T\} > t t\~\ z}&$6.0\cdot10^{-2}$ $^{+155 \%}_{-50 \%}$&$6.2\cdot10^{-1}$ $^{+134 \%}_{-45 \%}$&$9.5\cdot10^{-1}$ $^{+127 \%}_{-43 \%}$\\
		$V_0 V'_T\to t\bar t\; Z$&\texttt{vxp\{0\} vxm\{T\} > t t\~\ z}&$7.6\cdot10^{-2}$ $^{+62 \%}_{-31 \%}$&$7.5\cdot10^{-1}$ $^{+54 \%}_{-27 \%}$&$1.1\cdot10^{0}$ $^{+52 \%}_{-26 \%}$\\
		$V_0 V'_0\to t\bar t\; Z$&\texttt{vxp\{0\} vxm\{0\} > t t\~\ z}&$4.9\cdot10^{-3}$ &$5.8\cdot10^{-2}$ &$8.3\cdot10^{-2}$\\
 		\hline
		$\sum V_{\lambda_A} V'_{\lambda_B}\to t\bar t\; W^+$&\texttt{vxp vxm > t t\~\ w+}&$1.7\cdot10^{-1}$ $^{+68 \%}_{-28 \%}$&$1.4\cdot10^{0}$ $^{+61 \%}_{-25 \%}$&$2.0\cdot10^{0}$ $^{+59 \%}_{-24 \%}$\\
		$V_T V'_T\to t\bar t\; W^+$&\texttt{vxp\{T\} vxm\{T\} > t t\~\ w+}&$1.0\cdot10^{-1}$ $^{+99 \%}_{-39 \%}$&$8.0\cdot10^{-1}$ $^{+88 \%}_{-35 \%}$&$1.1\cdot10^{0}$ $^{+85 \%}_{-34 \%}$\\
		$V_0 V'_T\to t\bar t\; W^+$&\texttt{vxp\{0\} vxm\{T\} > t t\~\ w+}&$7.0\cdot10^{-2}$ $^{+29 \%}_{-14 \%}$&$6.0\cdot10^{-1}$ $^{+28 \%}_{-14 \%}$&$8.4\cdot10^{-1}$ $^{+27 \%}_{-14 \%}$\\
		$V_0 V'_0\to t\bar t\; W^+$&\texttt{vxp\{0\} vxm\{0\} > t t\~\ w+}&$2.6\cdot10^{-3}$ &$2.9\cdot10^{-2}$ &$4.2\cdot10^{-2}$\\
		\hline
		$\sum V_{\lambda_A} V'_{\lambda_B}\to \to t\bar t \;\gamma$&\texttt{vxp vxm > t t\~\ a}&$4.2\cdot10^{-2}$ $^{+88 \%}_{-34 \%}$&$4.3\cdot10^{-1}$ $^{+78 \%}_{-31 \%}$&$6.4\cdot10^{-1}$ $^{+76 \%}_{-30 \%}$\\
		$V_T V'_T\to t\bar t\; \gamma$&\texttt{vxp\{T\} vxm\{T\} > t t\~\ a}&$1.9\cdot10^{-2}$ $^{+133 \%}_{-43 \%}$&$1.8\cdot10^{-1}$ $^{+122 \%}_{-42 \%}$&$2.7\cdot10^{-1}$ $^{+117 \%}_{-40 \%}$\\
		$V_0 V'_T\to t\bar t\; \gamma$&\texttt{vxp\{0\} vxm\{T\} > t t\~\ a}&$2.0\cdot10^{-2}$ $^{+61 \%}_{-31 \%}$&$2.2\cdot10^{-1}$ $^{+54 \%}_{-27 \%}$&$3.2\cdot10^{-1}$ $^{+52 \%}_{-26 \%}$\\
		$V_0 V'_0\to t\bar t\; \gamma$&\texttt{vxp\{0\} vxm\{0\} > t t\~\ a}&$3.3\cdot10^{-3}$ &$3.1\cdot10^{-2}$ &$4.5\cdot10^{-2}$\\
		\hline
		$\sum V_{\lambda_A} V'_{\lambda_B}\to  t\bar t\;H$&\texttt{vxp vxm > t t\~\ h}&$1.5\cdot10^{-2}$ $^{+55 \%}_{-23 \%}$&$1.5\cdot10^{-1}$ $^{+45 \%}_{-19 \%}$&$2.2\cdot10^{-1}$ $^{+44 \%}_{-19 \%}$\\
		$V_T V'_T\to t\bar t\;H$&\texttt{vxp\{T\} vxm\{T\} > t t\~\ h}&$5.9\cdot10^{-3}$ $^{+86 \%}_{-31 \%}$&$4.5\cdot10^{-2}$ $^{+85\%}_{-31 \%}$&$6.5\cdot10^{-2}$ $^{+83 \%}_{-31 \%}$\\
		$V_0 V'_T\to t\bar t\;H$&\texttt{vxp\{0\} vxm\{T\} > t t\~\ h}&$6.4\cdot10^{-3}$ $^{+47 \%}_{-23 \%}$&$7.0\cdot10^{-2}$ $^{+44 \%}_{-22 \%}$&$1.0\cdot10^{-1}$ $^{+43 \%}_{-21 \%}$\\
		$V_0 V'_0\to t\bar t\;H$&\texttt{vxp\{0\} vxm\{0\} > t t\~\ h}&$2.6\cdot10^{-3}$ &$3.5\cdot10^{-2}$ &$5.1\cdot10^{-2}$\\
		\hline
		$\sum V_{\lambda_A} V'_{\lambda_B}\to t\bar t\;t\bar t_\textrm{\;QCD+EW}$&\texttt{vxp\{T\} vxm\{T\} > t t\~\ t t\~\ }&$7.1\cdot10^{-4}$ $^{+47 \%}_{-31 \%}$&$1.4\cdot10^{-2}$ $^{+36 \%}_{-26 \%}$&$2.3\cdot10^{-2}$ $^{+35 \%}_{-26 \%}$\\
		$V_T V'_T\to  t\bar t\;t\bar t_\textrm{\;QCD+EW}$&\texttt{vxp\{T\} vxm\{T\} > t t\~\ t t\~\ }&$4.9\cdot10^{-4}$ $^{+57 \%}_{-33 \%}$&$8.4\cdot10^{-3}$ $^{+45 \%}_{-29 \%}$&$1.4\cdot10^{-2}$ $^{+43 \%}_{-29 \%}$\\
		$V_0 V'_T\to  t\bar t\;t\bar t_\textrm{\;QCD+EW}$&\texttt{vxp\{0\} vxm\{T\} > t t\~\ t t\~\ }&$1.9\cdot10^{-4}$ $^{+35 \%}_{-27 \%}$&$4.5\cdot10^{-3}$ $^{+30 \%}_{-24 \%}$&$7.9\cdot10^{-3}$ $^{+30 \%}_{-23 \%}$\\
		$V_0 V'_0\to  t\bar t\;t\bar t_\textrm{\;QCD+EW}$&\texttt{vxp\{0\} vxm\{0\} > t t\~\ t t\~\ }&$2.6\cdot10^{-5}$ &$7.7\cdot10^{-4}$ &$1.4\cdot10^{-3}$\\
		\hline
		$\sum V_{\lambda_A} V'_{\lambda_B}\to t\bar t\;t\bar t_\textrm{\;EW}$&\texttt{vxp\{T\} vxm\{T\} > t t\~\ t t\~\  QCD=0}&$2.5\cdot10^{-4}$ $^{+45 \%}_{-19 \%}$&$5.1\cdot10^{-3}$ $^{+41 \%}_{-18 \%}$&$8.8\cdot10^{-3}$ $^{+40 \%}_{-17 \%}$\\
		$V_T V'_T\to  t\bar t\;t\bar t_\textrm{\;EW}$&\texttt{vxp\{T\} vxm\{T\} > t t\~\ t t\~\   QCD=0}&$1.6\cdot10^{-4}$ $^{+53 \%}_{-21 \%}$&$2.7\cdot10^{-3}$ $^{+53 \%}_{-21 \%}$&$4.4\cdot10^{-3}$ $^{+54 \%}_{-22 \%}$\\
		$V_0 V'_T\to  t\bar t\;t\bar t_\textrm{\;EW}$&\texttt{vxp\{0\} vxm\{T\} > t t\~\ t t\~\   QCD=0}&$7.1\cdot10^{-5}$ $^{+38 \%}_{-19 \%}$&$1.8\cdot10^{-3}$ $^{+35 \%}_{-18 \%}$&$3.3\cdot10^{-3}$ $^{+34 \%}_{-17 \%}$\\
		$V_0 V'_0\to  t\bar t\;t\bar t_\textrm{\;EW}$&\texttt{vxp\{0\} vxm\{0\} > t t\~\ t t\~\   QCD=0}&$1.8\cdot10^{-5}$ &$6.4\cdot10^{-4}$ &$1.1\cdot10^{-3}$\\
		\hline
		\hline
	\end{tabular}
	} 
	\caption{Same as Table~\ref{tab:eva_nH} but for $t\overline{t}$ pair production and $t\overline{t}$ associated production.}
	\label{tab:eva_TTX}
\end{center}
\end{table}

Focusing first on $V_{0} V_{0}'$ scattering in Fig.~\ref{fig:eva_TTX}(b), we find that the $t\overline{t}V$  and  $t\overline{t}H$ channels all have highly comparable rates. This is essentially due to all these channels being driven by either $W^+_0 W^-_0$ or $W^+_0 Z_0$ scattering. (There is no $\gamma_0$ PDF.) Hence, appreciable differences in rate are due to differences in coupling constants. Moreover, unlike other polarization configurations, there is no $\log(\mu_f^2/m_\mu^2)$ enhancement since this is associated with $\gamma_T$ PDFs. 

More explicitly, for $V_{T} V_{0}'$ scattering in Fig.~\ref{fig:eva_TTX}(c) and $V_{T} V_{T}'$ scattering in Fig.~\ref{fig:eva_TTX}(d), we find appreciably larger cross sections than for pure $V_{0} V_{0}'$ scattering. Again, we attribute this to the opening of $\gamma_T V_\lambda$ scattering and logarithmic enhancements in transverse PDFs. Interestingly, we find that the hierarchy of $t\overline{t}\gamma/Z/W^+/H$ depends on the precise polarization configuration of initial-state EW bosons. For example: $t\overline{t}Z$  has a larger rate than $t\overline{t}W^+$ in $V_{T} V_{0}'$ scattering, but the opposite is true in $V_{T} V_{T}'$ scattering. Both configurations lead to larger rates than for $t\overline{t}\gamma$, which is not the case for $V_{0} V_{0}'$ scattering.

We summarize these results for representative $\sqrt{s}$ in Table~\ref{tab:eva_TTX}.

\subsection{Diboson production}\label{sec:leptons_diboson}

\begin{figure*}[!t]
\begin{center} 
    \subfigure[]{\includegraphics[width=0.48\textwidth]{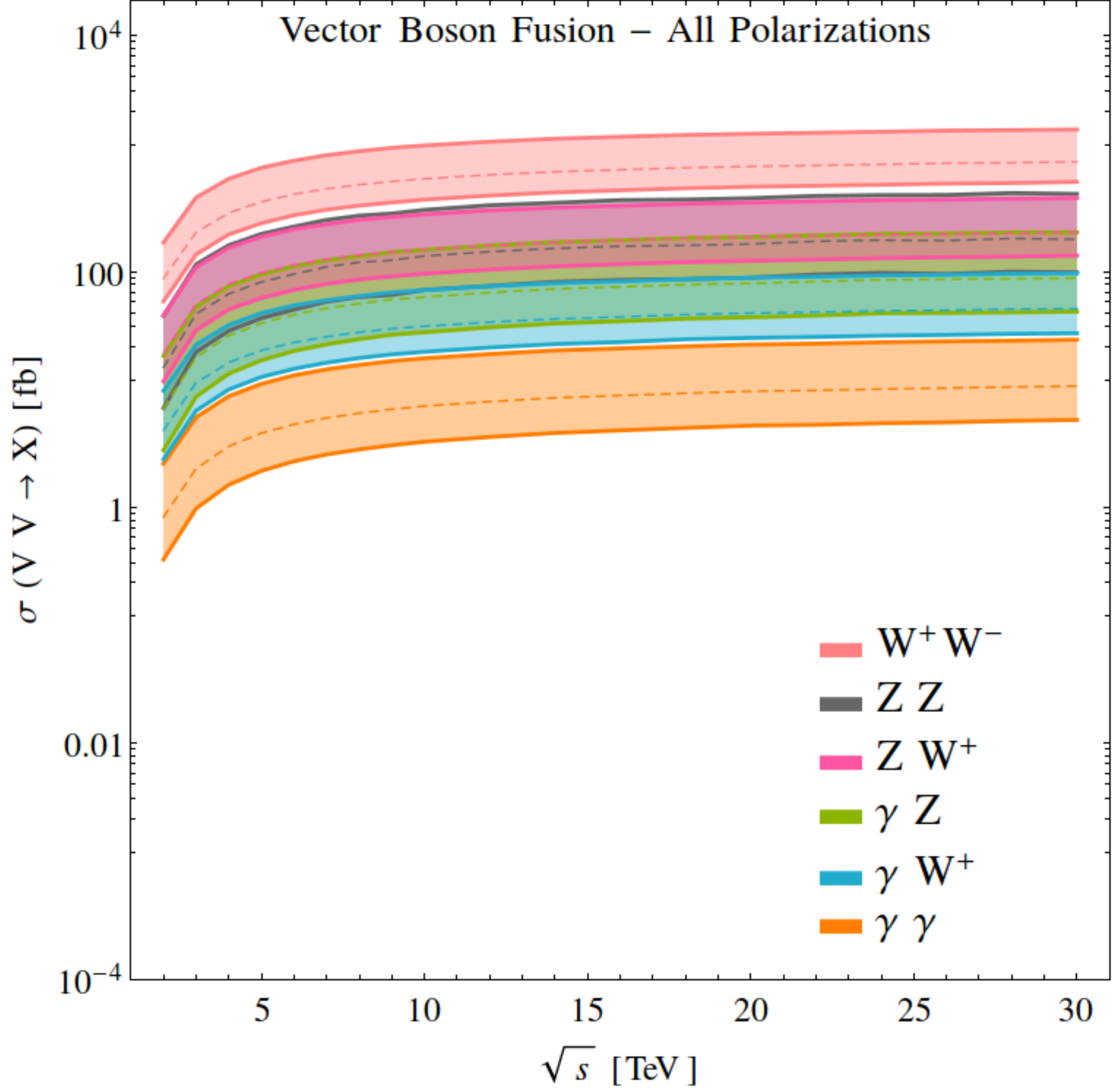}}
    \subfigure[]{\includegraphics[width=0.48\textwidth]{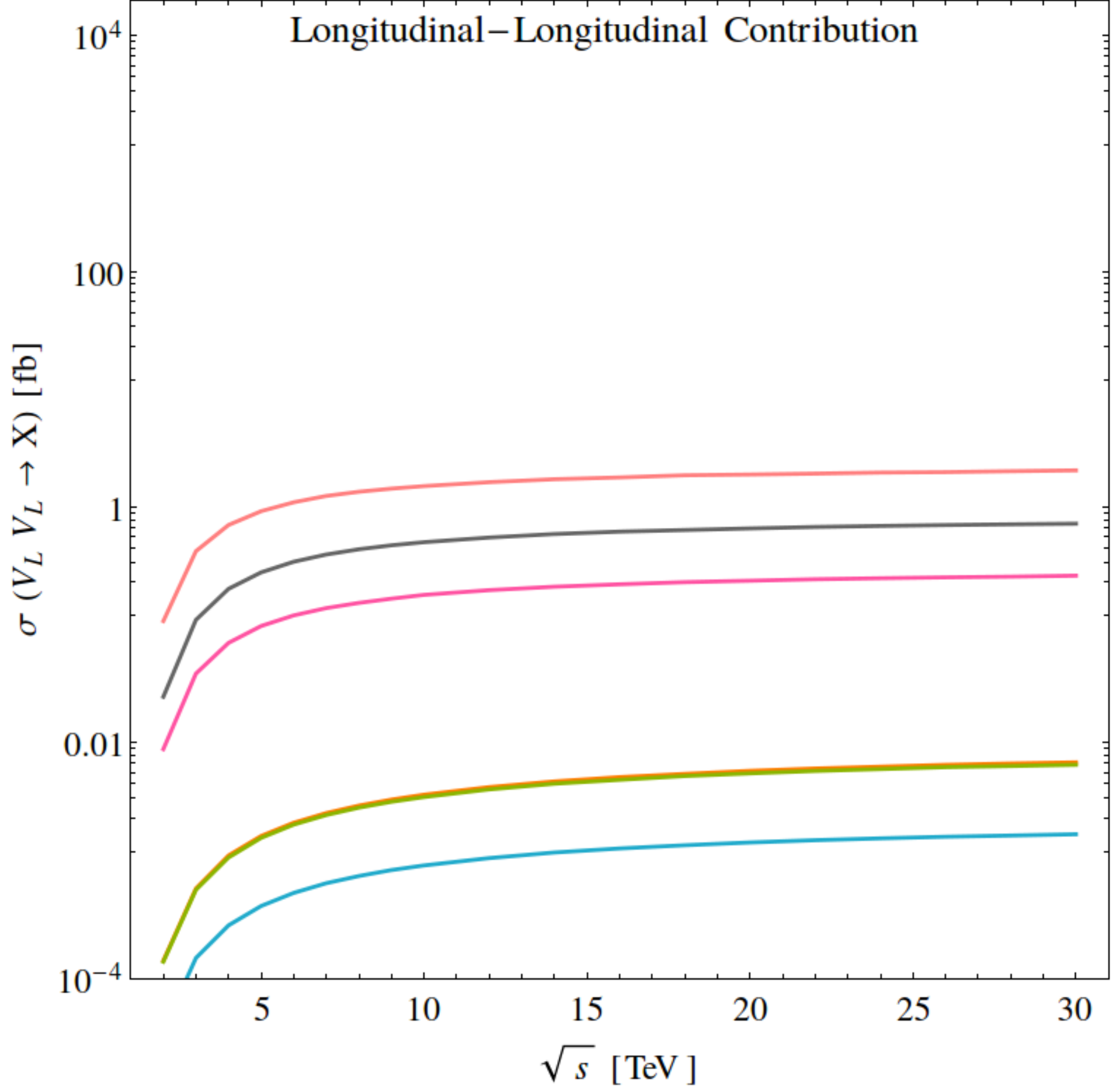}}
    \\
    \subfigure[]{\includegraphics[width=0.48\textwidth]{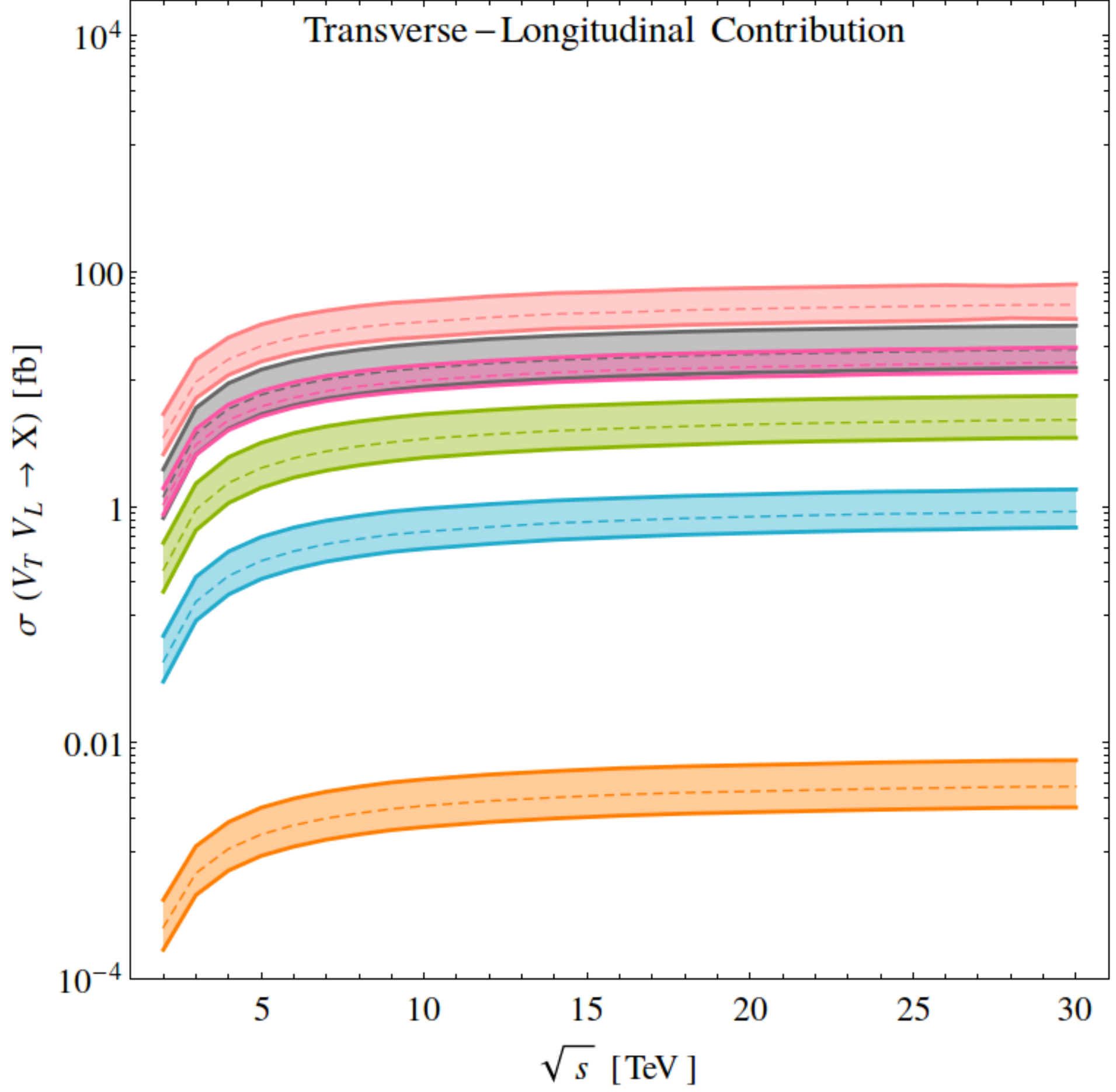}}   
    \subfigure[]{\includegraphics[width=0.48\textwidth]{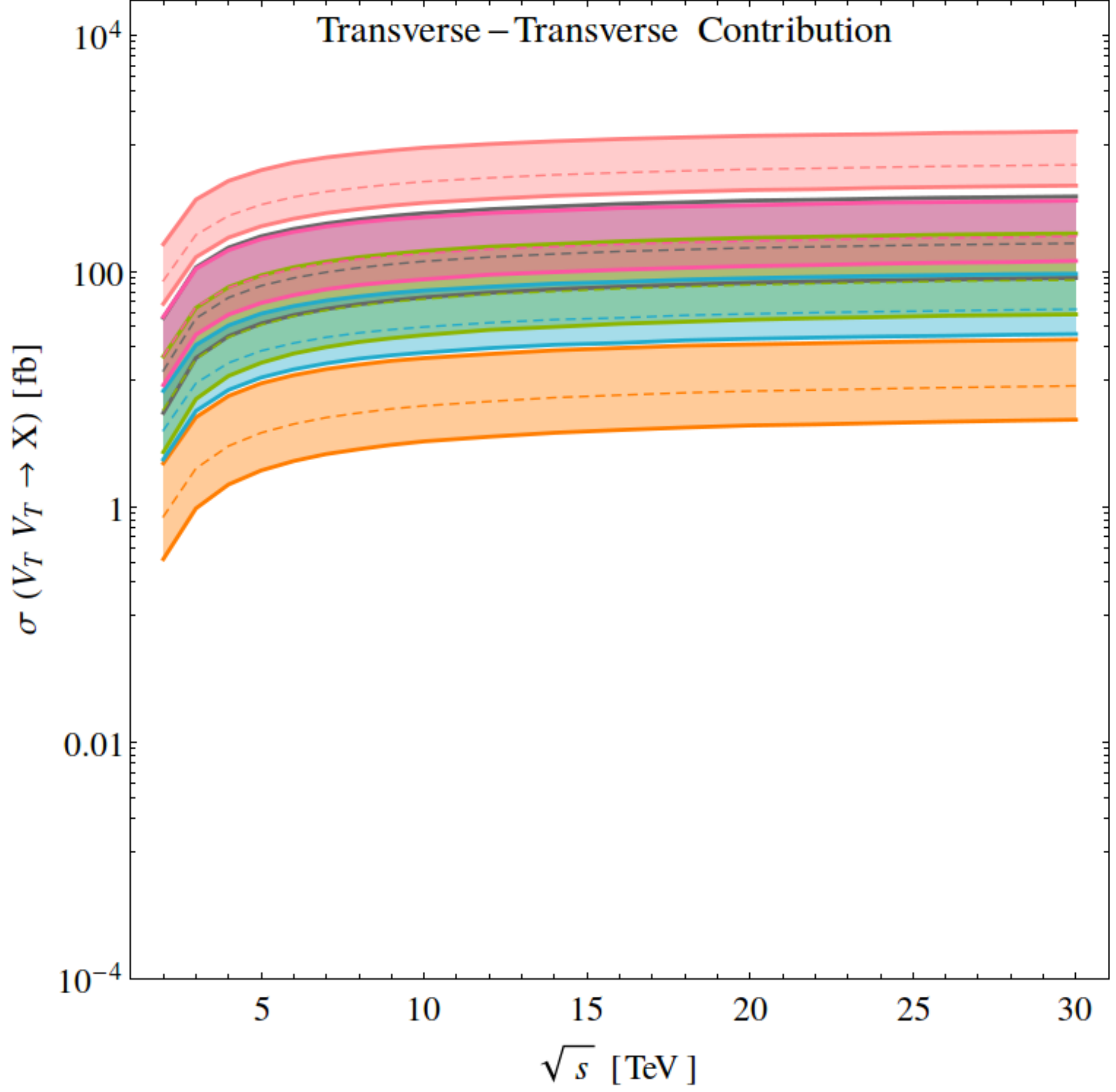}}
\end{center}
        \caption{Same as Fig.~\ref{fig:eva_nH} but for $VV$ production.}
    \label{fig:eva_VV}
\end{figure*}

We now turn to diboson production in EVA. In Fig.~\ref{fig:eva_VV}, we plot again the (a) polarization-summed and (b-d) polarized production cross section with their respective factorization scale uncertainties,  as a function of collider energy for the following six  processes
\begin{equation}
    V_{\lambda_A}V_{\lambda_B}' \quad \to \quad 
    W^+W^-, \quad ZZ, \quad ZW^+, \quad \gamma Z, \quad \gamma W^+, \quad\text{and}\quad \gamma\gamma.
\end{equation}
As in previous cases, there are several global features that one can infer. 
Foremost is that unlike Higgs (Fig.~\ref{fig:eva_nH}) and top quark (Fig.~\ref{fig:eva_TTX}) processes, we find a clear hierarchy among initial-state EW boson polarizations. 
More specifically, we find that $V_{T} V_{T}'$ scattering is categorically the dominant production vehicle of EW boson pairs.
The role of $V_{T} V_{0}'$ scattering is about one-to-two orders of magnitude smaller than $V_{T} V_{T}'$ scattering, and rate of $V_{0} V_{0}'$ scattering is yet another decade smaller.
As of $V_{0} V_{0}'$ scattering is negligible, unpolarized diboson cross sections in the EVA exhibit a relatively larger scale uncertainty than Higgs and top quark production, which feature a larger dependence on $V_{0} V_{0}'$ scattering.

Another consequence of the strong polarization dependence is that the hierarchy of fiducial cross sections shown in the unpolarized case largely mirrors the hierarchy in $V_{T} V_{T}'$ scattering.
That said, we find that this same hierarchy is mostly preserved with other helicity configurations.
This suggests a larger dependence on available partonic channels and gauge couplings than logarithmic enhancements from soft and collinear regions of phase space.
For example: the $W^+W^-$ production cross section dominates for all helicity configurations but also can be produced via the most number of partonic configurations, i.e., $W_{\lambda_A}^+W_{\lambda_B}^-,Z_{\lambda_A}Z_{\lambda_B},Z_{\lambda_A}\gamma_{\lambda_B}$, and $\gamma_{\lambda_A}\gamma_{\lambda_B}$ fusion. The reverse can be said for $\gamma\gamma$ and $\gamma W^+$ production. These processes exhibit the lowest diboson cross sections but can only proceed through one or two partonic channels, namely $W_{\lambda_A}^+W_{\lambda_B}^-$ and $W_{\lambda_A}^+ \gamma_{\lambda_B}$. We also note that helicity suppression stemming from angular momentum conservation also plays a role in this hierarchy. For example: while $W^+_0 Z_0 \to W^+_0 \gamma_T$ exhibits a larger longitudinal polarization enhancement than $W^+_0 Z_0 \to W^+_T \gamma_T$, the former (later) is disfavored (favored) since it must proceed through a high-wave (low-wave) angular momentum configuration.

\begin{table}[!t]
\begin{center}
\resizebox{\textwidth}{!}{ 
\renewcommand*{\arraystretch}{1.1}\setlength{\tabcolsep}{10pt}
	\begin{tabular}{r|l l l l}
		\hline \hline
		& 
		&\multicolumn{3}{c}{$\sigma$ [fb]}\\
		& {\mgamc} syntax
		&$\sqrt{s}=3$ TeV&$\sqrt{s}=14$ TeV&$\sqrt{s}=30$ TeV\\
		\hline
		$\sum V_{\lambda_A} V'_{\lambda_B}\to W^+ W^-$&\texttt{vxp vxm > w+ w-}&$2.2\cdot10^{2}$ $^{+98 \%}_{-35 \%}$&$7.0\cdot10^{2}$ $^{+91 \%}_{-33 \%}$&$8.6\cdot10^{2}$ $^{+88 \%}_{-32 \%}$\\
		$V_T V'_T\to W^+ W^-$&\texttt{vxp\{T\} vxm\{T\} > w+ w-}&$2.0\cdot10^{2}$ $^{+99 \%}_{-35 \%}$&$6.6\cdot10^{2}$ $^{+93 \%}_{-34 \%}$&$8.0\cdot10^{2}$ $^{+92 \%}_{-33 \%}$\\
		$V_0 V'_T\to W^+ W^-$&\texttt{vxp\{0\} vxm\{T\} > w+ w-}&$1.2\cdot10^{1}$ $^{+54 \%}_{-27 \%}$&$4.4\cdot10^{1}$ $^{+50 \%}_{-25 \%}$&$5.2\cdot10^{1}$ $^{+49 \%}_{-24 \%}$\\
		$V_0 V'_0\to W^+ W^-$&\texttt{vxp\{0\} vxm\{0\} > w+ w-}&$4.2\cdot10^{-1}$ &$1.7\cdot10^{0}$ &$2.0\cdot10^{0}$\\
 		\hline
 		$\sum V_{\lambda_A} V'_{\lambda_B}\to W^+ Z$&\texttt{vxp vxm > w+ z}&$5.3\cdot10^{1}$ $^{+105 \%}_{-40 \%}$&$1.8\cdot10^{2}$ $^{+97 \%}_{-37 \%}$&$2.2\cdot10^{2}$ $^{+95 \%}_{-37 \%}$\\
 		$V_T V'_T\to W^+ Z$&\texttt{vxp\{T\} vxm\{T\} > w+ z}&$5.0\cdot10^{1}$ $^{+111 \%}_{-42 \%}$&$1.6\cdot10^{2}$ $^{+103 \%}_{-39 \%}$&$2.0\cdot10^{2}$ $^{+100 \%}_{-38 \%}$\\
 		$V_0 V'_T\to W^+ Z$&\texttt{vxp\{0\} vxm\{T\} > w+ z}&$3.4\cdot10^{0}$ $^{+36 \%}_{-18 \%}$&$1.4\cdot10^{1}$ $^{+34 \%}_{-17 \%}$&$1.7\cdot10^{1}$ $^{+34 \%}_{-17 \%}$\\
 		$V_0 V'_0\to W^+ Z$&\texttt{vxp\{0\} vxm\{0\} > w+ z}&$3.9\cdot10^{-2}$ &$2.1\cdot10^{-1}$ &$2.6\cdot10^{-1}$\\
		\hline
		$\sum V_{\lambda_A} V'_{\lambda_B}\to Z Z$&\texttt{vxp vxm > z z}&$4.4\cdot10^{1}$ $^{+164 \%}_{-52 \%}$&$1.6\cdot10^{2}$ $^{+144 \%}_{-48 \%}$&$1.9\cdot10^{2}$ $^{+143 \%}_{-48 \%}$\\
		$V_T V'_T\to Z Z$&\texttt{vxp\{T\} vxm\{T\} > z z}&$4.0\cdot10^{1}$ $^{+171 \%}_{-54 \%}$&$1.4\cdot10^{2}$ $^{+153 \%}_{-50 \%}$&$1.7\cdot10^{2}$ $^{+150 \%}_{-49 \%}$\\
		$V_0 V'_T\to Z Z$&\texttt{vxp\{0\} vxm\{T\} > z z}&$4.2\cdot10^{0}$ $^{+66 \%}_{-33 \%}$&$1.8\cdot10^{1}$ $^{+61 \%}_{-30 \%}$&$2.2\cdot10^{1}$ $^{+60 \%}_{-30 \%}$\\
		$V_0 V'_0\to Z Z$&\texttt{vxp\{0\} vxm\{0\} > z z}&$1.1\cdot10^{-1}$ &$6.0\cdot10^{-1}$ &$7.2\cdot10^{-1}$\\
		\hline
		$\sum V_{\lambda_A} V'_{\lambda_B}\to \gamma Z$&\texttt{vxp vxm > a z}&$1.9\cdot10^{1}$ $^{+169 \%}_{-53 \%}$&$7.1\cdot10^{1}$ $^{+149 \%}_{-49 \%}$&$8.8\cdot10^{1}$ $^{+145 \%}_{-48 \%}$\\
		$V_T V'_T\to \gamma Z$&\texttt{vxp\{T\} vxm\{T\} > a z}&$1.8\cdot10^{1}$ $^{+172 \%}_{-54 \%}$&$6.8\cdot10^{1}$ $^{+153 \%}_{-50 \%}$&$8.4\cdot10^{1}$ $^{+149 \%}_{-49 \%}$\\
		$V_0 V'_T\to \gamma Z$&\texttt{vxp\{0\} vxm\{T\} > a z}&$9.5\cdot10^{-1}$ $^{+67 \%}_{-33 \%}$&$4.4\cdot10^{0}$ $^{+61 \%}_{-30 \%}$&$5.5\cdot10^{0}$ $^{+60 \%}_{-30 \%}$\\
		$V_0 V'_0\to \gamma Z$&\texttt{vxp\{0\} vxm\{0\} > a z}&$5.6\cdot10^{-4}$ &$4.5\cdot10^{-3}$ &$6.5\cdot10^{-3}$\\
		\hline
		$\sum V_{\lambda_A} V'_{\lambda_B}\to \gamma W^+$&\texttt{vxp vxm > a w+}&$1.1\cdot10^{1}$ $^{+111 \%}_{-42 \%}$&$4.0\cdot10^{1}$ $^{+101 \%}_{-39 \%}$&$4.9\cdot10^{1}$ $^{+99 \%}_{-38 \%}$\\
		$V_T V'_T\to \gamma W^+$&\texttt{vxp\{T\} vxm\{T\} > a w+}&$1.1\cdot10^{1}$ $^{+111 \%}_{-42 \%}$&$3.9\cdot10^{1}$ $^{+102 \%}_{-39 \%}$&$4.8\cdot10^{1}$ $^{+100 \%}_{-38 \%}$\\
		$V_0 V'_T\to \gamma W^+$&\texttt{vxp\{0\} vxm\{T\} > a w+}&$1.6\cdot10^{-2}$ $^{+62 \%}_{-31 \%}$&$7.3\cdot10^{-1}$ $^{+56 \%}_{-28 \%}$&$9.2\cdot10^{-1}$ $^{+54 \%}_{-27 \%}$\\
		$V_0 V'_0\to \gamma W^+$&\texttt{vxp\{0\} vxm\{0\} > a w+}&$1.5\cdot10^{-4}$ &$1.2\cdot10^{-3}$ &$1.7\cdot10^{-3}$\\
		\hline
		$\sum V_{\lambda_A} V'_{\lambda_B}\to \gamma \gamma$&\texttt{vxp vxm > a a}&$2.1\cdot10^{0}$ $^{+172 \%}_{-54 \%}$&$8.5\cdot10^{0}$ $^{+152 \%}_{-50 \%}$&$1.1\cdot10^{1}$ $^{+147 \%}_{-48 \%}$\\
		$V_T V'_T\to \gamma \gamma$&\texttt{vxp\{T\} vxm\{T\} > a a}&$2.1\cdot10^{0}$ $^{+172 \%}_{-54 \%}$&$8.5\cdot10^{0}$ $^{+152 \%}_{-50 \%}$&$1.1\cdot10^{1}$ $^{+147 \%}_{-48 \%}$\\
		$V_0 V'_T\to \gamma \gamma$&\texttt{vxp\{0\} vxm\{T\} > a a}&$7.8\cdot10^{-4}$ $^{+70 \%}_{-35 \%}$&$3.4\cdot10^{-3}$ $^{+67 \%}_{-34 \%}$&$4.2\cdot10^{-3}$ $^{+67 \%}_{-33 \%}$\\
		$V_0 V'_0\to \gamma \gamma$&\texttt{vxp\{0\} vxm\{0\} > a a}&$5.8\cdot10^{-4}$ &$4.7\cdot10^{-3}$ &$6.8\cdot10^{-3}$\\
		\hline
		\hline
	\end{tabular}
	} 
	\caption{Same as Table~\ref{tab:eva_nH} but for $VV$ production.}
	\label{tab:eva_VV}
\end{center}
\end{table}

We summarize these results for representative $\sqrt{s}$ in Table~\ref{tab:eva_VV}.

\subsection{Triboson production}\label{sec:leptons_triboson}

\begin{figure*}[!t]
\begin{center} 
    \subfigure[]{\includegraphics[width=0.48\textwidth]{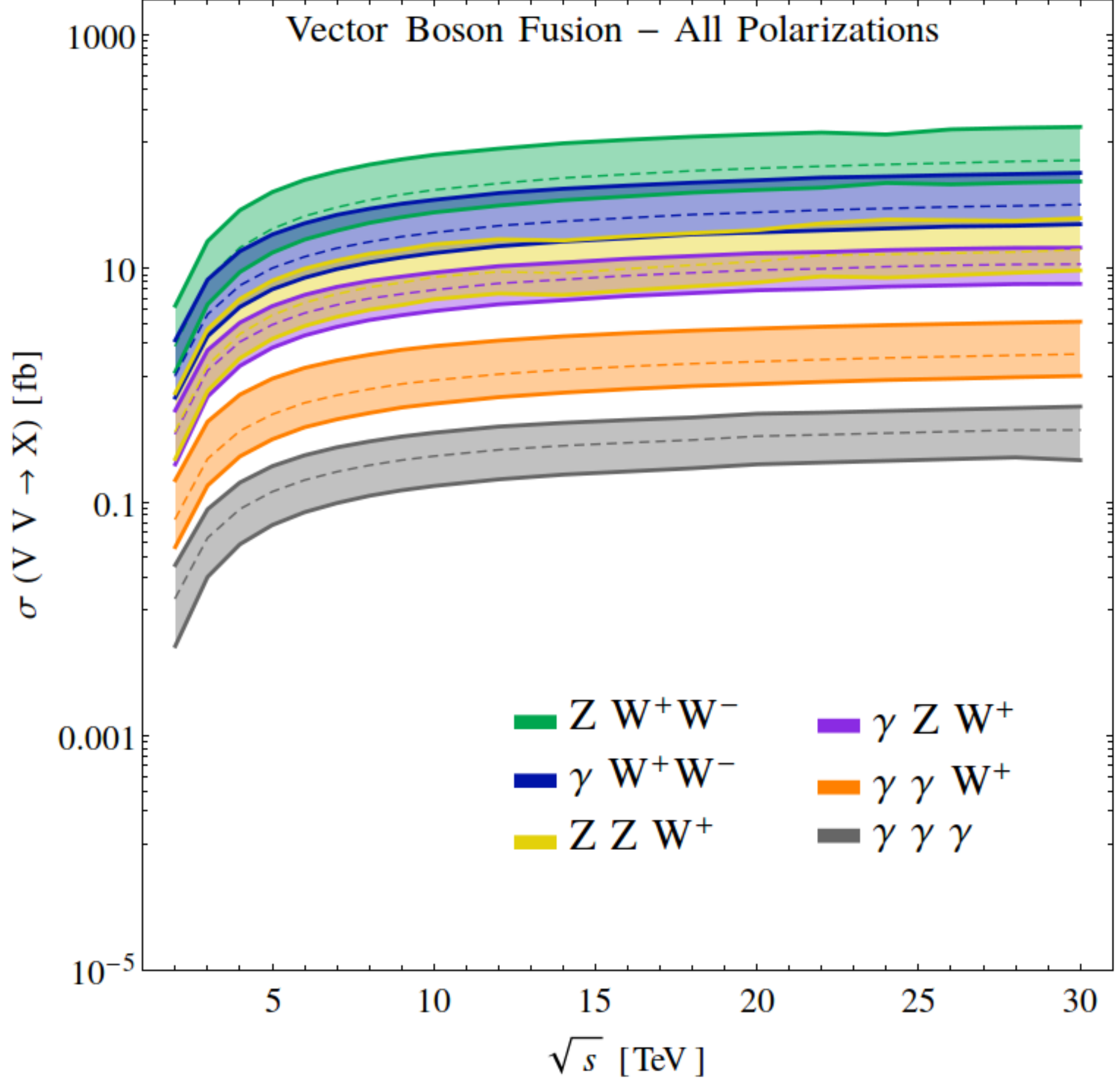}}
    \subfigure[]{\includegraphics[width=0.48\textwidth]{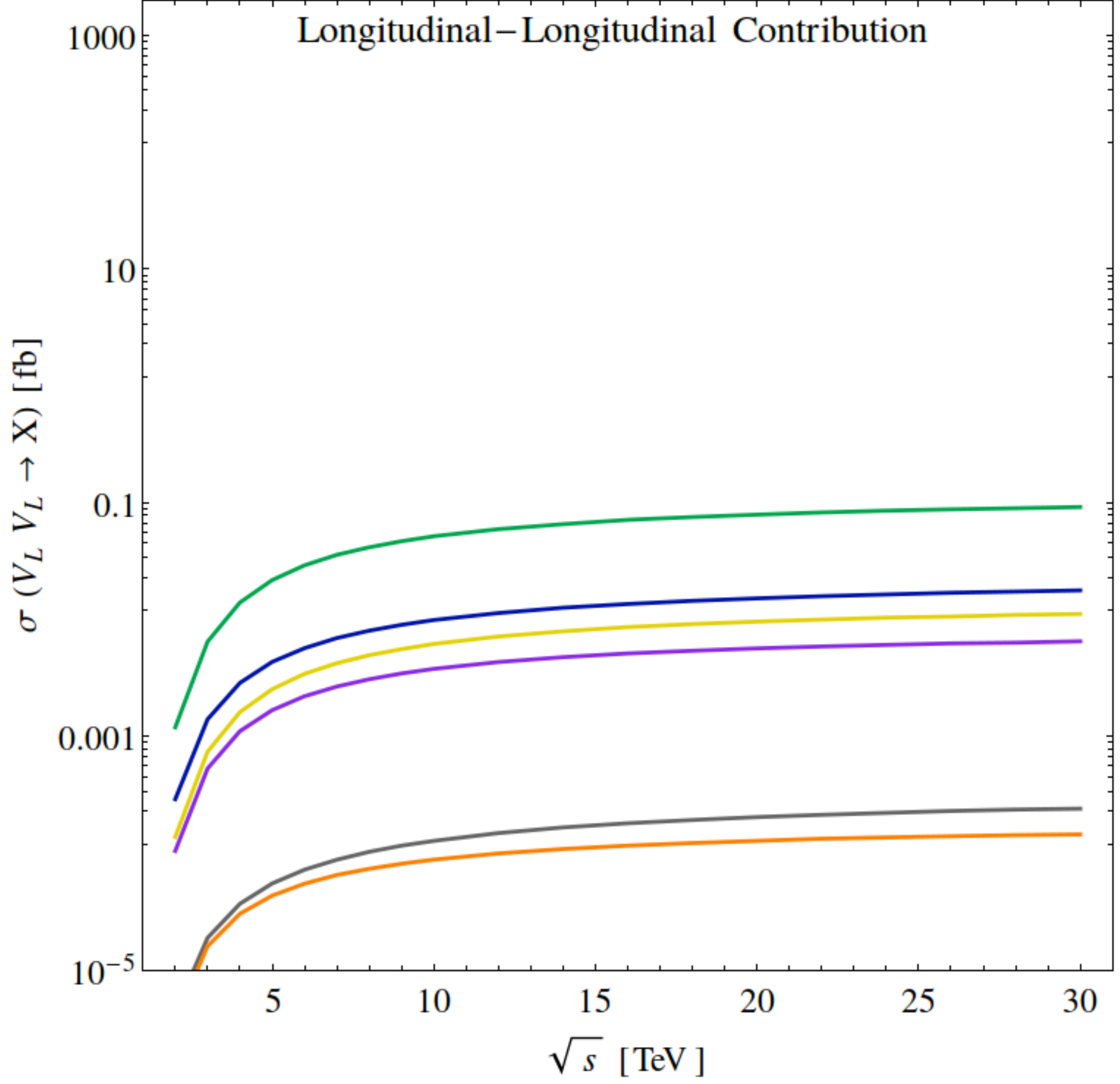}}
    \\
    \subfigure[]{\includegraphics[width=0.48\textwidth]{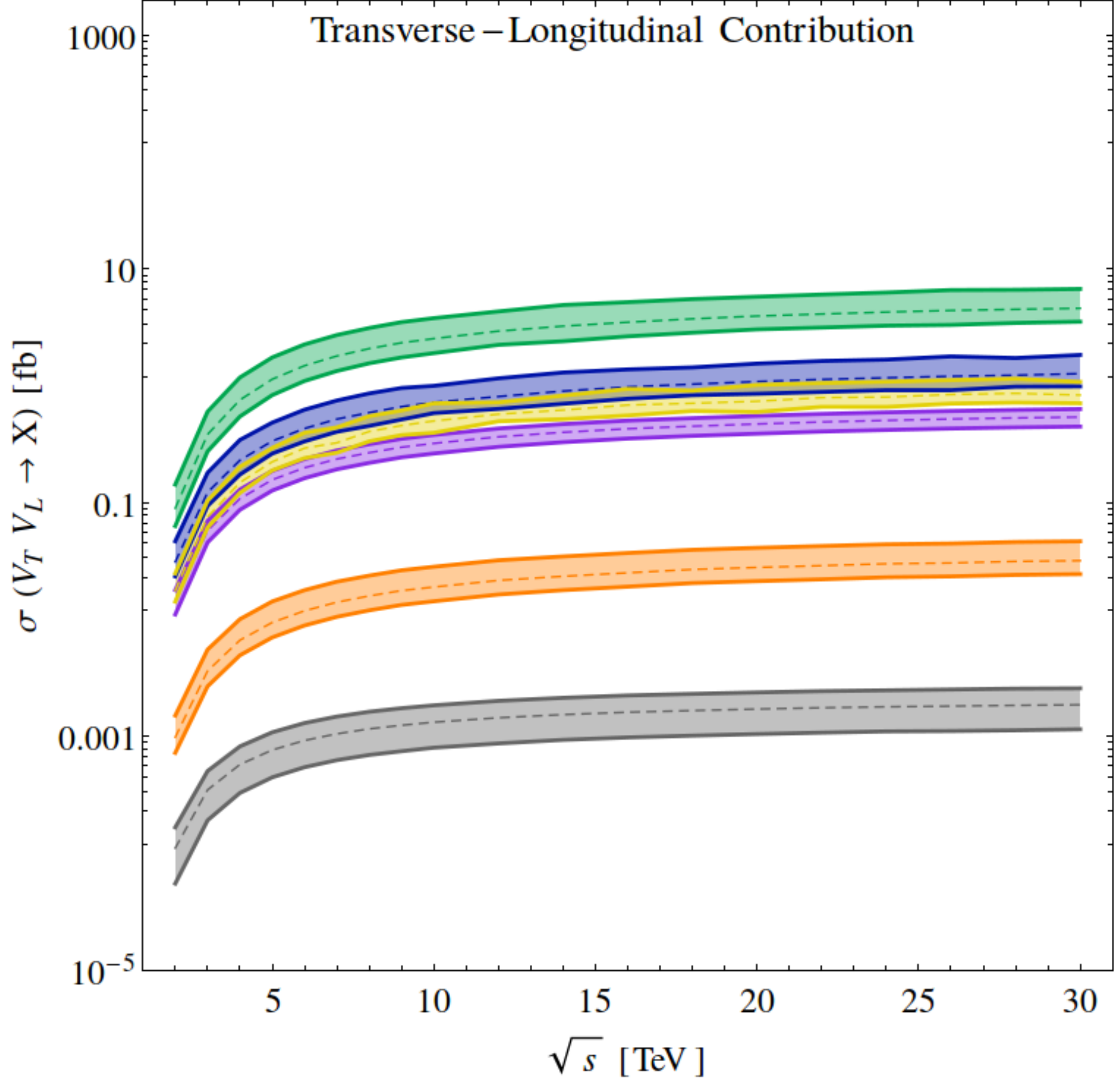}}   
    \subfigure[]{\includegraphics[width=0.48\textwidth]{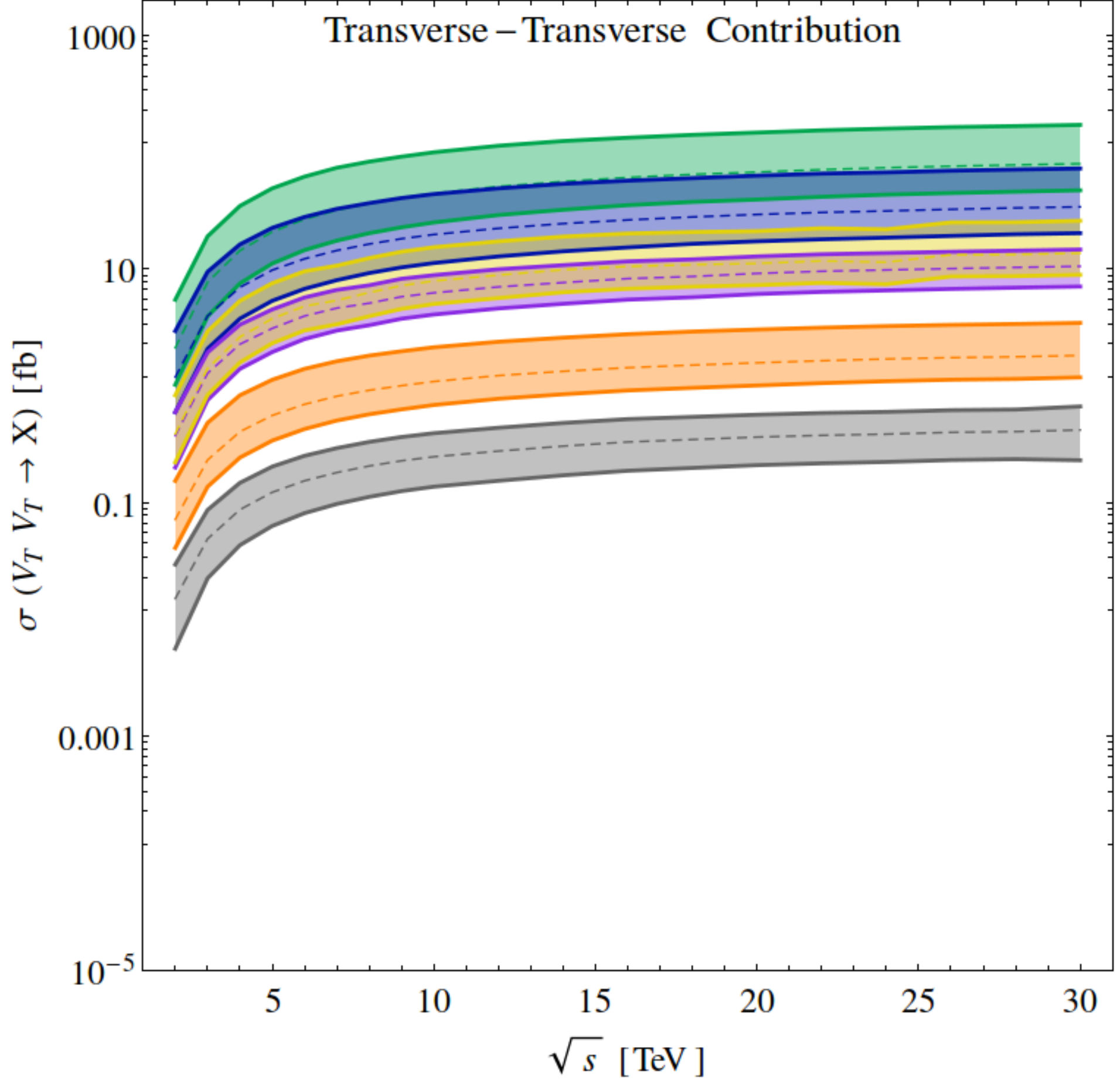}}
\end{center}
        \caption{Same as Fig.~\ref{fig:eva_nH} but for $VVV$ production.}
    \label{fig:eva_VVV}
\end{figure*}

In the final part of our survey, we show in Fig.~\ref{fig:eva_VVV} the fiducial cross sections for triboson production (a) after summing over all initial-state helicity polarizations and (b-d) for individual $(\lambda_A,\lambda_B)$ configurations. For conciseness, we focus on the representative channels
\begin{equation}
V_{\lambda_A} V'_{\lambda_B} \to Z W^+ W^-, \quad \gamma W^+ W^-, \quad Z Z Z, \quad Z Z W^+, \quad \gamma Z W^+, \quad \text{and} \quad \gamma \gamma W^+. 
\end{equation}
An immediate observation we can make is the qualitative similarities between triboson and diboson production in Fig.~\ref{fig:eva_VV}.
In particular, we find that triboson production is driven by largely $V_{T} V_{T}'$ scattering.
The $V_{0} V_{T}'$ rate is an order of magnitude or two smaller, and the $V_{0} V_{0}'$ rate is smaller by about one or two additional decades.

The precise polarization composition has a slight dependence on the underlying process:
The production of $Z W^+ W^-$ categorically exhibits the largest cross section, with $\sigma(ZW^+W^-)\sim\mathcal{O}(10-100\fb)$ for $\sqrt{s}\sim5-30\TeV$.
Over this same range, the $V_{0} V_{T}'$ component is about $\mathcal{O}(1-10\fb)$, and the 
$V_{0} V_{0}'$ component is about $\mathcal{O}(0.01\fb)$.
On the other hand, for $\gamma\gamma W^+$ production, which categorically exhibits the second smallest cross section with $\sigma(\gamma\gamma W^-)\sim\mathcal{O}(1\fb)$,
the $V_{0} V_{T}'$ component is about $\mathcal{O}(10^{-2}\fb)$, and the 
$V_{0} V_{0}'$ component is about $\mathcal{O}(10^{-5}-10^{-4}\fb)$.
In other words, $Z W^+ W^-$ production is very roughly $\mathcal{O}(90\%)$ $V_{T} V_{T}'$ scattering while $\gamma\gamma W^+$ production is roughly $\mathcal{O}(99\%)$ $V_{T} V_{T}'$ scattering.

\begin{table}[!t]
\begin{center}
\resizebox{\textwidth}{!}{ 
\renewcommand*{\arraystretch}{1.1}\setlength{\tabcolsep}{10pt}
	\begin{tabular}{r|l l l l}
		\hline \hline
		& 
		&\multicolumn{3}{c}{$\sigma$ [fb]}\\
		& {\mgamc} syntax
		&$\sqrt{s}=3$ TeV&$\sqrt{s}=14$ TeV&$\sqrt{s}=30$ TeV\\
		\hline
		$\sum V_{\lambda_A} V'_{\lambda_B}\to Z W^+ W^-$&\texttt{vxp vxm > z w+ w-}&$2.1\cdot10^{1}$ $^{+48 \%}_{-38 \%}$&$1.0\cdot10^{2}$ $^{+41 \%}_{-33 \%}$&$1.4\cdot10^{2}$ $^{+39 \%}_{-32 \%}$\\
		$V_T V'_T\to Z W^+ W^-$&\texttt{vxp\{T\} vxm\{T\} > z w+ w-}&$2.0\cdot10^{1}$ $^{+49 \%}_{-38 \%}$&$9.9\cdot10^{1}$ $^{+42 \%}_{-34 \%}$&$1.3\cdot10^{2}$ $^{+40 \%}_{-33 \%}$\\
		$V_0 V'_T\to Z W^+ W^-$&\texttt{vxp\{0\} vxm\{T\} > z w+ w-}&$9.4\cdot10^{-1}$ $^{+25 \%}_{-25 \%}$&$5.1\cdot10^{0}$ $^{+21 \%}_{-21 \%}$&$6.8\cdot10^{0}$ $^{+21 \%}_{-21 \%}$\\
		$V_0 V'_0\to Z W^+ W^-$&\texttt{vxp\{0\} vxm\{0\} > z w+ w-}&$1.6\cdot10^{-2}$ &$9.2\cdot10^{-2}$ &$1.2\cdot10^{-1}$\\
		\hline
		$\sum V_{\lambda_A} V'_{\lambda_B}\to \gamma W^+ W^-$&\texttt{vxp vxm > a w+ w-}&$1.2\cdot10^{1}$ $^{+44 \%}_{-35 \%}$&$4.8\cdot10^{1}$ $^{+40 \%}_{-32 \%}$&$6.3\cdot10^{1}$ $^{+38 \%}_{-31 \%}$\\
		$V_T V'_T\to \gamma W^+ W^-$&\texttt{vxp\{T\} vxm\{T\} > a w+ w-}&$1.2\cdot10^{1}$ $^{+45 \%}_{-36 \%}$&$4.6\cdot10^{1}$ $^{+40 \%}_{-33 \%}$&$6.1\cdot10^{1}$ $^{+40 \%}_{-32 \%}$\\
		$V_0 V'_T\to \gamma W^+ W^-$&\texttt{vxp\{0\} vxm\{T\} > a w+ w-}&$4.0\cdot10^{-1}$ $^{+24 \%}_{-24 \%}$&$1.6\cdot10^{0}$ $^{+21 \%}_{-21 \%}$&$2.1\cdot10^{0}$ $^{+17 \%}_{-17 \%}$\\
		$V_0 V'_0\to \gamma W^+ W^-$&\texttt{vxp\{0\} vxm\{0\} > a w+ w-}&$8.4\cdot10^{-3}$ &$2.9\cdot10^{-2}$ &$3.6\cdot10^{-2}$\\
		\hline
		$\sum V_{\lambda_A} V'_{\lambda_B}\to Z Z Z$&\texttt{vxp vxm > z z z}&$3.5\cdot10^{0}$ $^{+59 \%}_{-45 \%}$&$1.9\cdot10^{1}$ $^{+49 \%}_{-39 \%}$&$2.5\cdot10^{1}$ $^{+46 \%}_{-37 \%}$\\
		$V_T V'_T\to Z Z Z$&\texttt{vxp\{T\} vxm\{T\} > z z z}&$3.3\cdot10^{0}$ $^{+61 \%}_{-46 \%}$&$1.7\cdot10^{1}$ $^{+50 \%}_{-40 \%}$&$2.3\cdot10^{1}$ $^{+48 \%}_{-38 \%}$\\
		$V_0 V'_T\to Z Z Z$&\texttt{vxp\{0\} vxm\{T\} > z z z}&$2.2\cdot10^{-1}$ $^{+28 \%}_{-28 \%}$&$1.3\cdot10^{0}$ $^{+23 \%}_{-23 \%}$&$1.7\cdot10^{0}$ $^{+22 \%}_{-22 \%}$\\
		$V_0 V'_0\to Z Z Z$&\texttt{vxp\{0\} vxm\{0\} > z z z}&$4.2\cdot10^{-3}$ &$2.6\cdot10^{-2}$ &$3.5\cdot10^{-2}$\\
		\hline
		$\sum V_{\lambda_A} V'_{\lambda_B}\to Z Z W^+$&\texttt{vxp vxm > z z w+}&$3.5\cdot10^{0}$ $^{+45 \%}_{-37 \%}$&$1.8\cdot10^{1}$ $^{+38 \%}_{-32 \%}$&$2.3\cdot10^{1}$ $^{+37 \%}_{-31 \%}$\\
		$V_T V'_T\to Z Z W^+$&\texttt{vxp\{T\} vxm\{T\} > z z w+}&$3.3\cdot10^{0}$ $^{+46 \%}_{-38 \%}$&$1.7\cdot10^{1}$ $^{+39 \%}_{-33 \%}$&$2.2\cdot10^{1}$ $^{+38 \%}_{-32 \%}$\\
		$V_0 V'_T\to Z Z W^+$&\texttt{vxp\{0\} vxm\{T\} > z z w+}&$1.7\cdot10^{-1}$ $^{+19 \%}_{-19 \%}$&$9.3\cdot10^{-1}$ $^{+17 \%}_{-17 \%}$&$1.2\cdot10^{1}$ $^{+16 \%}_{-16 \%}$\\
		$V_0 V'_0\to Z Z W^+$&\texttt{vxp\{0\} vxm\{0\} > z z w+}&$1.9\cdot10^{-3}$ &$1.1\cdot10^{-2}$ &$1.5\cdot10^{-2}$\\
		\hline
		$\sum V_{\lambda_A} V'_{\lambda_B}\to \gamma Z W^+$&\texttt{vxp vxm > a z w+}&$3.5\cdot10^{0}$ $^{+46 \%}_{-38 \%}$&$1.5\cdot10^{1}$ $^{+39 \%}_{-33 \%}$&$1.9\cdot10^{1}$ $^{+38 \%}_{-32 \%}$\\
		$V_T V'_T\to \gamma Z W^+$&\texttt{vxp\{T\} vxm\{T\} > a z w+}&$3.4\cdot10^{0}$ $^{+47 \%}_{-38 \%}$&$1.4\cdot10^{1}$ $^{+40 \%}_{-34 \%}$&$1.8\cdot10^{1}$ $^{+39 \%}_{-33 \%}$\\
		$V_0 V'_T\to \gamma Z W^+$&\texttt{vxp\{0\} vxm\{T\} > a z w+}&$1.6\cdot10^{-1}$ $^{+20 \%}_{-20 \%}$&$6.7\cdot10^{-1}$ $^{+17 \%}_{-17 \%}$&$8.5\cdot10^{-1}$ $^{+17 \%}_{-17 \%}$\\
		$V_0 V'_0\to \gamma Z W^+$&\texttt{vxp\{0\} vxm\{0\} > a z w+}&$2.0\cdot10^{-3}$ &$8.3\cdot10^{-3}$ &$1.0\cdot10^{-2}$\\
		\hline
		$\sum V_{\lambda_A} V'_{\lambda_B}\to \gamma \gamma W^+$&\texttt{vxp vxm > a a w+}&$6.9\cdot10^{-1}$ $^{+48 \%}_{-39 \%}$&$2.6\cdot10^{0}$ $^{+41 \%}_{-34 \%}$&$3.3\cdot10^{0}$ $^{+40 \%}_{-33 \%}$\\
		$V_T V'_T\to \gamma \gamma W^+$&\texttt{vxp\{T\} vxm\{T\} > a a w+}&$6.8\cdot10^{-1}$ $^{+48 \%}_{-40 \%}$&$2.5\cdot10^{0}$ $^{+41 \%}_{-34 \%}$&$3.2\cdot10^{0}$ $^{+40 \%}_{-34 \%}$\\
		$V_0 V'_T\to \gamma \gamma W^+$&\texttt{vxp\{0\} vxm\{T\} > a a w+}&$1.5\cdot10^{-2}$ $^{+23 \%}_{-23 \%}$&$5.0\cdot10^{-2}$ $^{+21 \%}_{-21 \%}$&$6.2\cdot10^{-2}$ $^{+21 \%}_{-21 \%}$\\
		$V_0 V'_0\to \gamma \gamma W^+$&\texttt{vxp\{0\} vxm\{0\} > a a w+}&$1.8\cdot10^{-4}$ &$4.6\cdot10^{-3}$ &$5.3\cdot10^{-3}$\\
		\hline
		\hline
	\end{tabular}
	} 
	\caption{Same as Table~\ref{tab:eva_nH} but for $VVV$ production.}
	\label{tab:eva_VVV}
\end{center}
\end{table}

Other similarities between triboson and diboson production include the cross section hierarchy that one observes in $V_{T} V_{T}'$ scattering
largely appears also in $V_{0} V_{T}'$ scattering and $V_{0} V_{0}'$ scattering.
One exception is $\gamma\gamma W^+$ and $\gamma\gamma\gamma$ production in $V_0 V_0'$ scattering, where the ordering inverts.
We attribute this to the opening of orbital angular momentum configurations in $2\to3$ scattering, which can spoil the cancellations described in Sec.~\ref{sec:leptons_diboson}.
While it is beyond our immediate scope, one can, in principle, investigate the hierarchy of triboson processes by considering the relative important of specific partonic and helicity channels, 
e.g., $\gamma_- \gamma_+$ scattering or $Z_+ Z_0$ scattering. One can also further decompose the final-state triboson system into its helicity components, thereby allowing one to investigate the relative importance of high- and low-wave angular momentum configurations. Finally, due to the similarities between diboson and triboson production, we conjecture that comparable behavior will be observed for the production for four or more EW vector bosons.

We summarize these results for representative $\sqrt{s}$ in Table~\ref{tab:eva_VVV}.


\section{Discussion, Outlook, and Conclusions}\label{sec:conclusions}

As a weakly coupled, non-Abelian gauge theory, the weak sector of the SM naturally exhibits similarities to the electromagnetic and strong sectors, e.g., coupling universality. However, as the weak sector is also spontaneously broken, scattering and decay rates involving weak bosons feature a power-law dependence on the scale ratio $(M_V^2/Q^2)^k$ or $(v^2/Q^2)^k$, with $k>0$. In the absence of such contributions, i.e., when momentum-transfer scales are much larger than the EW scale, meaning that $\mathcal{O}(M_V^2/Q^2)$ terms are negligible, scattering and decay rates involving weak gauge bosons resemble the analogous expressions for massless gauge bosons in QED and pQCD. Importantly, this resemblance also holds for the factorization of weak-boson emission in the collinear and soft limits. 

Precisely how factorization (and resummation) operates in the weak sector, and in particular how it differs from QED and pQCD, is no longer an academic intrigue as multi-TeV $\mu^+\mu^-$ colliders and 100 TeV $pp$ colliders are being seriously discussed as eventual successors of the HL-LHC program~\cite{EuropeanStrategyforParticlePhysicsPreparatoryGroup:2019qin,EuropeanStrategyGroup:2020pow}. At these colliders, typical parton collisions readily satisfy known criteria for collinear factorization of weak bosons. Even at the $\sqrt{s}=13\TeV$ LHC, such hard-scattering scales have already been observed in measurements of VBF/S.  Therefore, establishing a fuller picture of the colliders' physics potential requires a better understanding of how collinear and soft weak bosons behave in multi-TeV collisions.

Motivated by the fact that multi-TeV $\mu^+\mu^-$ colliders are effectively ``high-luminosity weak boson colliders''~\cite{Costantini:2020stv}, we have revisited the treatment of weak bosons as perturbative constituents of high-energy leptons, the so-called Effective $W/Z$ Approximation~\cite{Dawson:1984gx,Kane:1984bb}. To conduct this investigation, we have implemented  PDFs at LO for helicity-polarized $\gamma$, $W^\pm$, and $Z$ bosons from $e^\pm$ and $\mu^\pm$ into the MC event generator {\mgFull}.\footnote{These features were released publicly in version 3.3.0, and available from the {\mgamc} repository.} This allows for the fully differential simulation of  scattering processes that are initiated by one or two initial-state EW bosons. Starting from a formula for scattering partons from a muon, which we state in Eq.~\eqref{eq:factTheorem},
we systematically explored the limitations of the EWA in Sec.~\ref{sec:validity}. Novelties of our comparative investigation are the focus on: 
(i) universal and quasi-universal power-law corrections,
(ii) universal and quasi-universal logarithmic corrections,
(iii) phase space dependence, (iv) helicity polarization, and (v) many-body processes with VBF.
Past studies were mostly restricted to $2\to1$ and $2\to2$ scattering processes, which possess special kinematics, at a single choice of factorization scale. We stress that some important issues, such as $Z_T/\gamma_T$ mixing and Sudakov resummation, were not investigated and are left to future work.

As documented in Sec.~\ref{sec:validity_mass}, a key conclusion is that the EWA is acutely sensitivity to power corrections of the form $(p_T^2/M_{VV'}^2)^k\sim (M_V^2/M_{VV'}^2)^k$ and $(M_V^2/M_{VV'}^2)^k$ for $k>0$. These corrections originate in the derivation of weak boson PDFs and are related to the accuracy of collinear factorization and the Goldstone Equivalent Theorem. Our results suggest that using the EWA to describe VBF requires $M_{VV'}>\mathcal{O}(1\TeV)$, or $(M_V^2/M_{VV'}^2)\lesssim0.01$. This is in addition to the typical assumptions needed to justify collinear factorization. When VBF systems carry larger invariant masses, we find that the difference between the full computation and the EWA are within factorization scale uncertainties. As documented in and around Eq.~\eqref{eq:scaleUncRatio}, these uncertainties can be very large and demonstrate a need for RG evolution in order to achieve precise results with weak boson PDFs. We caution that we restricted our attention to dynamic factorization scales that are proportional to $M_{VV'}$ for consistency across the \textit{many} processes that we surveyed. As in pQCD, more ``optimal'' choices probably exist but are also (probably) process dependent and should be investigated thoroughly. Our implementation of EW boson PDFs in {\mgamc} can facilitate such studies.

Importantly, we show that the size of non-universal power corrections and the size of factorization scale ambiguities in multi-TeV $\mu^+\mu^-$ collisions are due to the largeness of the $W$ and $Z$ masses. At first, this may seem at odds with collinear factorization in pQCD, where above even a few GeV, PDFs can describe full matrix elements involving light quarks. However, using the operator product expansion, one can show that the phenomenon of ``precocious scaling,'' i.e., the emergence of asymptotic freedom at moderate energies, is due to the smallness of \textit{parton} masses~\cite{Georgi:1976ve,Georgi:1976vf,DeRujula:1976baf}. Power corrections associated with quark masses $m_q$ are of the form $(m_q^2/Q^2)^k$. For a scattering scale of $Q\sim2-3\GeV$, these reach at most $\mathcal{O}(m_q^2/Q^2)\lesssim10^{-5}-10^{-2}$. Likewise, heavy quark PDFs become adequate and reliable tools for $\mathcal{O}(m_q^2/Q^2)\lesssim0.01$~\cite{Witten:1975bh,Barnett:1987jw,Olness:1987ep,Collins:1998rz,Aivazis:1993pi}.
Both are consistent with requiring that $(M_V^2/M_{VV'}^2)\lesssim0.01$ for the EWA to adequately describe full matrix elements. 

For $M_{VV'}\lesssim\mathcal{O}(1\TeV)$, we find contrasting behaviors between longitudinal and transverse weak boson PDFs:
whereas longitudinal PDFs overestimate full scattering amplitudes, transverse PDFs underestimate them.
As documented in Sec.~\ref{sec:validity_collider}, increasing the collider energy does not necessarily improve the accuracy of the EWA for $M_{VV'}\lesssim\mathcal{O}(1\TeV)$.
Whereas the presence of soft logarithms slightly improve the accuracy of transverse PDFs when $\sqrt{s}$ is  increased, these same logarithms worsen the accuracy for longitudinal PDFs.
It is clear that a matching, subtraction, or re-weighting scheme akin to those already available for pQCD and QED is needed to correct EWA matrix elements in this region. To facilitate such developments,  we also report the availability in {\mgamc} of both $q^2$- and $p_T^2$-dependent PDFs for transversely polarized EW bosons. As shown in Sec.~\ref{sec:validity_evo}, $p_T^2$-dependent PDFs consistently lead to larger cross sections in EWA, but converge to the $q^2$-dependent results when $\sqrt{s}$ increases. To further strengthen the parallels with pQCD, we give a proof-of-principle demonstration in Sec.~\ref{sec:validity_matching} of matrix-element matching of $W_T$ PDFs with the full matrix elements. Despite its formally large scale uncertainty band, the matched result shows significant independence on the matching scale.
Broadly speaking, these capabilities provide a starting point for matching EWA matrix elements to EW parton showers and more sophisticated EW boson PDFs that involve RG evolution.

Given these considerations,  we cataloged in Sec.~\ref{sec:leptons} a litany of processes of the form 
\begin{equation}
\sum_{V_\lambda\in\{\gamma_\lambda,\,Z_\lambda,\,W^\pm_\lambda\}} V_{\lambda_A}\, V' _{\lambda_B} \to \mathcal{F}, 
\end{equation}
where $\mathcal{F}$ contains up to $n_{\mathcal{F}}=4$ states from the collection $\{H, t, \overline{t}, W^+, W^-, Z, \gamma \}$.
In comparing polarized and polarization-summed cross sections, we find an intriguing interplay between helicity polarizations and hard scattering processes. For example: whereas multiboson and many-boson production is driven by the scattering of initial-states in the $(\lambda_A,\lambda_B)=(T,T)$ configuration, top quark and associated top quark production features a large $(0,T)$ and $(T,0)$ component. In further contrast, multi-Higgs processes are dominated by $(T,T)$ and $(0,0)$ helicity configurations but receive only a marginal contribution from mixed configurations. It is worth noting that scale uncertainties at $\sqrt{s}=3-30\TeV$ reach about $\delta \sigma/\sigma \sim 20\%-50\%$ for many processes, which is beyond expectations based on coupling order. As we remained inclusive with respect to the helicities of final-state particles, the processes we surveyed and their differential behavior can all be further investigated.

\subsection{Recommendations for using $W/Z$ PDFs in high-energy lepton collisions}\label{sec:conclusions_recs}

Finally, while much about PDFs for polarized and unpolarized weak bosons remains to be investigated, we believe this work helps clarify quantitatively when the $W$ and $Z$ bosons can be treated as partonic constituents of high-energy leptons. 
To further this prerogative, we provide a set of recommendations on using weak boson PDFs in many-TeV scattering calculations. 
These guidelines draw heavily from the findings in Sec.~\ref{sec:validity}, are supported by analytic derivations of weak boson PDFs, 
and are applied to our survey in Sec.~\ref{sec:leptons}. 
For details on the usage of EW boson PDFs in {\mgamc}, see App.~\ref{app:usage}.

\begin{itemize}
 \item To minimize power corrections of the form $\left(M_V^2/M_{VV'}^2\right)^k$, for $k>0$, and which spoil the accuracy of the Goldstone Equivalence Theorem, require that $M_{VV'}>\mathcal{O}(1)\TeV$.
 \item To minimize power corrections of the form $\left(p_T^{l~2}/M_{VV'}^2\right)^k$, for $k>0$, and which spoil the accuracy of collinear factorization, require that $\mu_f <\mathcal{O}(1)\TeV$.
 \item To minimize corrections associated with gauge bosons at $x\to1$, restrict the use of $p_T^2$-evolved PDFs  to $\sqrt{s}\gtrsim\mathcal{O}(10\TeV)$, or choose small $\mu_f$ for $\sqrt{s}\lesssim\mathcal{O}(10\TeV)$.
 \item While not discussed in detail, we find notably improved numerical stability for computations throughout Sec.~\ref{sec:leptons} when evaluating matrix elements in the Feynman gauge.
 \end{itemize}

\section*{Acknowledgments}
The authors are grateful to 
Carlos Alishaan-Harz, 
Federico De Lillo, 
Tao Han, 
Aleksander Kusina, 
Fred Olness,
Yang Ma,
Luca Mantani,
Keping Xie, and
Xiaoran Zhao 
for enlightening discussions.

AC received funding from FRS-FNRS agency (grant no. T.04142.18). 
FM and OM received funding from the European Union's Horizon 2020 research and innovation programme as part of the Marie Sk{\l}odowska-Curie Innovative Training Network MCnetITN3 (grant agreement no. 722104) and from FRS-FNRS agency via the IISN maxlhc convention (4.4503.16).
FM also received funding from FNRS ``Excellence of Science'' EOS be.h Project
No. 30820817. 
RR acknowledges the support of Narodowe Centrum Nauki under Grant No. 2019/34/E/ST2/00186,
and the contribution of the VBSCan COST Action CA16108. RR also acknowledges the support of the Polska Akademia Nauk (grant agreement PAN.BFD.S.BDN. 613. 022. 2021 - PASIFIC 1, POPSICLE). This work has received funding from the European Union's Horizon 2020 research and innovation program under the Sk{\l}odowska-Curie grant agreement No.  847639 and from the Polish Ministry of Education and Science

Computational resources were provided by the supercomputing facilities of the Universit{\'e} catholique de Louvain (CISM/UCL) and the Consortium des {\'E}quipements de Calcul Intensif en Fédération Wallonie Bruxelles (C{\'E}CI) funded by the Fond de la Recherche Scientifique de Belgique (FRS-FNRS) under convention 2.5020.11 and by the Walloon Region.

\appendix

\section{How to use EVA in MadGraph5\_aMC@NLO}\label{app:madgraph}

The simulation of polarized and unpolarized EW PDFs in high-energy charged lepton collisions, i.e., EVA, is possible using a series of commands  inside the {\mgamc} interface. Instruction on how to run/setup {\mgFull} for various configurations involving unpolarized matrix elements can be found in Ref.~\cite{Alwall:2014hca}; for the setup of polarized matrix elements, see Ref.~\cite{BuarqueFranzosi:2019boy}. In this appendix, we describe how to setup a EVA computation in {\mgamc} and particularly focus the new options and syntax introduced for this mode.

As a concrete example, we consider the hard scattering process $W^+ W^- \to h h $ in $\mu^+\mu^-$ collisions. A typical set of {\mgamc} commands to simulate a process like this  is 
\begin{verbatim}
set group_subprocesses False
generate w+ w- > h h 
output DIRECTORY_OUTPUT
launch DIRECTORY_OUTPUT 
\end{verbatim}
The command ``\texttt{set group\_subprocesses False}'' is currently mandatory and deactivates some internal optimization mechanisms that are not (yet) compatible with EW boson PDFs as implemented into {\mgamc}. The second command corresponds to the hard process, and operates at the level of initial-state weak bosons. (In this sense, EW bosons are treated as partons of $e^\pm$ and $\mu^\pm$.) Note that EVA is only implemented here at LO in perturbation theory, without $Z_T/\gamma_T$ mixing, and without EW-DGLAP evolution. Such corrections have a nontrivial impact on numerical results~\cite{Chen:2016wkt,Bauer:2017isx,Fornal:2018znf,Bauer:2018xag,Bauer:2018arx,Han:2020uid,Han:2021kes}. The ``\texttt{output}'' command defines the directory where the code containing MEs and phase space integration routines, i.e., \texttt{MadEvent}~\cite{Maltoni:2002qb}, are physically written on disk. The ``\texttt{launch}'' command activates an interface to configure, compile, and execute this code. As {\mgamc} works by numerically evaluating helicity amplitudes, the command above syntax is equivalent~\cite{BuarqueFranzosi:2019boy} to the syntax
\begin{verbatim}
set group_subprocesses False
generate    w+{+} w-{+} > h h 
add process w+{+} w-{-} > h h 
add process w+{+} w-{0} > h h 
add process w+{-} w-{+} > h h 
add process w+{-} w-{-} > h h 
add process w+{-} w-{0} > h h 
add process w+{0} w-{+} > h h 
add process w+{0} w-{-} > h h 
add process w+{0} w-{0} > h h 
output DIRECTORY_OUTPUT
\end{verbatim}
In both cases, the particle species $W^\pm_{\lambda}$ is paired with the polarized PDF $f_{W^\pm_\lambda/\mu^\pm}(\xi,\mu_f)$.

When the user interface is initiated, i.e., just after the ``\texttt{launch}'' command,  the user is prompted with the ability to edit multiple configuration files.  To run in EVA mode, a user will need to edit the file \texttt{DIRECTORY\_OUTPUT/Cards/run\_card.dat}. This file contains all configuration details related to the beam, factorization scales, phase space restrictions (cuts), etc.  The list of the important and new parameters are summarized in Table \ref{tab:eva_config}, and are described below.
It is important to stress that in our implementation of EVA, initial- and final-state $W$ and $Z$ bosons retain their masses in all helicity amplitudes; nowhere do we set $M_W, M_Z = 0\GeV$. As a consequence, other {\mgamc} modules, such as \texttt{MadSpin}~\cite{Artoisenet:2012st,Alwall:2014bza}, can be employed in conjunction with EVA computations. This allows one to study, for example, the full process, $W^+_{T}W^-_{T} \to h(\to c\overline{c}) h(\to b\overline{b})$.

Investigating new physics remains possible through the interface~\cite{deAquino:2011ub} to \textit{Universal FeynRules Object} (UFO) libraries~\cite{Degrande:2011ua}. We caution, however, that EW boson PDFs are hard-coded into files \texttt{ElectroweakFlux.f} and \texttt{ElectroweakFluxDriver.f} in the directory \texttt{LO/Source/PDF/}. This means that modifications to the $\ell-\ell-\gamma/Z$ and $\ell-\nu-W$ vertices introduced by a UFO will not propagate into the PDFs. We have designed and organized the calling of EW boson PDFs in {\mgamc} such that the $W$ and $Z$ boson masses as well as the EW couplings are automatically  set to those values listed in the configuration file \texttt{DIRECTORY\_OUTPUT/Cards/param\_card.dat}. The values of the electron and muon masses are not read from the \texttt{param\_card.dat}. Instead, the values listed in Eq.~\eqref{eq:leptonMasses} are hard-coded into the file 
\texttt{DIRECTORY\_OUTPUT/Source/PDF/ElectroweakFlux.inc}.

In order to initiate a computation with the EVA, the most important parameter that must be set in the file \texttt{run\_card.dat}
 is the  PDF set.
Choosing EW boson PDFs for both beams can be done via the ``\texttt{pdlabel}'' parameter, which now accepts three additional modes: 
``\texttt{eva}'' for the EW boson PDFs described in Sec.~\ref{sec:formalism};
``\texttt{iww}'' for the so-called Improved Weizs\"acker-Williams (IWW) $\gamma$ PDF of Ref.~\cite{Frixione:1993yw};
and
``\texttt{mixed}'' for enabling different PDF configurations for beams 1 and 2. 
A fourth option ``\texttt{none}'' deactivates  PDFs for both beams.
The $\gamma$ PDF described in Sec.~\ref{sec:formalism} is known historically as the Weizs\"acker-Williams approximation~\cite{vonWeizsacker:1934nji,Williams:1934ad,Peskin:1995ev}, and is analogous to the gluon PDF in QCD at LO.
Setting ``\texttt{pdlabel=iww}'' calls the $\gamma$ PDF derived using the IWW approximation~\cite{Frixione:1993yw}; this PDF is sometimes mislabeled in the literature. Simply put, the IWW $\gamma$ PDF augments the original Weizs\"acker-Williams $\gamma$ PDF by terms that correspond to operators in the operator product expansion with a  twist larger than 2, i.e., are relatively suppressed by powers of $(m_\ell^2/Q^2)$.
Just like for partons in hadron PDFs, the appropriate PDF and ME are paired automatically by the routines of \texttt{pdg2pdf.f} and \texttt{pdg2pdf\_lhapdf6.f} in the directory \texttt{LO/Source/PDF/}. To make these options clearer, we have updated instructions within \texttt{run\_card.dat} to read:
\begin{verbatim}
#*********************************************************************
# PDF CHOICE: this automatically fixes alpha_s and its evol.         *
# pdlabel: lhapdf=LHAPDF (installation needed) [1412.7420]           *
#          iww=Improved Weizsaecker-Williams Approx.[hep-ph/9310350] *
#          eva=Effective W/Z/A Approx.       [21yy.zzzzz]            *
#          none=No PDF, same as lhapdf with lppx=0                   *
#*********************************************************************
\end{verbatim}

\begin{table}[!t]
\begin{center}
\resizebox{\textwidth}{!}{
\renewcommand*{\arraystretch}{0.95}
\begin{tabular}{c|c|c|c}
\hline\hline
 & \texttt{run\_card} variable & Newly allowed values & Comments 
\\\hline
\multirow{4}{*}{PDF config. for beams 1 \& 2} & 
\multirow{4}{*}{\texttt{pdlabel}} & 
\texttt{none} & Deactivates PDF for beams 1 \& 2 \\
    & & \texttt{eva} & Activates EVA for beams 1 \& 2\\
    & & \texttt{iww} & Activates Improved Weizs\"acker-Williams $\gamma$ PDF~{\cite{Frixione:1993yw}}\\
    & & \texttt{mixed} & 
    Allows beams 1 \& 2 to be configured differently
    \\\hline
\multirow{3}{*}{PDF config. for beam 1} & 
\multirow{3}{*}{\texttt{pdlabel1}} & 
\texttt{none} & Deactivates PDF for beam 1 \\
    & & \texttt{eva} & Activates EVA for beam 1\\
    & & \texttt{iww} & Activates IWW $\gamma$ PDF for beam 1
\\\hline
PDF config. for beam 2 & \texttt{pdlabel2} & same as \texttt{pdlabel1} & Analogous to \texttt{pdlabel1} 
\\\hline
Fixed factorization scale & 
\multirow{3}{*}{\texttt{fixed\_fac\_scale1}} 
              & \texttt{True}  & Set $\mu_f$ for beam 1 to be static \\
 for beam 1             & & \texttt{False} & Sets $\mu_f$ for beam 1 to be dynamic \\
                        & &  & This option overrides global variable \texttt{fixed\_fac\_scale}
\\\hline
Fixed fact. scale for beam 2 & \texttt{fixed\_fac\_scale2} & same as \texttt{fixed\_fac\_scale1} & Analogous to \texttt{fixed\_fac\_scale1} 
\\\hline
\multirow{2}{*}{PDF evolution variable}    & \multirow{2}{*}{\texttt{ievo\_eva}} &    \texttt{0} (default) & Sets EVA PDF evolution variable to  $q^2$ \textbf{(only for EVA)} \\
    & & \texttt{1} & Sets EVA PDF evolution variable to $p_T^2$ \textbf{(only for EVA)}    
\\\hline
\multirow{2}{*}{Treatment of external helicities} & \multirow{2}{*}{\texttt{nhel}} & \texttt{0} & Summation over all helicities \\
    &   &   \texttt{1} & Importance sampling over helicities \textbf{(req. for EVA)}
\\\hline
New cut & dsqrt\_shat & default:0 & Min. invariant mass cut on the scattering process (in GeV)\\
\hline\hline
\end{tabular}
}
\caption{
The list of the important and/or new parameters introduce into {\mgamc} to support PDFs for polarized EW gauge bosons (EVA) from high-energy leptons. See text for further details. 
}
\label{tab:eva_config}
\end{center}
\end{table}

As described in Sec.~\ref{sec:formalism}, it is possible to derive EW boson PDFs that are functions of either the virtuality carried by the incoming EW boson or the transverse momentum carried away by the outgoing lepton in $\ell \to V \ell'$ splittings. The two are related but can lead to numerical differences (see Sec.~\ref{sec:validity_evo}). In {\mgamc}, both have been implemented and can be selected using the \texttt{run\_card.dat} parameter ``\texttt{ievo\_eva}.'' If this parameter is set to ``\texttt{0}'' (default), $q^2$-based PDFs will be called, otherwise $p_T^2$-based PDFs will be called. 

For asymmetric process, it is also possible, for the first time in {\mgamc}, to employ a different PDF configuration for each beam. This development makes it possible to call both EW boson PDFs and QCD parton PDFs for processes such as $W^- \overline{b} \to \overline{t}$ in $e^-p$ or $\mu^-p$ collisions. To enable this, a user must set ``\texttt{pdlabel=mixed}'' and configure two new \texttt{run\_card.dat} parameters ``\texttt{pdlabel1}'' and ``\texttt{pdlabel2}.'' These two parameters operate in the same manner as ``\texttt{pdlabel}'' but are limited to only one beam. 
We stress that it is up to the user to ensure that ``\texttt{pdlabel1}'' and ``\texttt{pdlabel2}'' align with the parameters ``\texttt{lpp1}'' and ``\texttt{lpp2},'' as well as align with the \texttt{1 2 > 3 4 \dots} ordering of ``\texttt{generate}'' command above.

Presently, the EVA is only enabled for $e^\pm$ and $\mu^\pm$ beams; other capabilities are under development. Therefore, for the EVA to work in {\mgamc}, one must stipulate which types of beams are colliding. 
In \texttt{run\_card.dat}, this is specified via the parameters ``\texttt{lpp1}'' for beam 1 ``\texttt{lpp2}'' for beam 2.
The allowed values of these parameters retain the same meaning as in previous versions of {\mgamc}. Explicitly, setting ``\texttt{lppX=0}'' corresponds to no PDF for beam \texttt{X}, where \texttt{X} is \texttt{1} or \texttt{2}. Likewise, ``\texttt{lppX=+1(-1)}'' corresponds to a PDF for a proton (antiproton), and ``\texttt{lppX=+2(-2)}'' calls 
the Equivalent Photon Approximation\footnote{We report the correction of a long-standing labeling ambiguity of this PDF in the file \texttt{LO/Source/PDF/ PhotonFlux.f}, which contains the implementation of this PDF. We also reiterate that this PDF describes a photon from a proton in the elastic limit~\cite{Alva:2014gxa} and therefore does not include DGLAP evolution. For inelastic emissions of photons from protons, set ``\texttt{lppX=+1(-1)}'' and use a proton PDF set with QED evolution. 
} $\gamma$ PDF of Ref.~\cite{Budnev:1975poe}, which describes the elastic emission of a photon from a proton (antiproton).
Setting ``\texttt{lppX=+3(-3)}'' means employing a PDF for an electron (positron) beam, and ``\texttt{lppX=+4(-4)}'' means a PDF for a muon (antimuon) beam. Presently, setting ``\texttt{pdlabel=eva}'' or ``\texttt{pdlabelX=eva}'' requires setting ``\texttt{lppX=$\pm$3}'' or ``\texttt{lppX=$\pm$4}.'' (The same is true for setting ``\texttt{pdlabel=iww}'' or ``\texttt{pdlabelX=iww}.'') To make these options clearer, we have updated the relevant instructions within \texttt{run\_card.dat} to now read:
\begin{verbatim}
#*********************************************************************
# Collider type and energy                                           *
# lpp: 0=No PDF, 1=proton, -1=antiproton, 2=elastic photon of proton,*
#             +/-3=PDF of electron/positron beam                     *
#             +/-4=PDF of muon/antimuon beam                         *
#*********************************************************************
\end{verbatim}

It is possible to use the EVA in {\mgamc} with same-sign lepton beams, e.g., ``\texttt{lpp1=+4}'' and ``\texttt{lpp2=+4},'' as well as mixed-flavor lepton beams, e.g.,  ``\texttt{lpp1=-3}'' and ``\texttt{lpp2=+4}.'' At this point, we reminder potential users that due to electric and weak isospin charge assignments, polarized weak boson PDFs are not charge symmetric, e.g., $\tilde{f}_{W^-_{-1}/\mu^-_L}(\xi,\mu) \neq \tilde{f}_{W^+_{-1}/\mu^+_L}(\xi,\mu)$. Likewise, the EW boson PDF implemented here are only LO accurate. This means that the $W^+$ content of $\ell^-$ is zero since such splittings first appear first at $\mathcal{O}(\alpha \alpha_W)$.

Presently, it is possible in {\mgamc} to polarize electron and muon beams in lepton-lepton and lepton-hadron collisions in the absence of lepton PDFs, i.e., ``\texttt{lppX=0}'' \cite{Alwall:2014hca}. We have extended this capability and it is now also possible to polarize electron and muon beams when ``\texttt{lppX=$\pm$3,$\pm$4}'' and ``\texttt{pdlabel=eva}'' or ``\texttt{pdlabelX=eva}.''
This is done via the \texttt{run\_card.dat} parameters ``\texttt{polbeam1}'' and ``\texttt{polbeam2}.''
 A value of ``\texttt{polbeamX=0}'' (default) corresponds to an unpolarized beam, while ``\texttt{-100}'' and ``\texttt{+100}'' indicate, respectively, that 100\% of beam \texttt{X} is polarized in the LH and RH helicity state.\footnote{
We note for clarity that setting ``\texttt{polbeamX=-70}'' indicates that 70\% of beam \texttt{X} is polarized in the LH state while the remaining 30\% is unpolarized. This implies that 85\%~(15\%) of beam \texttt{X} consists of leptons in their LH (RH) helicity state.}
 Note that the helicity polarization of an EW boson cannot be changed at run time via \texttt{run\_card.dat}. It can only be specified when executing the ``\texttt{generate}'' command; see Ref.~\cite{BuarqueFranzosi:2019boy} for details.

To implement beam polarization with the EVA, we have augmented Eq.~\eqref{eq:pdfDef_unpolarizedMuon}, which describes a polarized EW boson $V_\lambda$ from an unpolarized muon by the expression
\begin{equation}
 \tilde{f}_{V_\lambda/\mu^\pm}(\xi,\mu_f) = (\beta_L+\frac{1-\beta_L}{2}) \times \tilde{f}_{V_\lambda/\mu^\pm_L}(\xi,\mu_f)   + \frac{(1-\beta_L)}{2} \times \tilde{f}_{V_\lambda/\mu^\pm_R}(\xi,\mu_f).   
\end{equation}
Here, $-1\leq \beta_L \leq 1$ is a parameter describing the degree of LH polarization of the parent beam. 
For $\beta_L=0$, which corresponds to setting ``\texttt{polbeamX=0},''
one recovers Eq.~\eqref{eq:pdfDef_unpolarizedMuon}.
Likewise, setting $\beta_L=-1~(1)$, and implies that the muon beam itself is purely in the LH~(RH) helicity state corresponds to setting ``\texttt{polbeamX=-100~(100)}.''

The collinear factorization scale $\mu_f$ that enters into EW boson PDFs in the EVA can be either dynamical, i.e., determined on an event-by-event basis, or fixed. This is chosen in \texttt{run\_card.dat} via the Boolean parameter ``\texttt{fixed\_fac\_scale}.'' Setting this parameter to ``\texttt{true (false)}'' activates a fixed (dynamical) $\mu_f$.
Like ``\texttt{pdlabel},'' the dynamical/static scale scheme can be set separately for each beam using the two Boolean parameters ``\texttt{fixed\_fac\_scale1}'' and ``\texttt{fixed\_fac\_scale2}.''
If a static $\mu_f$ is selected, then its value is set in units of GeV by the parameters ``\texttt{dsqrt\_q2fact1}'' and ``\texttt{dsqrt\_q2fact2}.'' For dynamical choices of $\mu_f$, a user can choose from predefined or user-defined scale schemes using the parameter ``\texttt{dynamical\_scale\_choice}.'' For details on this, see Ref.~\cite{Alwall:2014hca}. It is possible to rescale $\mu_f$ by the scale factor ``\texttt{scalefact}''.\footnote{This new release of {\mgamc} is the first to allow both ``\texttt{scalefact}'' to be set different from unity and simultaneously allow ``\texttt{use\_syst=True}.'' However, one should note that the deprecated implementation of scale computation (SysCalc \cite{Kalogeropoulos:2018cke}) is not compatible with either the EVA or the new ``\texttt{scalefact}'' capability.
} Importantly, automated scale variation of EW boson PDFs is possible using the ``\texttt{systematics}'' feature and setting ``\texttt{use\_syst=True}.'' Users are reminded that $W/Z$ boson PDFs are not defined for values of $\mu_f < M_{W/Z}$.


\subsection{Example usage of EVA in \mgFull}\label{app:usage}

To reproduce the cross sections for the process $V_{\lambda_A}V_{\lambda_B}\to HH$ in Table~\ref{tab:eva_nH}, we use the following syntax to generate our (polarized) matrix elements and work environments:
\begin{verbatim}
set group_subprocesses false
set gauge Feynman
define vxp = w+ z a
define vxm = w- z a

generate vxp vxm > h h
output vxvx_hh

generate vxp{T} vxm{T} > h h
output vtvt_hh

generate    vxp{T} vxm{0} > h h
add process vxp{0} vxm{T} > h h
output vtv0_hh

generate vxp{0} vxm{0} > h h
output v0v0_hh
\end{verbatim}

Fiducial cross sections, scale uncertainties, and events at $\sqrt{s}=3\TeV$ can then be obtained in accordance to the setup in Sec.~\ref{sec:leptons} through the following run time commands
\begin{verbatim}
launch vxvx_hh
set width all 0
set lpp1 -4
set lpp2  4
set pdlabel eva
set fixed_fac_scale false
set dynamical_scale_choice 4 # muf = scalefact * dsqrt(shat)
set scalefact 0.5
set ievo_eva 0 # (0=q^2 or 1=pT^2)
set ebeam 1500
set no_parton_cut
set nevents 1k
set dsqrt_shat 1000
set pt_min_pdg  {25:50}
set eta_max_pdg {25:3}
set use_syst true
set nhel 1
\end{verbatim}
The new and important commands for using the EVA in {\mgamc} are documented in App.~\ref{app:madgraph} and summarized in Table~\ref{tab:eva_config}.

\subsection{Validation of EVA in {\mgFull}}\label{app:validation}

\begin{table}[!t]
\begin{center}
\resizebox{\textwidth}{!}{
\renewcommand*{\arraystretch}{0.95}
\begin{tabular}{c|c|c|c|c|c|c|c|c|c}
\hline\hline
        & \multicolumn{3}{c|}{$\sqrt{s}=$10 TeV} & \multicolumn{3}{c|}{$\sqrt{s}=$14 TeV} & \multicolumn{3}{c}{$\sqrt{s}=$30 TeV} \\
Channel & $\sigma^{\rm Ref.}$ [fb] & $\sigma^{\rm EVA}$ [fb] &         
        $\frac{\Delta\sigma}{\delta\sigma_{\rm Stat.}}$
        & $\sigma^{\rm Ref.}$ [fb] & $\sigma^{\rm EVA}$ [fb] & $\frac{\Delta\sigma}{\delta\sigma_{\rm Stat.}}$
        & $\sigma^{\rm Ref.}$ [fb] & $\sigma^{\rm EVA}$ [fb] & $\frac{\Delta\sigma}{\delta\sigma_{\rm Stat.}}$ \\
\hline
Unpolarized         &17.23  &17.21   &-0.8 &21.17  &21.15   &-0.6  &30.71  &30.68   &-0.7 \\
$W_0^+W_0^-$        &7.702  &7.694   &-0.6 &9.244  &9.235   &-0.6  &12.85  &12.84   &-0.7 \\
$W_0^\pm W_T^\mp$   &7.713  &7.704   &-0.7 &9.595  &9.582   &-0.8  &14.18  &14.17   &-0.6 \\
$W_T^+ W_T^-$ 	    &1.811  &1.810   &-0.3 &2.329  &2.327   &-0.6  &3.682  &3.682   &-0.1 \\

\hline\hline
\end{tabular}
}
\caption{
For the process $W^+_{\lambda_A}W^-_{\lambda_B}\to t \overline{t}$ in multi-TeV $\mpmm$ collisions, EWA-level cross sections [fb] as reported by Ref.~\cite{Han:2020uid} $(\sigma^{\rm Ref.})$, the cross section computed with {\mgamc} $(\sigma^{\rm EVA})$, and the statistical pull $(\Delta \sigma / \delta\sigma_{\rm Stat.})$,  for
(top row) unpolarized $W^+W^-$,
(second row) $W_0^+W_0^-$,
(third row) $W_0^\pm W_T^\mp$,
and (bottom row) $W_T^+ W_T^-$ scattering, at $\sqrt{s}=10\TeV$ (left), 14\TeV (center), and 30 TeV (right).
The {\mgamc} statistical uncertainty corresponds to 400k events, or $\delta\sigma_{\rm Stat.}\approx\pm0.16\%$.
}
\label{tab:eva_validation}
\end{center}
\end{table}

As one high-level check (of several) of our implementation of the EVA, we consider the process $W^+_{\lambda_A}W^-_{\lambda_B}\to t \overline{t}$ in multi-TeV $\mpmm$ collisions as studied in Ref.~\cite{Han:2020uid}. To simulate this process for various polarization configurations, we use the {\mgamc} syntax
\begin{verbatim}
set group_subprocesses false
set gauge Feynman

generate w+ w- > t t~ QED=2 QCD=0
output wxwx_tt_XLO

generate w+{0} w-{0} > t t~ QED=2 QCD=0
output w0w0_tt_XLO

generate w+{0} w-{T} > t t~ QED=2 QCD=0
add process w+{T} w-{0} > t t~ QED=2 QCD=0
output w0wT_tt_XLO

generate w+{T} w-{T} > t t~ QED=2 QCD=0
output wTwT_tt_XLO
\end{verbatim}
The commands above correspond to
(i) unpolarized $W^+W^-$ scattering,
(ii) $W_0^+W_0^-$ scattering,
(iii) $W_0^\pm W_T^\mp$ scattering for $T=\pm1$,
and (iv) $W_T^+ W_T^-$ scattering.

To avoid potential instabilities, the authors of Ref.~\cite{Han:2020uid} require final-state top quarks to have a nonzero polar angle in the hard-scattering frame. Specifically, they require
\begin{equation}
    \cos\theta_{t~(\overline{t})} = \frac{p_z^{t~(\overline{t})}}{\vert \vec{p}^{~t~(\overline{t})}}\vert < \frac{m_t^2}{m_{t\overline{t}}^2}.
\end{equation}
Here $p_z^{t(\overline{t})}$ is the $z$ momentum of the (anti)top quark in the hard frame, $\vec{p}$ is its three-momentum in the same frame, and $m_{t\overline{t}}$ is the invariant mass of the $(t\overline{t})$-system. We implement this cut by adding the following lines in their appropriate locations to the \texttt{dummy\_cuts} function in the auxiliary file \texttt{DIRECTORY\_OUTPUT/SubProcesses/dummy\_fct.f}:
\begin{verbatim}
integer ff
double precision mtop2,sHat,rat,cosTh
double precision SumDot
external SumDot

mtop2 = (173.0d0)**2
sHat = SumDot(p(0,1), p(0,2), 1d0)
rat = 1.d0 - mtop2 / sHat

do ff=nincoming+1,nexternal
c         =      pz / dsqrt(px2 + py2 + pz2)
    cosTh = p(3,ff) / dsqrt(p(1,ff)**2 + p(2,ff)**2 + p(3,ff)**2)
    if(cosTh.gt.rat) then
        dummy_cuts=.false.
        return
    endif
enddo
\end{verbatim}

To steer phase space integration, we use the following commands at $\sqrt{s}=10\TeV$:
\begin{verbatim}
launch wxwx_tt_XLO
set no_parton_cut
set ebeam 5000
set nevents 40k
set lpp1  -4 # setup beam1 as anti-muon
set lpp2   4 # setup beam2 as muon
set pdlabel eva
set fixed_fac_scale1 false
set fixed_fac_scale2 false
set dynamical_scale_choice 4 # muf = scalefact * dsqrt(shat)
set scalefact 0.5
set ievo_eva 0 # (0=q^2 or 1=pT^2)
set use_syst true
done    
\end{verbatim}
We make the necessary modification to this script for $\sqrt{s}=14$ and $30\TeV$.

We report in Table~\ref{tab:eva_validation} the EVA-level cross section [fb] as reported by Ref.~\cite{Han:2020uid}, denoted by $\sigma^{\rm Ref.}$, the cross section computed with {\mgamc}, denoted by $\sigma^{\rm EVA}$, and the statistical pull, defined by $\Delta \sigma / \delta\sigma_{Stat.} \equiv (\sigma^{\rm EVA} - \sigma^{\rm Ref.}) / \delta\sigma_{Stat.}$, for $\sqrt{s}=10, 14$, and 30 TeV, for
(top row) unpolarized $W^+W^-$ scattering,
(second row) $W_0^+W_0^-$ scattering,
(third row) $W_0^\pm W_T^\mp$ scattering for $T=\pm$,
and (bottom row) $W_T^+ W_T^-$ scattering.
Due to \texttt{MadEvent}'s multi-channel integration routines, a precise determination of our MC statistical uncertainty is complicated. Therefore, as a conservative estimate of our statistical uncertainty for 400k events, we take $\delta\sigma_{Stat.}=1/\sqrt{4\cdot10^5}\approx0.16\%$. While we find that differences are all negative, meaning $\sigma^{\rm EVA}$ is always larger, we report that differences only span $\Delta\sigma\approx-0.13\%$ to $-0.01\%$. This translates to a statistical pull ranging from $\Delta\sigma/\delta\sigma_{\rm Stat.}\approx-0.1$ to $-0.8$.

\section{Effective $W, Z, \gamma$ Approximation}\label{app:formalism}
In this appendix, we derive the (polarized) $W/Z$ PDFs at LO and construct Eq.~\eqref{eq:factTheorem} as implemented in the  MC event generator {\mgamc}. We include this for completeness and to make this work more self-contained for the non-expert reader.
In making these mechanics explicit, we hope to minimize ambiguities in definitions of PDF that sometimes appear in the literature. Such differences are due to equally reasonable choices of, e.g., evolution variables, but can lead to sizable numerical differences at LO.
We direct expert readers to derivations of the EVA in the axial gauge~\cite{Kunszt:1987tk,Borel:2012by,Chen:2016wkt} and a fully covariant formalism~\cite{Cuomo:2019siu} for considerations of more nuanced issues, such as Goldstone mixing and renormalization.

The construction proceeds in the following manner: after establishing notation and conventions in App.~\ref{app:formalism_notation}, MEs for polarized collinear splitting functions are derived in App.~\ref{app:formalism_splitting}, with the corresponding phase space decomposition/factorization given in App.~\ref{app:phaseSpace}. The unrenormalized, transverse-momentum-dependent distribution functions at LO are then derived in App.~\ref{app:tmd}, and finally the analogous collinear PDFs are given in App.~\ref{app:collinearPDF}. While the construction follow closely that of Ref.~\cite{Peskin:1995ev} for parton splitting in QED, the identification of a factorization scale $\mu_f$ as an ultraviolet regulator of a phase space integral is more closely aligned with Ref.~\cite{Kunszt:1987tk}. This identification makes clearer that the procedures for implementing matrix-element matching (see Sec.~\ref{sec:validity_matching}) and (potential) RG evolution with renormalized PDFs are similar to those in pQCD.

\subsection{Helicity Amplitude Notation and Conventions}\label{app:formalism_notation}
To derive polarized, weak boson splitting functions and PDFs, we work at the helicity amplitude level, in the so-called \texttt{HELAS} basis~\cite{Murayama:1992gi}. We assume the spacetime metric signature $ g_{\mu\nu}=\text{diag}(+,-,-,-)$, and work in the unitary gauge at dimension $d=4$.

\subsubsection*{Spin-$1/2$ Particles}
For a fermion with mass $m=\sqrt{E^2 - \vert\vec{p}\vert^2}$, and a spin axis aligned with its momentum, its 4-momentum $(p^\mu)$ and transverse momentum $(p_T)$ can be generically parameterized by 
\begin{align}
p^\mu &= (E, p_x, p_y, p_z) = (E,\vert\vec{p}\vert \sin\theta\cos\phi,\vert\vec{p}\vert\sin\theta\sin\phi,\vert\vec{p}\vert\cos\theta), \\
		p_T^2 &= p_x^2 + p_y^2 = \vert \vec{p}\vert^2 \sin^2\theta.
\end{align}
The two-component helicity eigenstates with respect to the 3-momentum direction $\hat{p}$ are 
\begin{eqnarray}
 \chi(\hat{p},\lambda=+1)&=&\frac{1}{\sqrt{2\vert\vec{p}\vert(\vert\vec{p}\vert+p_z)}}
\begin{pmatrix} \vert\vec{p}\vert+p_z \\ p_x + ip_y \end{pmatrix}
=\begin{pmatrix} \cos\frac{\theta}{2} \\ e^{i\phi}\sin\frac{\theta}{2} \end{pmatrix},
\\
\chi(\hat{p},\lambda=-1)&=&\frac{1}{\sqrt{2\vert\vec{p}\vert(\vert\vec{p}\vert+p_z)}}
\begin{pmatrix} -p_x + ip_y \\ \vert\vec{p}\vert+p_z \end{pmatrix}
=\begin{pmatrix} -e^{-i\phi}\sin\frac{\theta}{2}  \\ \cos\frac{\theta}{2} \end{pmatrix},
\end{eqnarray}
RH/LH helicities are normalized such that $\lambda=+1/-1$.
Given this, the four-component helicity spinors for fermions $(u)$ and antifermions $(v)$ are:
\begin{eqnarray}
 u(p,\lambda) =  \left(\begin{matrix}
 \sqrt{E-\lambda\vert\vec{p}\vert} ~\chi(\hat{p},\lambda)
\\ 
\sqrt{E+\lambda\vert\vec{p}\vert} ~\chi(\hat{p},\lambda)
\end{matrix}\right)
\quad\text{and}\quad
v(p,\lambda) = \left(\begin{matrix}
-\lambda\sqrt{E+\lambda\vert\vec{p}\vert} \chi(\hat{p},-\lambda)
\\ 
\lambda\sqrt{E-\lambda\vert\vec{p}\vert} ~\chi(\hat{p},-\lambda)
\end{matrix}\right).
\end{eqnarray}
In the high energy limit, where $(m/E)\ll1$, the Dirac spinors simplify to
\begin{eqnarray}
u(p,\lambda=+1) \approx \sqrt{2E} \left(\begin{matrix} 0 \\  0 \\ \cos\frac{\theta}{2} \\ e^{i\phi}\sin\frac{\theta}{2}\end{matrix}\right),
&\quad&
u(p,\lambda=-1) \approx \sqrt{2E} \left(\begin{matrix} -e^{-i\phi}\sin\frac{\theta}{2}  \\ \cos\frac{\theta}{2} \\ 0 \\ 0 \end{matrix}\right),
\\
v(p,\lambda=+1) \approx \sqrt{2E} \left(\begin{matrix} e^{-i\phi}\sin\frac{\theta}{2}  \\ -\cos\frac{\theta}{2} \\ 0 \\ 0 \end{matrix}\right),
&\quad&
v(p,\lambda=-1) \approx  -\sqrt{2E} \left(\begin{matrix} 0 \\ 0 \\ \cos\frac{\theta}{2} \\ e^{i\phi}\sin\frac{\theta}{2} \end{matrix}\right).
\end{eqnarray}
As such, the RH/LH chiral projection operators $P_{R/L}$ are respectively,
\begin{equation}
P_R = \frac{1}{2}(1+\gamma^5) = \begin{pmatrix}
0 & 0 & 0 & 0 \\ 
0 & 0 & 0 & 0 \\ 
0 & 0 & 1 & 0 \\ 
0 & 0 & 0 & 1
\end{pmatrix}
\quad\text{and}\quad
P_L = \frac{1}{2}(1-\gamma^5) = \begin{pmatrix}
1 & 0 & 0 & 0 \\ 
0 & 1 & 0 & 0 \\ 
0 & 0 & 0 & 0 \\ 
0 & 0 & 0 & 0
\end{pmatrix}.
\end{equation}
Immediately one sees that the projection operators satisfy the identities
\begin{equation}
1=P_R + P_L \quad\text{and}\quad \gamma^5 = P_R - P_L.
\end{equation}

\subsubsection*{Massive Spin-1 Particles}
For a vector boson with mass $M_V=\sqrt{E_V^2 - \vert\vec{q}\vert^2}$ and a spin axis aligned with its momentum, its 4-momentum $(q^\mu)$ and transverse momentum $(q_T)$ can be  parameterized by
\begin{align}
 q^\mu &= (E_V, q_x, q_y, q_z) = (E_V, \vert\vec{q}\vert \sin\bar\theta\cos\bar\phi, \vert\vec{q}\vert \sin\bar\theta\sin\bar\phi, \vert\vec{q}\vert \cos\bar\theta),\\ 
 		q_T^2 &= q_x^2 + q_y^2 = \vert \vec{q}\vert^2 \sin^2\bar\theta.
\end{align}
In the Cartesian representation, its polarization vectors are given by
\begin{eqnarray}
 \varepsilon^\mu (q,x) =& \cfrac{1}{\vert\vec{q}\vert q_T} (0, q_x q_z, q_y q_z, -q_T^2 ) &= (0, \cos\bar\theta \cos\bar\phi, ~\cos\bar\theta \sin\bar\phi, ~-\sin\bar\theta),
  \label{eq:formalism_polX}
 \\
 \varepsilon^\mu (q,y) =& \cfrac{1}{q_T} (0, - q_y, q_x, 0 ) &= (0, -\sin\bar\phi, ~\cos\bar\phi, 0),
  \label{eq:formalism_polY}
 \\
 \varepsilon^\mu (q,z) =& \cfrac{E_V}{M_V\vert\vec{q}\vert } \left( \frac{\vert\vec{q}\vert^2}{E}, q_x , q_y , q_z \right) &= \gamma ( \beta, ~\sin\bar\theta \cos\bar\phi, ~\sin\bar\theta \sin\bar\phi, ~\cos\bar\theta ).
 \label{eq:formalism_polZ}
\end{eqnarray}
In the last line we used the Lorentz factor $\gamma = E_V/M_V$ and the relationship $\vert\vec{q}\vert = \beta E_V$.
In this representation, the above polarization vectors satisfy the orthogonality relationships:
\begin{equation}
q_\mu \varepsilon^\mu (q,\lambda)=0 
\quad\text{and}\quad
\varepsilon^\mu (q,\lambda) \varepsilon_\mu (q,\lambda')=-\delta_{\lambda,\lambda'} \quad\text{for}\quad \lambda=x,y,z.
\end{equation}
Using the above, we can build the polarization vectors in the polar representation.
For the RH $(\lambda=+1)$, LH $(\lambda=-1)$, and longitudinal $(\lambda=0)$ polarizations, these are given by
\begin{eqnarray}
 \varepsilon^\mu (q,\lambda=\pm1) 	&=& \frac{1}{\sqrt{2}}\left( -\lambda\varepsilon^\mu(q,x) - i\varepsilon^\mu(q,y)\right) \\
 							&=& \frac{1}{\sqrt{2}}\left(0,	-\lambda\cos\bar\theta\cos\bar\phi + i\sin\bar\phi, 
													-\lambda\cos\bar\theta\sin\bar\phi -  i\cos\bar\phi,
														       \lambda\sin\bar\theta \right),
 \label{eq:formalism_polT}
 \\
 \varepsilon^\mu (q,\lambda=0) &=& \varepsilon^\mu (q,z).
 \label{eq:formalism_polL}
\end{eqnarray}
For $\lambda=\pm1$, the orthogonality relationships are modified such that 
\begin{equation}
q_\mu \varepsilon^\mu (q,\lambda)=0,
\quad 
\varepsilon^\mu (q,\lambda) \varepsilon_\mu (q,\lambda')=\delta_{\lambda,-\lambda'}
\quad\text{and}\quad 
\varepsilon^\mu (q,\lambda) \varepsilon_\mu^* (q,\lambda')=-\delta_{\lambda,\lambda'}.
\end{equation}

In the EVA it is useful~\cite{Dawson:1984gx} to rewrite the $\lambda=0$ vector in the following exact form:
\begin{align}
 \varepsilon^\mu (q,\lambda=0) &= \frac{q^\mu}{M_V} + \tilde{\varepsilon}^\mu (q), \quad\text{where}
 \label{eq:formalism_polA}
   \\
\tilde{\varepsilon}^\mu (q) &\equiv \frac{M_V}{E_V+\vert \vec{q}\vert}\left(-1,\frac{q_x}{\vert \vec{q}\vert},\frac{q_y}{\vert \vec{q}\vert},\frac{q_z}{\vert \vec{q}\vert}\right)\\
&= \frac{M_V}{E_V+\vert \vec{q}\vert}\left(-1,\sin\bar\theta\cos\bar\phi,\sin\bar\theta\sin\bar\phi,\cos\bar\theta\right).
\end{align}
Explicit computation shows that the auxiliary  vector $\tilde{\varepsilon}$ obeys the following inner products:
\begin{eqnarray}
q_\mu \tilde{\varepsilon}^\mu(q) = -M_V, &\quad&	\tilde{\varepsilon}^\mu (q) \tilde{\varepsilon}_\mu (q) = 0, \quad \tilde{\varepsilon}^\mu (q) \tilde{\varepsilon}_\mu^* (q) = 0, \\  
\tilde{\varepsilon}^\mu (q)  \varepsilon_\mu (q,\lambda=\pm1) = 0, &\quad& \tilde{\varepsilon}^\mu (q)  \varepsilon_\mu^* (q,\lambda=\pm1) = 0.
\end{eqnarray}
The purpose of this decomposition is two-fold:
The first is to make manifest that the Goldstone contribution, i.e., the term that scales as $ \varepsilon^\mu(\lambda=0) \sim q^\mu/M_V$,  vanishes when contracted with external parton currents via the Dirac equation.  Ultimately, this is due to SU$(2)_L$ current conservation of massless leptons,  meaning that massless leptons do not participate in helicity-inverting couplings of longitudinally weak boson. The second purpose is to make manifest that the non-vanishing part of $ \varepsilon^\mu(\lambda=0)$ is formally a quasi-universal, beyond-twist-two term that scales as $\varepsilon^0 \sim M_V / E$. {Massless vector bosons have identical transverse polarization vectors but do not possess a longitudinal polarization vector.}

\subsection{Polarized Collinear Splitting Functions for Massive Vector Bosons}\label{app:formalism_splitting}

\begin{figure}[!t]
\centering
\includegraphics[width=.7\textwidth]{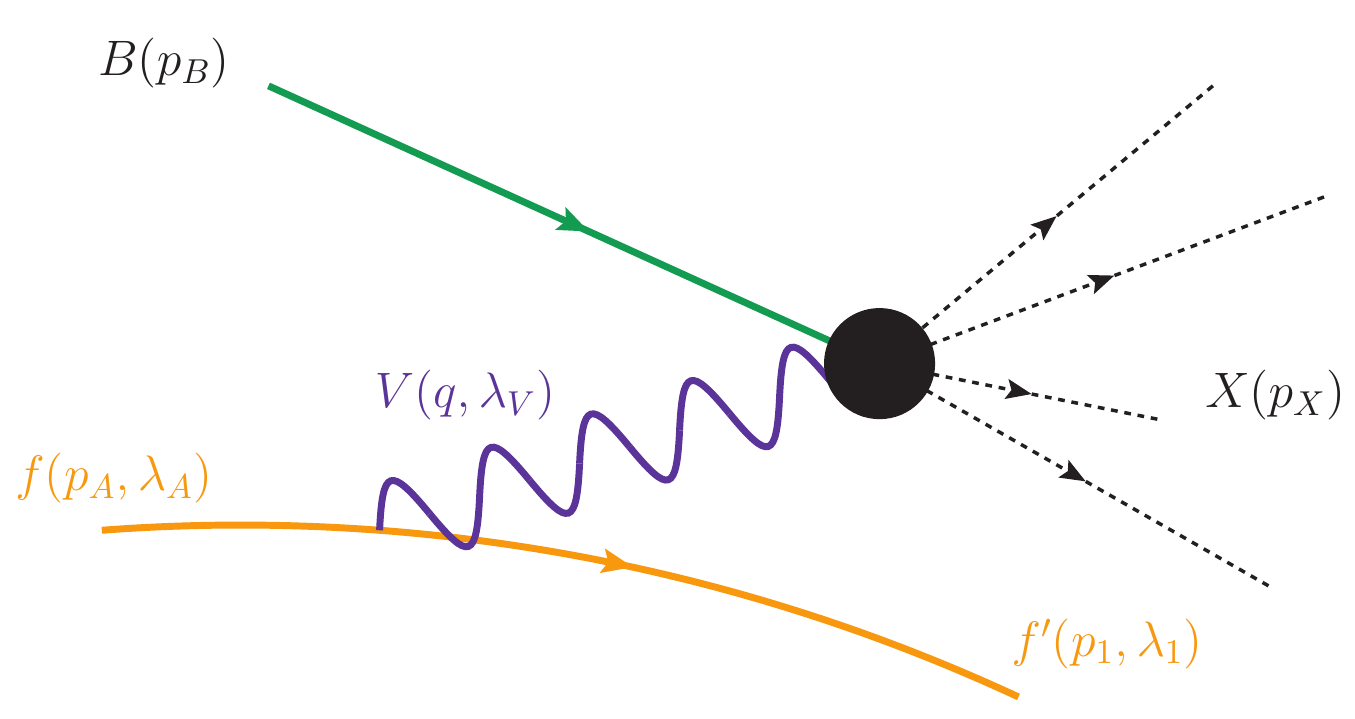}
\caption{
Diagrammatic representation of $f(p_A,\lambda_A)\to V(q,\lambda_V) f'(p_1,\lambda_1)$ splitting.
} 
\label{fig:diagram_EVA_DIS}
\end{figure}

\subsubsection*{Setup and On-shell Decomposition}

To build collinear splitting functions for weak bosons from high-energy leptons, we consider the deeply inelastic scattering (DIS) process shown in Fig,~\ref{fig:diagram_EVA_DIS}, given by
\begin{equation}
f(p_A,\lambda_A) + B(p_B) ~\xrightarrow{V(q,\lambda_V)+B(p_B)\to X(p_X)}~ f'(p_1, \lambda_1) + X(p_X).
\label{eq:ewDIS}
\end{equation}
The process above describes $f(p_A) \to f'(p_1) + V(q)$ splitting with external fermions $f, f',$ and an internal weak boson $V$  with mass $M_V$. Throughout this section, we remain agnostic to the composition of $B$. In App.~\ref{app:collinearPDF}, we will extend our result and present a scattering formula involving a PDF for each lepton beam. As we are only working at LO in the EW theory for both the ``hard'' scattering $(VB\to X)$ and the splitting processes $(f\to f'V)$, we automatically work in the Single Boson Exchange Approximation. 
Furthermore, throughout this analysis, we neglect non-factorizable contributions. As we are working in the Unitary gauge (for clarity), the class of diagrams where this is an acceptable approximation is known to be limited in comparison to working in other gauges. For discussions on this, see Refs.~\cite{Libby:1978qf,Libby:1978bx,Kunszt:1987tk,Accomando:2006mc,Borel:2012by,Chen:2016wkt,Cuomo:2019siu}.

For the process in Eq.~\eqref{eq:ewDIS}, the associated ME is given by
\begin{equation}
 \mathcal{M}_{\rm EW-DIS}  = J^\mu(f_{\lambda_A} \rightarrow f'_{\lambda_1}) ~\cdot~ \Delta_{\mu\nu}^V(q = p_A-p_1) ~\cdot~ \mathcal{M}^\nu(V^*_{\lambda_V} B\to X).
\end{equation}
Here $\mathcal{M}^\nu(V^*_{\lambda_V} B\to X)$ describes the $V^*_{\lambda_V} B\to X$ hard scattering process that occurs at the scale $Q=\sqrt{(q+p_B)^2}\sim E_V\gg M_V\sim\sqrt{\vert q^2\vert}$.
The propagator for $V$ in the unitary gauge is denoted by $\Delta_{\mu\nu}^V$. Particle helicities  are denoted by  $\lambda$. For a $V-f-f'$ chiral interaction with LH (RH) coupling $g_L^f~(g_R^f)$ and a universal strength $\tilde{g}$, the $f_{\lambda_A} \rightarrow f'_{\lambda_1} V^*_{\lambda_V}$ fermion current can be written generically as (see Table \ref{tab:ewa_coup} in Sec.~\ref{sec:formalism_pdfs} for explicit expressions)
\begin{equation}
J^\mu(f_{\lambda_A} \rightarrow f'_{\lambda_1})  = \left[\overline{u}(p_1,\lambda_1) (-i\tilde{g})\gamma^\mu\left(g_L^f P_L + g_R^f P_R\right) u(p_A,\lambda_A)\right].
\label{eq:splitting_current_gen}
\end{equation}

An important caveat to this expression is that it assumes we are working solely within the confines of the SM's SU$(2)_L$ gauge group (or some similarly broken gauge theory). It does not account for the fact that quarks and leptons in the SM  both reside in the SU$(3)_c\otimes$SU$(2)_L$ product group. While working in this larger structure has trivial implications for EW splitting functions themselves, it does impact
d.o.f. counting and cancellations at the level of cross sections.
More precisely, for the $f_{\lambda_A} \rightarrow f'_{\lambda_1} V^*_{\lambda_V}$ color-singlet splitting process, the $V-f-f'$ vertex is modified by a Kronecker $\delta$-function that ensures color conservation.
In equation \eqref{eq:splitting_current_gen}, this causes the modification
\begin{equation}
{\rm SU}(2)_L \to {\rm SU}(3)_c\otimes{\rm SU}(2)_L \quad:\quad (-i\tilde{g}) \to (-i\tilde{g})\delta_{JI},
\label{eq:ewa_colorFR}
\end{equation}
where the indices $I$ and $J$ denote the colors of $f$ and $f'$, respectively, and run over $I,J=1,\dots,N_c=3$. For quarks, $\delta_{II}=N_c$, whereas $I,J=1$ and $\delta_{II}=1$ for leptons.

The treatment of $V$'s propagator is subtle. In the language of Ref.~\cite{Peskin:1995ev}, we are considering the kinematic regime where $V$ is ``almost real.'' For weak bosons, means we are assuming that the norm of $V$'s virtuality, its mass, and the difference of these two quantities, are all small compared to the hard scattering scale. We are also assuming that $V$ propagates only a single helicity between the $(ff')$ and $B$ systems since helicity inversion is suppressed by $(M_V^2/Q^2)\ll1$.
Following Ref.~\cite{Peskin:1995ev}, this entails making the replacement
\begin{align}
 \Delta_{\mu\nu}^V(q) =  \frac{(-i)\left(g_{\mu\nu} - q_\mu q_\nu/M_V^2\right)}{(q^2 - M_V^2)} \to  \Delta_{\mu\nu}^V(q) =\frac{i\sum_{\lambda_V\in\{0,\pm\}}\varepsilon_\mu^*(q,\lambda_V) \varepsilon_\nu(q,\lambda_V)}{(q^2 - M_V^2)}.
\end{align}
The idea of this (heuristic) replacement is that if $V$, which possesses an invariant lifetime $\tau_V=\hbar/\Gamma_V$,  is ``almost'' on-shell and if the hard scattering scale is sufficiently large such that $Q\gg \Gamma_V$, then $V^*$ is so long-lived compared to the hard process that it can be approximated as an asymptotic state with definite quantum numbers, e.g., with fixed helicity. 
Since $V$'s helicity $\lambda_V$ remains unchanged during propagation, the contribution from the unphysical / auxiliary polarization, which scales as $\varepsilon_\mu(q,A) \sim q_\mu\sqrt{q^2-M_V^2}$, vanishes when contracted with $J^\mu(f_{\lambda_A}\to f'_{\lambda_1})$.
That is to say, since we assume  $f$ and $f'$ are massless, $J(f_{\lambda_A}\to f'_{\lambda_1})\cdot q=0$ by the Dirac equation.
The net contribution of the propagator $ \Delta_{\mu\nu}^V(q)$ reduces to a \textit{coherent} sum over physical polarizations $\lambda_V$, i.e., a summation over $\lambda_V$ at the squared ME level. We note that this replacement can be made stronger when working in other gauges
~\cite{Kunszt:1987tk,Borel:2012by,Chen:2016wkt,Cuomo:2019siu}.

In terms of bras and kets, this decomposition is equivalent to writing the ME as
\begin{align}
  -i\mathcal{M}_{\rm EW-DIS} & ~=~  \langle X f' \vert  f B \rangle  
  ~=~  
  \langle X \vert \langle f' \vert
  \left(\frac{i}{q^2 - M_V^2} \sum_\lambda  \vert V_\lambda \rangle\langle V_\lambda\vert \right)
  \vert f \rangle \vert B \rangle
  \\
    & ~\approx~ \frac{i}{q^2 - M_V^2} \sum_\lambda 
   \langle X \vert V_\lambda B\rangle
   ~
   \langle V_\lambda f'\vert f\rangle ,
\end{align}
which translates to a squared ME of
\begin{equation}
  \vert \mathcal{M}_{\rm EW-DIS}\vert^2 ~\approx~ \frac{1}{(q^2 - M_V^2)^2} 
  \sum_{\lambda,\lambda'} 
  \langle f \vert V_{\lambda'} f'\rangle
  \langle V_\lambda f'\vert f\rangle 
  ~
  \langle V_{\lambda'} B \vert X \rangle
  \langle X \vert V_\lambda B\rangle.
\end{equation}
Na\"ively, this suggests a double summation over the helicities of $V$. However, when one is totally inclusive over the final-state $X$, then by unitarity $\mathbb{1} = \sum_X \vert X \rangle \langle X\vert$. This implies that the squared ME is only non-vanishing when the polarization of $V$ in $f\to f' V_\lambda$ is the same as the polarization for $V$ in $V_{\lambda'}B\to X$. Other $\langle V_\lambda\vert V_{\lambda'}\rangle$ combinations are orthogonal:
\begin{equation}
\sum_X  \langle V_{\lambda'} B \vert X \rangle
  \langle X \vert V_\lambda B\rangle = \langle V_{\lambda'} B \vert V_\lambda B\rangle = 0, \quad\text{for}~\lambda\neq\lambda'.
\end{equation}
The unitarity of final-state $X$ acts as an effective Kronecker $\delta$-function $\delta_{\lambda\lambda'}$. This reduce the double summation over helicities into a single summation over helicities, resulting in
\begin{equation}
\sum_X   \vert \mathcal{M}_{\rm EW-DIS}\vert^2 ~\approx~ \frac{1}{(q^2 - M_V^2)^2} 
  \sum_{X,\lambda} 
  \langle f \vert V_{\lambda} f'\rangle
  \langle V_\lambda f'\vert f\rangle 
  ~
  \langle V_{\lambda} B \vert X \rangle
  \langle X \vert V_\lambda B\rangle.
\end{equation}

Returning to our derivation, after enjoining the outgoing $(\varepsilon^*_\mu)$ and incoming $(\varepsilon_\nu)$ polarizations vectors with the $f_{\lambda_A} \rightarrow f'_{\lambda_1} V_{\lambda_V}$ splitting amplitude and $V_{\lambda_V} B\to X$ scattering amplitude, respectively, one obtains the pair of MEs:
\begin{align}
  \mathcal{M}(f_{\lambda_A}\rightarrow f'_{\lambda_1} V_{\lambda_V}) &=  (-i\tilde{g}) \left[\overline{u}(p_1,\lambda_1)\not\!\varepsilon^*(q,\lambda_V) \left(g_L^f P_L + g_R^f P_R\right) u(p_A,\lambda_A)\right],
  \label{eq:mexx_splitting}
    \\
  \mathcal{M}(V_{\lambda_V} B\to X) &= \varepsilon_\nu(q,\lambda_V)\cdot \mathcal{M}^\nu(V_{\lambda_V} B\to X).
    \label{eq:mexx_hard}
\end{align}
For fixed polarizations of \textit{external} particles $f, f',$ and $V$,  the ME for the process in Eq.~\eqref{eq:ewDIS} can be factorized into two spin-correlated sub-amplitudes:
\begin{align}
 \mathcal{M}_{\rm EW-DIS}\Big\vert_{q^2 \to M_V^2}^{\lambda_A,\lambda_1,\lambda_V} 
 \approx \frac{i}{(q^2 - M_V^2)}  \mathcal{M}(f_{\lambda_A}\rightarrow f'_{\lambda_1} V_{\lambda_V}) ~\mathcal{M}(V_{\lambda_V}B\to X).
     \label{eq:mexx_factor}
 \end{align}
 Immediately, one can write at the squared ME level: 
\begin{eqnarray}
\sum_X \vert  \mathcal{M}_{\rm EW-DIS} \vert^2 \Big\vert_{q^2 \to M_V^2}^{\lambda_A,\lambda_1,\lambda_V} 
\approx \frac{1}{(q^2 - M_V^2)^2} \vert \mathcal{M}(f_{\lambda_A}\to f'_{\lambda_1} V_{\lambda_V}) \vert^2 \vert \mathcal{M}(V_{\lambda_V}B\to X)\vert^2.
\label{eq:mesq_factor}
\end{eqnarray}
This indicates that if $V$ goes on-shell and its mass is small compared to the scattering scale, then the probability density for $f_{\lambda_A} B \to f'_{\lambda_1} X$, when mediated by the space-like exchange of $V_{\lambda_A}$,  can be approximated as the spin-correlated product of the $f_{\lambda_A}\to f'_{\lambda_1} V_{\lambda_V}$ and $V_{\lambda_V} B\to X$ probability densities.


\subsubsection*{Helicity-dependent Collinear Splitting Amplitudes}
To enact the on-shell condition such that Eqs.~\eqref{eq:mexx_factor} and \eqref{eq:mesq_factor} are valid, we work in the kinematic configuration where $V$ and $f'$ are emitted at shallow angles in $f\to f' V$ splitting, i.e., the collinear limit.
Formally, this involves working to leading order in the parameter $\lambda\equiv p_T/E_A\ll1$, where $\vec{p}_T$ is the transverse momentum 2-vector of $f'(p_1)$ and $E_A$ is the energy of $f(p_A)$.
Defining $z=E_V/E_A$ as the energy fraction carried by  $V(q)$ in the hard scattering frame, we can parameterize the momenta of the process in Eq.~\eqref{eq:ewDIS} by
\begin{subequations}
\begin{align}
p_A^\mu 	&= E_A(1,0,0,+1), \qquad p_B^\mu = E_A(1,0,0,-1),\\
p_1^\mu 	&= \left( (1-z)E_A, ~\vec{p}_T, ~(1-z)E_A-\frac{p_T^2}{2(1-z)E_A}   \right),\\
		&\equiv \left( (1-z)E_A, ~\vert \vec{p}_1\vert \sin\theta_1\cos\phi_1, ~\vert \vec{p}_1\vert \sin\theta_1\sin\phi_1, ~(1-z)E_A-\frac{p_T^2}{2(1-z)E_A}   \right),	\\
q^\mu 	&= p_A^\mu - p_1^\mu 
		= \left( zE_A, ~-\vec{p}_T,  ~z E_A+\frac{p_T^2}{2(1-z)E_A}   \right)\\
		&\equiv \left( zE_A, ~\vert \vec{q}\vert \sin\theta_V\cos\phi_V, ~\vert \vec{q}\vert \sin\theta_V\sin\phi_V,  ~\vert \vec{q}\vert \cos\theta_V   \right).
\end{align}
\label{eq:ewa_kinematics}
\end{subequations}
For the momenta above,  $\sqrt{s_{AB}}=2E_A$ is the scale at which the $f+B\to f'+X$ process occurs. 
While explicit computation of individual invariant masses leads to 
\begin{equation}
p_A^2 = 0, \quad p_1^2 = -\frac{p_T^4}{4(1-z)^2E_A^2}, \quad q^2 = - \frac{p_T^2}{(1-z)} - \frac{p_T^4}{4(1-z)^2E_A^2}, \quad
\label{eq:ewa_limits}
\end{equation}
truncating wide-angle contributions at $\mathcal{O}(\lambda^2)$ results in the following approximations:
\begin{equation}
p_1^2 \approx 0, 
\quad
 q^2 \approx - \frac{p_T^2}{(1-z)},
\quad
\vert \vec{p}_1\vert^2\approx (1-z)^2E_A^2, 
 \quad
 \frac{\sin^2\theta_V}{\sin^2\theta_1}\approx\frac{(1-z)^2}{z^2}.
\end{equation}
Intuitively, this indicates that external  particles are massless or approximately massless (to $\mathcal{O}(\lambda^2)$). The internal $V$, on the  other hand, carries a virtuality $\sqrt{-q^2} \sim p_T$  much smaller than the  incoming energy $E_A$. As $V$ is also nearly on-shell, the scaling relationship $M_V \sim \sqrt{-q^2}\ll E_A$ must consistently hold.
Implicitly, we also work in the domain where the energy fraction $z$ is far from its boundaries at $z=0$ (the high-energy regime) and $z=1$ (the threshold regime), which would otherwise necessitate  resummation.

As $f,f'$ are massless, the only non-zero $f_{\lambda_A}(p_A)\to f'_{\lambda_1}(p_1)$ currents (as defined in Eq.~\eqref{eq:splitting_current_gen}) involving a vector emission  are those that conserve helicity, namely
\begin{align}
J^\mu(\lambda_A=L,\lambda_1=L) 	&= -i2\tilde{g} g_L^f \sqrt{E_A E_1} \left(\cos\frac{\theta_1}{2}, e^{i\phi_1}\sin\frac{\theta_1}{2}, -ie^{i\phi_1}\sin\frac{\theta_1}{2},\cos\frac{\theta_1}{2} \right), \\
J^\mu(\lambda_A=R,\lambda_1=R) 	&= -i2\tilde{g} g_R^f \sqrt{E_A E_1} \left(\cos\frac{\theta_1}{2}, e^{-i\phi_1}\sin\frac{\theta_1}{2}, ie^{-i\phi_1}\sin\frac{\theta_1}{2},\cos\frac{\theta_1}{2} \right). \qquad
\end{align}
Employing the transverse polarization vector in Eq.~\eqref{eq:formalism_polT} for $V_{\lambda_V=\pm}$, we obtain as the helicity  amplitude $(\mathcal{M})$ for the $f_L \to f'_L V_+$ splitting process 
\begin{align}
\mathcal{M}(f_L \to f'_L V_+) 	&= J^\mu(\lambda_A=L,\lambda_1=L)  \cdot \varepsilon_\mu^*(\lambda_V=+)  \\
						&= i2\sqrt{2} \tilde{g} g_L^f   E_A \sqrt{1-z} \cos\left(\frac{\theta_1+\theta_V}{2}\right) \sin\left(\frac{\theta_V}{2}\right) \\
						&= i \sqrt{2} \tilde{g} g_L^f   E_A \sqrt{1-z} ~\theta_V + \mathcal{O}\left(\theta_1^2, ~\theta_V^2\right)  \\
						&\approx i \sqrt{2} \tilde{g} g_L^f \frac{p_T\sqrt{1-z}}{z}.
\label{eq:me_fL_fLVp_splitting}						
\end{align}
In the second line of the above, we evaluated the ME exactly. In the third, we expanded the angular dependence to lowest order in the opening angles, and in the final line substituted $\theta_V$ for $p_T$.
Repeating the same steps for the  $f_L \to f'_L V_-$ splitting process, we obtain
\begin{align}
\mathcal{M}(f_L \to f'_L V_-) 	&= J^\mu(\lambda_A=L,\lambda_1=L)  \cdot \varepsilon_\mu^*(\lambda_V=-)  \\
						&= -i2\sqrt{2} \tilde{g} g_L^f   E_A \sqrt{1-z} \sin\left(\frac{\theta_1+\theta_V}{2}\right) \cos\left(\frac{\theta_V}{2}\right) \\
						&= -i \sqrt{2} \tilde{g} g_L^f   E_A \sqrt{1-z} ~(\theta_1+ \theta_V) + \mathcal{O}\left(\theta_1^2, ~\theta_V^2\right)  \\
						&\approx -i \sqrt{2} \tilde{g} g_L^f \frac{p_T}{z\sqrt{1-z}}.
\label{eq:me_fL_fLVm_splitting}												
\end{align}
Note  that  the Dirac equation for massless particles implies the following  orthogonality:
\begin{equation}
J^\mu(p_A,p_1)\cdot (p_A - p_1)_\mu = (-i \tilde{g}) \left[\overline{u}(p_1,\lambda_1) (\not\!\! p_A - \not\!\! p_1)\left(g_L^f P_L + g_R^f P_R\right) u(p_A,\lambda_A)\right] = 0.
\end{equation}
Employing this and the longitudinal polarization vector in Eq.~\eqref{eq:formalism_polA} for $V_{\lambda_V=0}$, then the helicity amplitude for the  $f_L \to f'_L V_0$ splitting process is 
\begin{align}
\mathcal{M}(f_L \to f'_L V_0) 	&= J^\mu(\lambda_A=L,\lambda_1=L)  \cdot \varepsilon_\mu^*(\lambda_V=0) \\
						&=   J^\mu(\lambda_A=L,\lambda_1=L)  \cdot \tilde{\varepsilon}_\mu^*  \\
						&= i4  \tilde{g} g_L^f \frac{M_V E_A \sqrt{1-z}}{z E_A + \sqrt{z^2E_A^2 - M_V^2}} \cos\left(\frac{\theta_V}{2}\right) \cos\left(\frac{\theta_1+\theta_V}{2}\right) \\
						&= i2  \tilde{g} g_L^f \frac{M_V \sqrt{1-z}}{z} + \mathcal{O}\left(\theta_1^2, ~\theta_V^2, ~\theta_1\theta_V, ~\frac{M_V^2}{E_V^2}\right).
\label{eq:me_fL_fLV0_splitting}
\end{align}
Importantly, the amplitudes for a transversely polarized $V$ exhibit a dependence on $p_T$, whereas the dependence is on $M_V$ for a longitudinally polarized $V$. As is well-documented throughout the literature, this distinction leads to qualitative differences in scale evolution for $\lambda_V=\pm$ and $\lambda_V=0$ states.
Moreover, as $\mathcal{M}(f_L \to f'_L V_0)$ vanishes as $(M_V/E_V)\to0$, one can interpret the forward emission of $V_0$ bosons as a ``beyond-twist-2'' phenomenon.

By parity inversion, the helicity amplitudes for the RH $f_R \to f'_R$ currents are: 
\begin{align}
\mathcal{M}(f_R \to f'_R V_+) 	&= J^\mu(\lambda_A=R,\lambda_1=R) \cdot \varepsilon_\mu^*(\lambda_V=+)  \\
						&= -\left(\frac{g_R^f}{g_L^f}\right) \mathcal{M}(f_L \to f'_L V_-),
\\
\mathcal{M}(f_R \to f'_R V_-) 	&= J^\mu(\lambda_A=R,\lambda_1=R) \cdot \varepsilon_\mu^*(\lambda_V=-)   \\ 
						&= -\left(\frac{g_R^f}{g_L^f}\right) \mathcal{M}(f_L \to f'_L V_+),
\\
\mathcal{M}(f_R \to f'_R V_0) 	&= J^\mu(\lambda_A=R,\lambda_1=R) \cdot \varepsilon_\mu^*(\lambda_V=0)  \\ 
						&= \left(\frac{g_R^f}{g_L^f}\right) \mathcal{M}(f_L \to f'_L V_0).
\end{align}
Explicit computation reveals that the relative minus sign in the transverse polarization cases can be traced to the relative phase difference that defines the $\lambda_V=\pm$  polarization vectors.
(This is also evident by the relative positive sign in the longitudinal case.)  
At the level of squared MEs $(\vert\mathcal{M}\vert^2)$, the above splitting helicity amplitudes are summarized in Table~\ref{tab:ewa_squaredME}.
As discussed under Eq.~\eqref{eq:splitting_current_gen}, the helicity amplitudes here neglect the complication of $f,f'$ carrying color.
We now account for this by nothing that at the squared-ME level, the $\delta_{IJ}$ in Eq.~\eqref{eq:ewa_colorFR} is squared and for quarks (leptons) sums to
\begin{equation}
{\rm SU}(2)_L \to {\rm SU}(3)_c\otimes{\rm SU}(2)_L \quad:\quad \tilde{g}^2 \to \tilde{g}^2\times\delta_{IJ}\delta_{IJ} =  \tilde{g}^2\times\delta_{II} =  \tilde{g}^2 \times N_c ~ (1).
\label{eq:ewa_colorSum}
\end{equation}

\begin{table}[!t]
\begin{center}
\resizebox{\textwidth}{!}{
\renewcommand*{\arraystretch}{0.95}
\begin{tabular}{c|c|c|c}
\hline\hline
Helicity Configuration	&	$\vert \mathcal{M}(f_{\lambda_A} \to f'_{\lambda_1} V_{\lambda_V}) \vert^2$ 		& Helicity Configuration	&	$\vert \mathcal{M}(f_{\lambda_A} \to f'_{\lambda_1} V_{\lambda_V}) \vert^2$ 		\\
\hline\hline
$f_L \to f'_L V_+$	&	$2 \left(\tilde{g} g_L^f\right)^2 \frac{p_T^2(1-z)}{z^2}$	&	$f_R \to f'_R V_+$	& 	$\left(\frac{g_R^f}{g_L^f}\right)^2 \vert \mathcal{M}(f_L \to f'_L V_-) \vert^2$ \\
$f_L \to f'_L V_-$	&	$2 \left(\tilde{g} g_L^f\right)^2 \frac{p_T^2}{z^2(1-z)}$	&	$f_R \to f'_R V_-$	&	$\left(\frac{g_R^f}{g_L^f}\right)^2 \vert \mathcal{M}(f_L \to f'_L V_+) \vert^2$ \\
$f_L \to f'_L V_0$	&	$4 \left(\tilde{g} g_L^f\right)^2  \frac{M_V^2 (1-z)}{z^2}$	&	$f_R \to f'_R V_0$	&	$\left(\frac{g_R^f}{g_L^f}\right)^2 \vert \mathcal{M}(f_L \to f'_L V_0) \vert^2$ \\
\hline\hline
\end{tabular}
}
\caption{
The squared helicity amplitude for the $f_{\lambda_A} \to f'_{\lambda_1} V_{\lambda_V}$ process in the collinear limit for each non-vanishing helicity permutation, and assuming the coupling normalization of table \ref{tab:ewa_coup}.
}
\label{tab:ewa_squaredME}
\end{center}
\end{table}

\subsection{Phase Space Decomposition}\label{app:phaseSpace}
For the $f(p_A) + B(p_B) \to f'(p_1)  + X(p_X)$ process, where $X$ is some $n_X$-body final-state system, the phase space volume of the $(n_X+1)$-body system can be organized to isolate the one-body phase space for $f'$. In doing so, we can factorize its contribution from the phase space for the hard $B+V\to X$ sub-process and remain inclusive with respect to the kinematics of $f'$. Noting the definition of $V$'s momentum $q=p_A-p_1$, decomposing the momentum of $X$ over its $n_X$ constituents,  and expressing the phase space volume for the $(p_A+p_B)$-system in terms of the $(q+p_B)$-system, we obtain 
\begin{eqnarray}
dPS_{n_X+1}(p_A+p_B ; p_1, \{ p_k\}) 	&=& (2\pi)^4 \delta\left(p_A + p_B - p_1 - \sum_{k=2}^{n_X+1} p_k\right)\prod^{n_X+1}_{k=1}\frac{d^3 p_k}{(2\pi)^3 2E_k} \\
								&=& (2\pi)^4 \delta\left(q + p_B  - \sum_{k=2}^{n_X+1} p_k\right)\frac{d^3 p_1}{(2\pi)^3 2E_1} \prod^{n_X+1}_{k=2}\frac{d^3 p_k}{(2\pi)^3 2E_k} \qquad \\
								&=& dPS_{n_X}(q + p_B; \{ p_k\}) \times \frac{d\phi_{f'} ~dz ~dq^2 }{4(2\pi)^3}.
\end{eqnarray}
In the last line, we exploited the limits of Eq.~\eqref{eq:ewa_limits}, which allows the one-body phase space for $f'$ to be written in terms of either evolution variable $p_T^2$ or $q^2$. Explicitly, this is 
\begin{equation}
\frac{d^3 p_1}{(2\pi)^3 2E_1} = 
(-1)\frac{d\phi_{f'} ~dz ~dp_T^2 }{4(2\pi)^3(1-z)} =
\frac{d\phi_{f'} ~dz ~dq^2 }{4(2\pi)^3}.
\end{equation}

\subsection{Transverse Momentum-Dependent Distribution Functions for EW Bosons}
\label{app:tmd}

To build a set of splitting functions that can be used with both spin-averaged and helicity-polarized MEs, we follow the formalism of Ref.~\cite{BuarqueFranzosi:2019boy}
and consider the $f + B \to f'  + X$ scattering process when the external states $f$ and $f'$ are in definite helicity states $\lambda_A$ and $\lambda_1$ but all other external states are unpolarized.
(This implies that helicities are averaged and summed for $B$ and $X$.)
We work in the rest frame of $X$, which one can eventually identify as the hard-scattering frame. In the collinear approximation, this frame is related to the beam c.m.~frame by a boost along the $z$ axis. For massless objects, which travel on the light cone, such longitudinal boosts cannot change the helicity; for massive objects, e.g., EW bosons $V$, helicity inversions are suppressed by factors of $(M_V/E_V)$, or some power thereof, and are strongly suppressed in our kinematic limit.
Using the results of App.~\ref{app:phaseSpace}, the semi-polarized, $2\to n_X+1$ cross section is subsequently given by
\begin{align}
\sigma\left(f_{\lambda_A} + B \to f'_{\lambda_1}  + X\right) 	= \int dPS_{n_X+1}(p_A+p_B; p_1, \{ p_k\})  &~ \frac{d\sigma_{fB}}{dPS_{n_X+1} } \Bigg\vert^{\{\lambda\}}  \\
					= \int \frac{d\phi_{f'} ~dz ~dq^2 }{4(2\pi)^3} \times \int dPS_{n_X}(q+p_B; \{ p_k\}) &~\frac{d\sigma_{fB}}{dPS_{n_X+1} }\Bigg\vert^{\{\lambda\}}, 
\label{eq:ewa_xsec_polar}					
\end{align}
where the color-averaged but helicity-dependent, totally differential cross section is
\begin{equation}
\frac{d\sigma_{fB}}{dPS_{n_X+1} }\Bigg\vert^{\{\lambda\}} = \frac{1}{2s_{AB} N_c^f N_c^B}\sum_{\rm dof} \Big\vert  \mathcal{M}_{\rm EW-DIS} \Big\vert^2 \Bigg\vert^{\{\lambda\}}.
\end{equation}

In this  expression, the summation runs over all extraneous dof, including color and possible multiplicities of $X$. $N_c^f$ and $N_c^B$ are the color factors for $f$ and $B$. Working now in the collinear limit, fixing the polarization of $V$ to be $\lambda_V$, and using  the factorized squared ME of Eq.~\eqref{eq:mesq_factor}, we rewrite the helicity-dependent, differential cross section as
\begin{align}
\frac{d\sigma_{fB}}{dPS_{n_X+1} }\Bigg\vert^{\{\lambda\}} &= \frac{1}{2s_{AB} N_c^f N_c^B}  \frac{1}{(q^2 - M_V^2)^2}  \nonumber\\
& \qquad~\qquad \times \sum_{\rm dof} \Big\vert \mathcal{M}(f_{\lambda_A}\to f'_{\lambda_1} V_{\lambda_V}) \Big\vert^2 \Big\vert \mathcal{M}(V_{\lambda_V}B_{\lambda_B}\to X_{\{\lambda_X\}}) \Big\vert^2 \\
&=
\frac{z}{(q^2 - M_V^2)^2}  ~ \Big\vert \mathcal{M}(f_{\lambda_A}\to f'_{\lambda_1} V_{\lambda_V}) \Big\vert^2   \nonumber\\
& \qquad~\qquad \times \frac{1}{2\lambda^{1/2}(Q^2,M_V^2,0) N_c^V N_c^B} \sum_{\rm dof} \Big\vert \mathcal{M}(V_{\lambda_V}B_{\lambda_B}\to X_{\{\lambda_X\}}) \Big\vert^2
\end{align}

In reaching the second line we exploited several observations.
First is that in the kinematic limit in which we are working, the invariant masses of the $(VB)$-system $(Q)$ and $(fB)$-system are related by $Q^2(1-M_V^2/Q^2)=\lambda^{1/2}(Q^2,M_V^2,0)\approx z s_{AB}$. Here, $\lambda(x,y,z)$ is the K\"allen function defined just below Eq.~\eqref{eq:partonXSec}. Second is that $f\to f' V$ splitting is a color-singlet process and that we are free to introduce color-averaging factors of $1/N_c^V=1$  for the $V+B\to X$ sub-process. This also implies that the squared amplitude for $f\to f' V$ passes through the summation after color indices have been counted  (see Eq.~\eqref{eq:ewa_colorSum}), i.e.,
\begin{equation}
\sum_{\rm color} \Big\vert \mathcal{M}(f_{\lambda_A}\to f'_{\lambda_1} V_{\lambda_V}) \Big\vert^2 = N_c^f \Big\vert \mathcal{M}(f_{\lambda_A}\to f'_{\lambda_1} V_{\lambda_V}) \Big\vert^2.
\end{equation}
Consequentially, we can consistently define the color-averaged but  spin-dependent, total and differential cross sections for the (partonic) $V+B\to X$  sub-process at $Q^2$ to be
\begin{align}
\hat{\sigma}\left(V_{\lambda_V} + B \to f'_{\lambda_1}  + X\right) 	&= \int dPS_{n_X}(q+p_B; \{ p_k\})  ~ \frac{d\hat{\sigma}_{VB}}{dPS_{n_X} } \Bigg\vert^{\{\lambda\}}, \quad\text{and}\\
\frac{d\hat{\sigma}_{VB}}{dPS_{n_X} }\Bigg\vert^{\{\lambda\}} &=   \frac{1}{2Q^2 N_c^V N_c^B} \sum_{\rm dof} \Big\vert \mathcal{M}(V_{\lambda_V}B\to X) \Big\vert^2.
\end{align}

Immediately, this allows us to express the total cross section for the full $f +  B \to f' + X$, i.e., Eq.~\eqref{eq:ewa_xsec_polar}, when mediated in the collinear limit by a sufficiently long-lived $V_\lambda$  as
\begin{align}
\sigma\left(f_{\lambda_A} + B \xrightarrow{V_{\lambda_V}} f'_{\lambda_1}  + X \right) =
\sum_{\lambda_V} &
 \int \frac{d\phi_{f'} ~dz ~dq^2 }{4(2\pi)^3} 
 \frac{z}{(q^2 - M_V^2)^2} \Big\vert \mathcal{M}(f_{\lambda_A}\to f'_{\lambda_1} V_{\lambda_V}) \Big\vert^2   \quad\nonumber\\
& \times
 \hat{\sigma} \left(V_{\lambda_V} + B \to f'_{\lambda_1}  + X \right)
\label{eq:ewa_xsec_polar_col}					
\\ 
\equiv \sum_{\lambda_V}
\int_{z_0}^1 dz \int^{0}_{-\mu_f^2} dq^2 ~\tilde{\mathcal{F}}_{V_{\lambda_V}/f_{\lambda_A}}(z,q^2) 
& \times \hat{\sigma}\left(V_{\lambda_V} + B \to f'_{\lambda_1}  + X\right),
\label{eq:ewa_xsec_polar_VDpdf}
\\ 
\equiv \sum_{\lambda_V}
\int_{z_0}^1 dz \int_{0}^{\mu_f^2} dp_T^2 ~\tilde{\mathcal{H}}_{V_{\lambda_V}/f_{\lambda_A}}(z,p_T^2) 
& \times \hat{\sigma}\left(V_{\lambda_V} + B \to f'_{\lambda_1}  + X\right).
\label{eq:ewa_xsec_polar_TMDpdf}					
\end{align}
In Eqs.~\eqref{eq:ewa_xsec_polar_VDpdf} and \eqref{eq:ewa_xsec_polar_TMDpdf} we extracted all ME factors associated with the polarized $f_{\lambda_A}\to f'_{\lambda_1} V_{\lambda_V}$ splitting process and any dependence on the azimuth angle to define the unintegrated, helicity-dependent, virtuality-dependent parton density function (VD PDF)
\begin{equation}
\tilde{\mathcal{F}}_{V_{\lambda_V}/f_{\lambda_A}}(z,q^2) \equiv  \int \frac{d\phi_{f'}}{4(2\pi)^3} 
 \frac{z}{(q^2 - M_V^2)^2} \Big\vert \mathcal{M}(f_{\lambda_A}\to f'_{\lambda_1} V_{\lambda_V}) \Big\vert^2,
 \end{equation}
and analogously the helicity-dependent, transverse momentum-dependent (TMD) PDF 
\begin{equation}
\tilde{\mathcal{H}}_{V_{\lambda_V}/f_{\lambda_A}}(z,p_T^2) \equiv  \int \frac{d\phi_{f'}}{4(2\pi)^3}
 \frac{z(1-z)}{(p_T^2 + (1-z)M_V^2)^2} \Big\vert \mathcal{M}(f_{\lambda_A}\to f'_{\lambda_1} V_{\lambda_V}) \Big\vert^2.
 \end{equation}
Here and below, we follow Ref.~\cite{Maltoni:2012pa} and adopt the $~\tilde{~}~$ notation to stress that the distribution functions here are neither resummed nor renormalized. 
However, these distribution functions are related to the resummed distribution functions $\mathcal{F}, \mathcal{H}$ by perturbative corrections:
\begin{subequations}
\begin{align}
{\mathcal{F}}_{V_{\lambda_V}/f_{\lambda_A}}(z,q^2)
&= \tilde{\mathcal{F}}_{V_{\lambda_V}/f_{\lambda_A}}(z,q^2) +  \mathcal{O}\left((\alpha_W(q^2)\right)\ ,
\\
{\mathcal{H}}_{V_{\lambda_V}/f_{\lambda_A}}(z,p_T^2)
&= \tilde{\mathcal{H}}_{V_{\lambda_V}/f_{\lambda_A}}(z,p_T^2) +  \mathcal{O}\left((\alpha_W(p_T^2)\right)\ .
\end{align}
\end{subequations}
 
 Noting that $V$ must carry a minimum amount of energy $V+B\to X$ sub-process to kinematically proceed, the lower boundary of the energy fraction integral is simply 
 \begin{equation}
 z_0 = \min(z)=\min\left(Q^2\right)/s_{AB} = \left(\sum_{k=2}^{n_X+1}m_k\right)^2/s_{AB}> \frac{M_V^2}{s_{AB}}.
\end{equation}
Importantly, in Eqs.~\eqref{eq:ewa_xsec_polar_VDpdf} and \eqref{eq:ewa_xsec_polar_TMDpdf}, the positive-definite cutoff scale $\mu_f > M_V$ was introduced by hand to regulate the ultraviolet limit of the virtuality and transverse momentum integrals.
Intuitively, $\mu_f^2\ll Q^2$ denotes the phase space boundary (cutoff) for which $\vert q^2\vert, p_T^2 < \mu_f^2$ correspond to kinematics that justify a collinear expansion of ME; in the same way, $\vert q^2\vert\sim p_T^2 > \mu_f^2$ correspond to wide-angle kinematics.
As discussed in Sec.~\ref{sec:conclusions}, stipulating that $\mu_f \ll Q^2$ appears na\"ively at odds with common use of quark and gluon PDFs in pQCD. However, it must be stressed that in pQCD, PDFs are typically RG-evolved with the DGLAP evolution equations. This impacts the interpretation and scaling of $\mu_f$.

\subsection{Collinear Distribution Functions for EW Bosons}\label{app:collinearPDF}

Taking the squared MEs for $f_{\lambda_A}\to f'_{\lambda_1} V_{\lambda_V}$ splitting in Table \ref{tab:ewa_squaredME} and evaluating the azimuth integral will generate explicit forms for the unintegrated PDFs $\tilde{\mathcal{F}}_{V_{\lambda_V}/f_{\lambda_A}}(z,q^2)$ and $\tilde{\mathcal{H}}_{V_{\lambda_V}/f_{\lambda_A}}(z,p_T^2)$. Importantly, these are functions of momentum fraction $z$ and the respective evolution variable. From these distributions, one can then construct collinear density functions by performing the evolution integral over $q^2$ in Eq.~\eqref{eq:ewa_xsec_polar_VDpdf} and over $p_T^2$ Eq.~\eqref{eq:ewa_xsec_polar_TMDpdf}:
\begin{align}
    \tilde{f}_{V_{\lambda_V}/f_{\lambda_A}}(z,\mu_f^2) & \equiv \int^{0}_{-\mu_f^2} dq^2 ~\tilde{\mathcal{F}}_{V_{\lambda_V}/f_{\lambda_A}}(z,q^2)
    \\
    & = \int^{0}_{-\mu_f^2} dq^2 \int \frac{d\phi_{f'}}{4(2\pi)^3} 
 \frac{z}{(q^2 - M_V^2)^2} \Big\vert \mathcal{M}(f_{\lambda_A}\to f'_{\lambda_1} V_{\lambda_V}) \Big\vert^2,
    \\
    \tilde{h}_{V_{\lambda_V}/f_{\lambda_A}}(z,\mu_f^2)  &\equiv
    \int_{0}^{\mu_f^2} dp_T^2 ~\tilde{\mathcal{H}}_{V_{\lambda_V}/f_{\lambda_A}}(z,p_T^2)
    \\
    & = \int_{0}^{\mu_f^2} dp_T^2 \int \frac{d\phi_{f'}}{4(2\pi)^3}
 \frac{z(1-z)}{(p_T^2 + (1-z)M_V^2)^2} \Big\vert \mathcal{M}(f_{\lambda_A}\to f'_{\lambda_1} V_{\lambda_V}) \Big\vert^2.
\end{align}
Importantly, these are functions of $z$ and the ultraviolet regulator $\mu_f$, which at this order of perturbation theory has a physical interpretation. We report in  Eqs.~\eqref{eq:formalism_pdfs_q2} and \eqref{eq:formalism_pdfs_pT} respectively, the full expressions for the $q^2$ PDFs $(\tilde{f})$ and the $p_T^2$ PDFs $(\tilde{h})$ of polarized weak bosons from high-energy charged leptons  in the collinear limit.  Using collinear PDFs, we can rewrite the total cross section for the full $f +  B \to f' + X$, i.e., Eq.~\eqref{eq:ewa_xsec_polar}, as
\begin{align}
\sigma\left(f_{\lambda_A} + B \to f'_{\lambda_1}  + X \right) 
\qquad ~ \qquad & \nonumber\\ 
 = \sum_{\lambda_V}
\int_{z_0}^1 dz ~ \tilde{f}_{V_{\lambda_V}/f_{\lambda_A}}(z,\mu_f^2) 
& \times \sigma\left(V_{\lambda_V} + B  \to f'_{\lambda_1}  + X \right)
\nonumber\\
&\underset{\rm Power~and~Logarithmic~Corrections}{\underbrace{
+ \mathcal{O}\left(\frac{p_T^2}{Q^2}\right)
+ \mathcal{O}\left(\frac{M_{V}^2}{Q^2}\right)
+\mathcal{O}\left(\log \frac{\mu_f^2}{M_V^2}\right)}},
\label{eq:ewa_xsec_polar_q2xpdf}
\\ 
 = \sum_{\lambda_V}
\int_{z_0}^1 dz ~ \tilde{h}_{V_{\lambda_V}/f_{\lambda_A}}(z,\mu_f^2) 
& \times \sigma\left(V_{\lambda_V} + B \to f'_{\lambda_1}  + X \right) 
\nonumber\\
&+ \mathcal{O}\left(\frac{p_T^2}{Q^2}\right)
+ \mathcal{O}\left(\frac{M_{V}^2}{Q^2}\right)
+\mathcal{O}\left(\log  \frac{\mu_f^2}{M_V^2}\right).
\label{eq:ewa_xsec_polar_pT2pdf}					
\end{align} 
In these expressions, we have made explicit the power and logarithmic  corrections that parameterize the uncertainty of collinear factorization of weak bosons in scattering computations.
The first correction is associated with how collinear $V_{\lambda_V}$ and $f'_{\lambda_1}$ are to the beam axis (or $f_{\lambda_A}$)
and originate from expanding the $f_{\lambda_A} \to V_{\lambda_V} f'_{\lambda_1}$
splitting amplitudes as seen in 
Eqs.~\eqref{eq:me_fL_fLVp_splitting}, \eqref{eq:me_fL_fLVm_splitting}, and \eqref{eq:me_fL_fLV0_splitting}.
The second correction is associated with the accuracy of the Goldstone Equivalent Theorem and originate from expanding the $f_{\lambda_A} \to V_{0} f'_{\lambda_1}$ amplitude as seen exclusively in Eq.~\eqref{eq:me_fL_fLV0_splitting}.
The third correction is associated with the $\mu_f$ dependence 
introduced to regulate the evolution integrals in Eqs.~\eqref{eq:ewa_xsec_polar_VDpdf} and \eqref{eq:ewa_xsec_polar_TMDpdf}.

For the high-energy process $\mu^+ \mu^- \to \mathcal{F} + X$, where the production of $\mathcal{F}$ is mediated by high-energy $V_{\lambda_A}V_{\lambda_B}$ scattering, and $X$ is some arbitrary, inclusive final state, we are able to extend the above results and write the scattering formula
\begin{align}
\sigma(\mu^+ \mu^- \to \mathcal{F} + X)
&=~
\tilde{f} \otimes \tilde{f} \otimes \hat{\sigma}
+ {\rm Power~and~Logarithmic~Corrections}
\\
&=~
\sum_{V_{\lambda_A},V'_{\lambda_B}} \int_{\tau_0}^1 d\xi_1
\int_{\tau_0/\xi_1}^1 d\xi_2
\int dPS_{n} ~ 
\nonumber\\
& ~\times~
\tilde{f}_{V_{\lambda_A}/\mu^+}(\xi_1,\mu_f)\
\tilde{f}_{V'_{\lambda_B}/\mu^-}(\xi_2,\mu_f)
\nonumber \\ 
& ~\times~ \frac{d\hat{\sigma}(V_{\lambda_A}V'_{\lambda_B} \to \mathcal{F})}{dPS_{n}}
\nonumber\\
&
~ + ~ \mathcal{O}\left(\frac{p_{T,l_k}^2}{M_{VV'}^2}\right)
+ \mathcal{O}\left(\frac{M_{V_k}^2}{M_{VV'}^2}\right)
+\mathcal{O}\left(\log  \frac{\mu_f^2}{M_{V_k}^2}\right)
.
\end{align}

\bibliography{madEVA_refs}

\end{document}